\PassOptionsToPackage{capitalize,noabbrev,nameinlink}{cleveref}
\PassOptionsToPackage{usenames,dvipsnames}{color}
\PassOptionsToPackage{numbers,compress,sort}{natbib}
\PassOptionsToPackage{final}{microtype}
\PassOptionsToPackage{scaled}{inconsolata}
\documentclass[sigconf,10pt]{acmart}

\newcommand{\paperTitle}{Designing Succinct Secondary Indexing Mechanism by Exploiting Column Correlations}
\newcommand{\shortTitle}{Succinct Secondary Indexing Mechanism}
\newcommand{\paperKeywords}{}
\newcommand{\paperAuthors}{Yingjun Wu, Jia Yu, Yuanyuan Tian, Richard Sidle, Ronald Barber}

\usepackage{enumerate}
\usepackage{amsmath}
\usepackage{amssymb}
\usepackage{boxedminipage}
\usepackage{xspace}
\usepackage{array}
\usepackage{epsfig}
\usepackage{calc}
\usepackage{multirow}
\usepackage{rotating}
\usepackage{enumitem}
\usepackage{tabularx}
\usepackage{balance}  
\usepackage{tikz}
\usepackage{wasysym }
\usepackage[hyphenbreaks]{breakurl}
\usepackage{xcolor}
\usepackage[font={small}]{caption}
\usepackage{alltt}
\usepackage{setspace}
\usepackage{etoolbox}
\usepackage{ctable} 
\usepackage[numbers,sort]{natbib}
\usepackage[capitalize,noabbrev,nameinlink]{cleveref}

\usepackage{graphicx}
\usepackage{subfig}
\usepackage{soul}
\usepackage{pifont}

\usepackage{makecell}

\hypersetup{
 pdfauthor = {\paperAuthors},
 pdftitle = {\paperTitle},
 pdfkeywords = {\paperKeywords},
 pdfborder={ 0 0 0 }
}

\usepackage{float}
\usepackage{bookmark}
\usepackage{amsmath,amsfonts,amssymb,stmaryrd}
\usepackage{latexsym}
\usepackage{multirow}
\usepackage{emptypage}
\usepackage{booktabs}
\usepackage{enumerate}
\usepackage{braket}
\usepackage[end]{algpseudocode}
\usepackage[ruled,linesnumbered,noend]{algorithm2e}
\usepackage{fancyvrb}

\usepackage[capitalize,nameinlink]{cleveref}

\Crefname{section}{Section}{Sections}
\Crefname{algocf}{Algorithm}{Algorithms}
\usetikzlibrary{
  decorations.pathreplacing,
  angles,
  quotes,
  arrows,
  positioning,
  shapes.symbols,
  shapes.callouts,
  patterns,
  fit
}

\DeclareCaptionType{copyrightbox}

\robustify{\cite}

\newcommand{\squishitemize}{
 \begin{list}{$\bullet$}
  { \setlength{\itemsep}{0pt}
     \setlength{\parsep}{3pt}
     \setlength{\topsep}{3pt}
     \setlength{\partopsep}{0pt}
     \setlength{\leftmargin}{1.95em}
     \setlength{\labelwidth}{1.5em}
     \setlength{\labelsep}{0.5em} } }

\newcounter{Lcount}
\newcommand{\squishlist}{
    \begin{list}{\arabic{Lcount}. }
   { \usecounter{Lcount}
        \setlength{\itemsep}{0pt}
        \setlength{\parsep}{3pt}
        \setlength{\topsep}{3pt}
        \setlength{\partopsep}{0pt}
        \setlength{\leftmargin}{2em}
        \setlength{\labelwidth}{1.5em}
        \setlength{\labelsep}{0.5em} } }

\newcommand{\squishend}{\end{list}}

\definecolor{todo-color}{rgb}{1,0,0}

\definecolor{yingjun-color}{rgb}{0,0,1}

\definecolor{jiayu-color}{rgb}{1,0,0.5}

\definecolor{comment-color}{rgb}{0.25,0.25,0.25}

\newcommand{\system}{\textsc{Hermit}\xspace}
\newcommand{\database}{\textsc{DBMS-X}\xspace}
\newcommand{\tree}{\textsc{TRS-Tree}\xspace}

\newcommand{\rev}[1]{{#1}}

\definecolor{my-blue}{RGB}{79,117,173}
\definecolor{my-green}{RGB}{93,166,108}
\definecolor{my-red}{RGB}{191,79,85}
\definecolor{my-red-2}{RGB}{187,28,43}
\definecolor{my-salmon}{RGB}{241,121,139}
\definecolor{my-purple}{RGB}{128,117,175}
\definecolor{bgblue}{RGB}{209,235,254}
\definecolor{bggreen}{RGB}{209,232,221}
\definecolor{bgred}{RGB}{254,212,210}

\begin{document}

\newcommand{\mail}[1]{\href{mailto:#1}{#1}}



\author{Yingjun Wu}
\affiliation{%
  \institution{IBM Research - Almaden}}
\email{yingjun.wu@ibm.com}

\author{Jia Yu}
\authornote{Work done during an internship at IBM Research - Almaden.}
\affiliation{%
  \institution{Arizona State University}}
\email{jiayu2@asu.edu}

\author{Yuanyuan Tian}
\affiliation{%
  \institution{IBM Research - Almaden}}
\email{ytian@us.ibm.com}

\author{Richard Sidle}
\affiliation{%
  \institution{IBM}}
\email{ricsidle@ca.ibm.com}

\author{Ronald Barber}
\affiliation{%
  \institution{IBM Research - Almaden}}
\email{rjbarber@us.ibm.com}


\title[\shortTitle]{\paperTitle}
\subtitle{(Extended Version)}

\begin{abstract}

Database administrators construct secondary indexes on data tables to 
accelerate 
query processing in relational database management systems (RDBMSs).
These indexes are built on top of the most frequently queried columns 
according to the data statistics.
Unfortunately, maintaining multiple secondary indexes in the same database can be 
extremely space consuming, causing significant performance
degradation due to the potential exhaustion of memory space.
In this paper, we demonstrate that there exist many opportunities
to exploit column correlations for accelerating data access.
We propose \system, a succinct secondary indexing mechanism for modern RDBMSs.
\system judiciously leverages the rich soft functional dependencies
hidden among columns to prune out redundant structures for
indexed key access.
Instead of building a complete index that stores every single 
entry in the key columns, 
\system navigates any incoming key access queries to an
existing index built on the correlated columns.
\rev{This is achieved through the Tiered Regression Search Tree (\tree), 
a succinct, ML-enhanced data structure 
that performs fast curve fitting to adaptively and dynamically capture both column correlations and outliers.}
We have developed \system in two different RDBMSs, and our 
extensive experimental study confirmed that \system can 
significantly reduce space consumption with limited performance 
overhead in terms of query response time and index maintenance time, especially when supporting complex range queries.


\end{abstract}

\maketitle
\section{Introduction}
\label{sec:introduction}

Modern relational database management systems (RDBMSs) support fast 
secondary indexes that help accelerate query processing in both transactional
and analytical workloads. These indexes, created either by database administrators
or automatically by query optimizers, are built on top of the most 
frequently queried columns, hence providing an efficient way to 
retrieve data tuples via these columns. 
However, managing multiple secondary indexes in the database can consume large amounts
of storage space, potentially causing severe performance degradation due to
the exhaustion of memory space. This problem is not uncommon especially 
in the context of modern main-memory RDBMSs, 
where memory space is a scarce resource.

Confronting this problem, researchers in the database community
have proposed various practical solutions to limit the space usage
for index maintenance.
From the database tuning perspective, 
some of the research works have introduced smart performance tuning 
advisors that can automatically select the most beneficial secondary indexes
given a fixed space budget~\cite{chaudhuri1997efficient,valentin2000db2,agrawal2000automated}. 
While satisfying the space constraints, these techniques essentially 
limit the number of secondary indexes built on the tables, consequently causing
poor performance for queries that lookup the unindexed columns.
From the structure design perspective,
a group of researchers has developed space-efficient index structures that consume 
less storage space compared to conventional indexes~\cite{graefe2011modern}.
These works either store only a subset of the column entries~\cite{stonebraker1989case} or use 
compression techniques to reduce space consumption~\cite{zhang2016reducing}.
However, such solutions save limited amount of space and can cause high overhead for 
lookup operations.

We attempt to address this problem in a third way. The main observation
that sparks our idea is that many columns in the data tables exhibit 
\textit{correlation} relations, or \textit{soft functional dependencies}, 
where the values of a column can be estimated by that of another column with 
approximation. Exploiting such a relation can greatly reduce the memory
consumption caused by secondary index maintenance.
Specifically, if we want to create an index on a column $M$ that is highly correlated 
with another column $N$ where an index has been built, we can simply construct a 
succinct, ML-enhanced data structure, called \textit{Tiered Regression Search Tree}, 
or \tree,
to capture the correlation between $M$ and $N$.
\tree exploits multiple simple statistical regression processes to 
fit the curve of the hidden correlation function. Different from existing machine learning-based indexing solutions, 
\tree efficiently handles inserts, deletes, and updates, 
\rev{and supports on-demand structure reorganization to re-optimize the index efficiency at system runtime.}
To perform a lookup query on $M$, the RDBMS retrieves a lookup range on $N$ from the 
newly constructed \tree and fetches the targeted tuples using $N$'s index.
We call this mechanism \system.

\system achieves competitive performance when supporting range queries, 
which are prevalent 
for secondary key column accesses.
It also presents a tradeoff between
computation and space consumption. While avoiding building a complete index
structure remarkably reduces space consumption, \system requires any incoming 
query to go through an additional hop before retrieving the targeted tuples.
However, as our experiments will show, this overhead does not substantially affect performance in practice, 
and it also brings in huge benefits when storage space is valuable and scarce, such as in main-memory RDBMSs.


We are not the first to exploit column correlations in RDMBSs. 
\rev{Several previous works have proposed correlation-based optimizations 
for query processing, database design, and data access. 
Different from these existing works, 
our work advocates a succinct \emph{ML-enhanced} tree-based structure, called 
\mbox{\tree}, that adaptively and dynamically
capture both complex correlations and outliers. 
Using \mbox{\tree}, we further show how \mbox{\system} 
exploits multiple correlations in both main-memory and disk-based RDBMSs.}


This paper is organized as follows.
\cref{sec:background} provides technical background.
\cref{sec:overview} gives an overview of \system's methodology. 
\cref{sec:index} presents the detailed design of \system's \tree structure,
and \cref{sec:design} shows how \system leverages \tree to perform 
tuple retrieval.
\cref{sec:discussion} provides detailed discussion on several issues. 
We report extensive experiment results in \cref{sec:evaluation}.
\cref{sec:related-work} reviews related works and \cref{sec:conclusions} concludes. \rev{We provide extended discussion on several issues in Appendix.}

\section{Background}
\label{sec:background}

In this section, we provide some technical background about index structures 
and column correlations in the context of RDBMSs.

\subsection{Index Structures}

Modern RDBMSs use secondary index structures to improve the speed of data retrieval
at the cost of additional writes and space consumption.
A secondary index is created either by a database administrator or automatically
by a query optimizer, and is built on top of one or more frequently accessed columns.
An index can be considered as a copy of the corresponding key columns organized 
in a format that provides fast mapping to the tuple identifiers, 
in terms of either tuple locations or primary keys~\cite{wu2017empirical}.
Maintaining multiple secondary indexes on the same database can be expensive, 
especially in
the context of main-memory RDBMSs.
This is confirmed by a recent study~\cite{zhang2016reducing} which showed that 
index maintenance can consume around 55\% of the total memory space.
Confronting this problem, researchers have proposed various 
space-efficient index structures to reduce space consumption.
In general, these works share two basic ideas: (1) using classic 
compression techniques such as Huffman encoding or dictionary encoding 
to reduce the index node size~\cite{graefe2011modern};
(2) storing only a subset of entries from the indexed columns to reduce 
the number of leaf nodes~\cite{stonebraker1989case}.
Despite the limited reduction in memory consumption, 
these techniques can incur
high overhead when processing lookup operations.

\subsection{Column Correlations}

A conventional RDBMS allows database administrators to set integrity constraints 
using SQL statements to express the functional dependencies among data columns.
These explicitly declared functional dependencies can be leveraged by query optimizers 
to provide a more accurate cost estimation during the query rewriting phase.  
In addition to these ``hard'' functional dependencies, 
modern query optimizers also attempt to explore ``soft'' functional dependencies 
to generate better query plans using the column correlation relations. 
Column correlations capture approximate dependencies, 
meaning that the value of a column can determine that of another approximately.
Following the definition in existing works, 
we define a column correlation relation 
as a triple $(M, N, Fn)$,
where $M$ and $N$ are data columns in the table exhibiting correlations, 
and $Fn$ is a \textit{correlation function} specifying how 
$N$'s values can be estimated from $M$. Besides simple
algebraic computation~\cite{brown2003bhunt} (e.g., $+$, $-$, $\times$, $/$) and
linear functions~\cite{gryz2001discovery,godfrey2001exploiting} (e.g., $N=\beta M+\alpha \pm \epsilon$),  
a correlation function $N=Fn(M)$ can be of any possible form.
This will allow us to capture the correlations 
in many modern database applications, 
such as environment monitoring (oxygen v.s. carbon dioxide),
stock market (Dow-Jones v.s. S\&P 500), 
and healthcare (glucagon v.s. insulin).
Existing RDBMSs have already exploited this kind of data characteristics to address 
system efficiency problems, including data compression, query rewriting, and database tuning~\cite{ilyas2004cords,kimura2009correlation,kimura2010coradd}. 
In the following sections, we will show how we can leverage correlations
to accelerate data access.

\section{Overview}
\label{sec:overview}

The correlation hidden among different columns in an RDBMS indicates 
a high similarity of their corresponding index structures. Observing that, we developed a succinct, 
yet fast secondary indexing mechanism, namely \system, 
which exploits the column correlations to answer queries.

To index a specified column $M$, \system requires two components: 
a succinct data structure called \textit{Tiered Regression Search Tree} (abbr., \tree) 
on the \textit{target} column $M$,
and a pre-existing complete index called \textit{host index} on the \textit{host} column $N$. 
\tree models the correlation between $M$ and $N$: 
it leverages a \textit{tiered regression} method to perform hierarchical curve fitting 
over the correlation function $Fn$ from $M$ to $N$, and uses a tree structure to index 
a set of regression functions 
each of which represents an approximate linear mapping from a value range of $M$ to that of $N$.

To process a query, \system runs a three-phase searching algorithm: 
(1) \tree search;
(2) host index search;
and (3) base table validation.
Specifically, \system uses the query predicate to search the \tree in order to retrieve 
the range mapping from $M$ to $N$. It then leverages the host index to find a set of candidate tuple identifiers. 
We note that this candidate set is approximate, and it contains false positives that fail to satisfy
the original predicates.
\system removes those false positives by directly validating the corresponding values on the base table.

\system also works for multi-column secondary indexes. Suppose that two columns $A$ and $M$ on a table are queried together frequently, 
so an index on $(A, M)$ is desirable. \system can utilize a host index on $(A, N)$ and the
correlation between $M$ and $N$, to answer queries on $A$ and $M$.

\rev{In the case where multi-column correlation exists (e.g., (W, X)->(Y,Z), although rarely detected by RDBMSs), \mbox{\system} can concatenate multiple keys and build \mbox{\tree} on them.}

We now use a running example to demonstrate how \system works.
Let us consider a data table \texttt{STOCK\_HISTORY} recording U.S. stock market
trading histories with four different columns: 
\texttt{TIME} (i.e., trading date), \texttt{DJ} (i.e., Dow Jones), \texttt{SP} (i.e., S\&P 500), and \texttt{VOL} (i.e., total trading volume). 
The database administrator has already created an index
on (\texttt{TIME}, \texttt{DJ}). 
Now she decides to create another index on (\texttt{TIME}, \texttt{SP}) due to the frequent occurrence of the queries like:

\begin{footnotesize}
\begin{Verbatim}[commandchars=\\\{\},codes={\catcode`$=3\catcode`_=8}]
\textbf{SELECT} * \textbf{FROM} STOCK\_HISTORY 
\textbf{WHERE} (TIME  \textbf{BETWEEN} ? \textbf{AND} ?) \textbf{AND} (SP \textbf{BETWEEN} ? \textbf{AND} ?)
\end{Verbatim}
\end{footnotesize}

On receiving the index creation statement, 
the RDBMS adopting \system first checks whether any column correlation involving \texttt{TIME} or \texttt{SP} has been detected via any correlation discovery algorithms.
If observing that the values in \texttt{SP} are highly correlated with 
those in \texttt{DJ} and that there is an existing index on (\texttt{TIME}, \texttt{DJ}),
the RDBMS then constructs a \tree to model the correlation mapping from 
\texttt{SP} to \texttt{DJ}. Given the query ranges $(T_{min}, T_{max})$ on column \texttt{TIME}  and $(S_{min}, S_{max})$ on column \texttt{SP},
\system first inputs the \texttt{SP} range $(S_{min}, S_{max})$ to the constructed \tree to fetch the corresponding 
range $(D_{min}, D_{max})$ on \texttt{DJ}. Then it searches the host index on (\texttt{TIME}, \texttt{DJ}), 
with \texttt{TIME} range $(T_{min}, T_{max})$  and \texttt{DJ} range $(D_{min}, D_{max})$, to find all the satisfying tuples.
To filter out false positives, the RDBMS reads the \texttt{SP} values from the base table and validates the correctness of the result.


\begin{figure}[t!]
    \centering
    \subfloat[Conventional index]{
        \includegraphics[width=0.49\columnwidth]
            {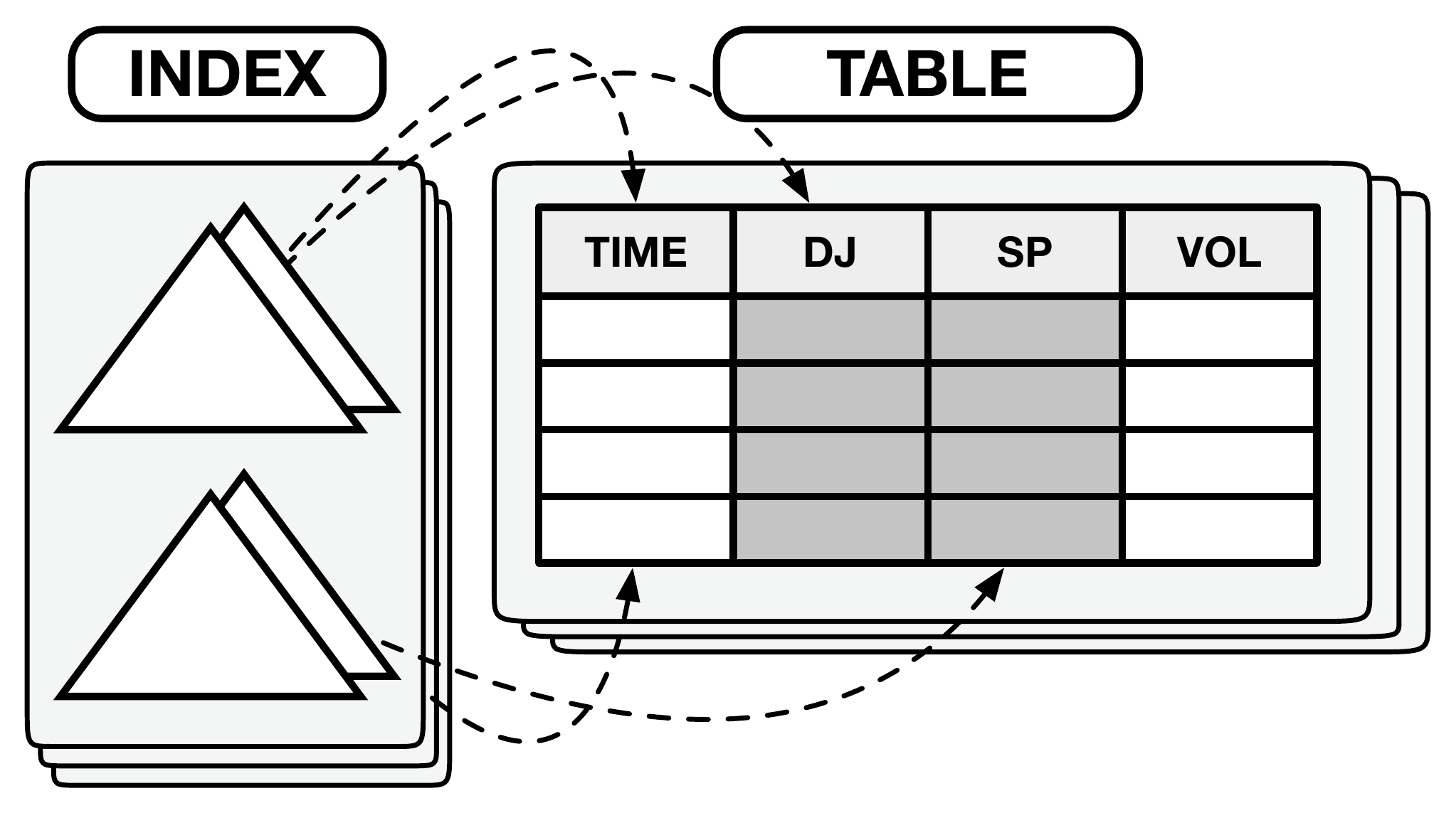}
        \label{figure:sindex-old}
    }
    \subfloat[\system]{
        \includegraphics[width=0.49\columnwidth]
            {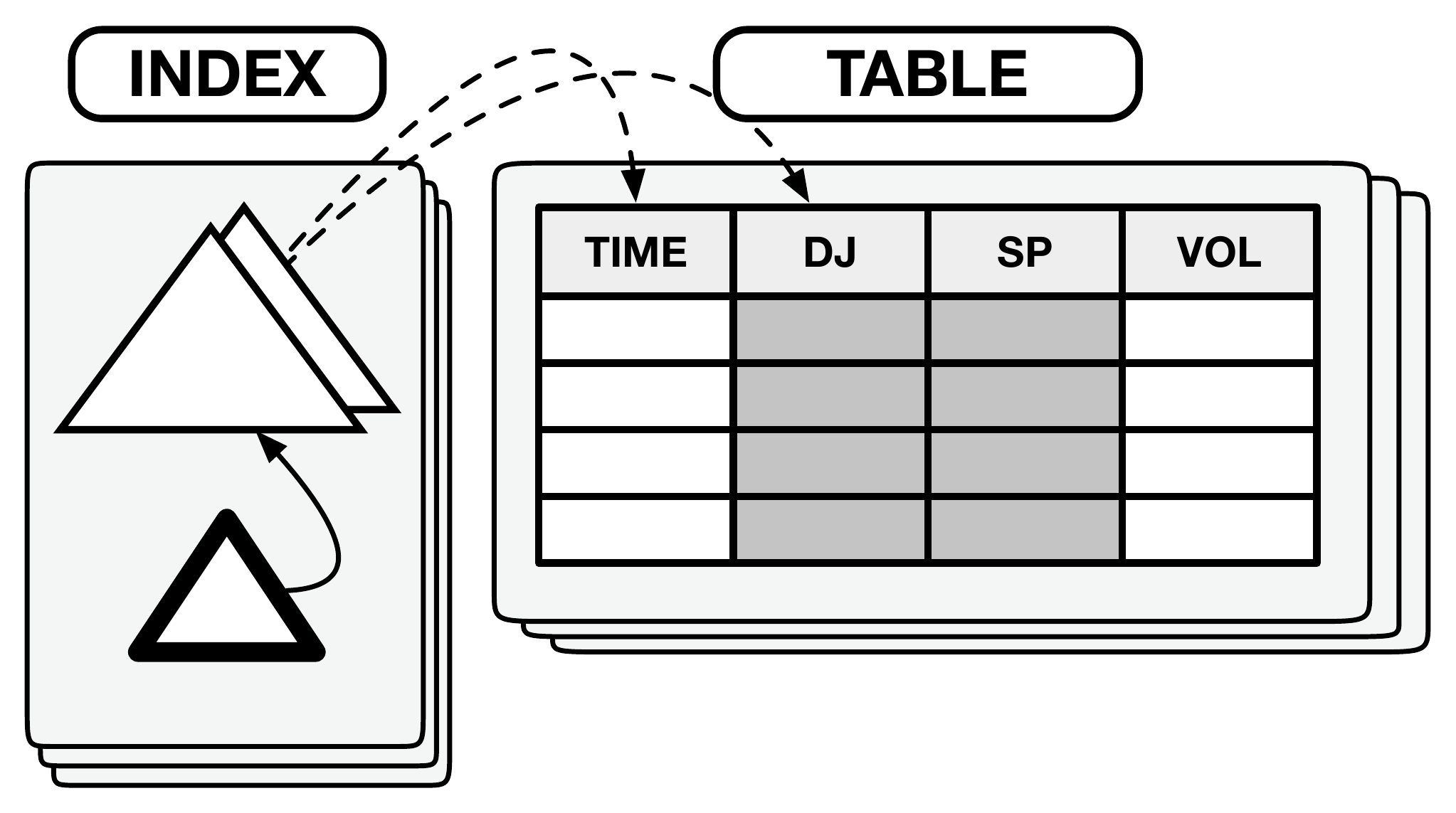}
        \label{figure:sindex-new}
    }
    \caption{
        A comparison between data retrieval via conventional secondary indexes and \system.
        (double triangles denotes conventional secondary index; small single 
        triangle denotes the proposed \tree structure)
    }
    \label{figure:comparison}
\end{figure}

\cref{figure:comparison} shows how \system is different from conventional secondary indexing mechanisms
when retrieving tuples in the running example. Unlike existing indexing techniques that 
provide direct accesses to the tuple identifiers, \system requires two-hop accesses. While
potentially causing higher access overhead for point queries, \system can achieve very competitive performance for range queries that are highly common for secondary indexes (as we will demonstrate in the experiments).
And, of course, \system can significantly reduce the space consumption for index maintenance.

\system can support
insert, delete, and update operations with correctness guarantees.
Due to its approximate characteristics, \system works best for range queries, 
which are quite common for secondary key columns, especially in data analytics. 
Furthermore, \system is extremely beneficial for main-memory RDBMSs, 
where memory space is scarce.


\section{\tree}
\label{sec:index}

\tree is a succinct tree structure that 
models data correlation between a target column $M$ and a host column $N$ 
within the same data table of a database. 
It leverages a tiered regression method to \textit{adaptively} and \textit{dynamically} 
perform the curve fitting over the 
correlation function $N=Fn(M)$. To be precise, \tree decomposes the complex curve-fitting 
problem into multiple simpler sub-problems and uses linear regression method to 
accurately address these sub-problems. \rev{\mbox{\tree} is adaptive, in the sense that it constructs
its internal structures based on the correlation complexity; it is also dynamic, meaning 
that it reorganizes its internals at runtime to ensure the best efficiency.}




In the following section, we first discuss \tree's internal structure, 
and then demonstrate its construction, lookup, maintenance, parameter setting, and optimization.

\subsection{Internal Structure}
\label{subsec:internal}
\begin{figure}[t!]
    \centering
    \subfloat{
        \includegraphics[width=0.9\columnwidth]
            {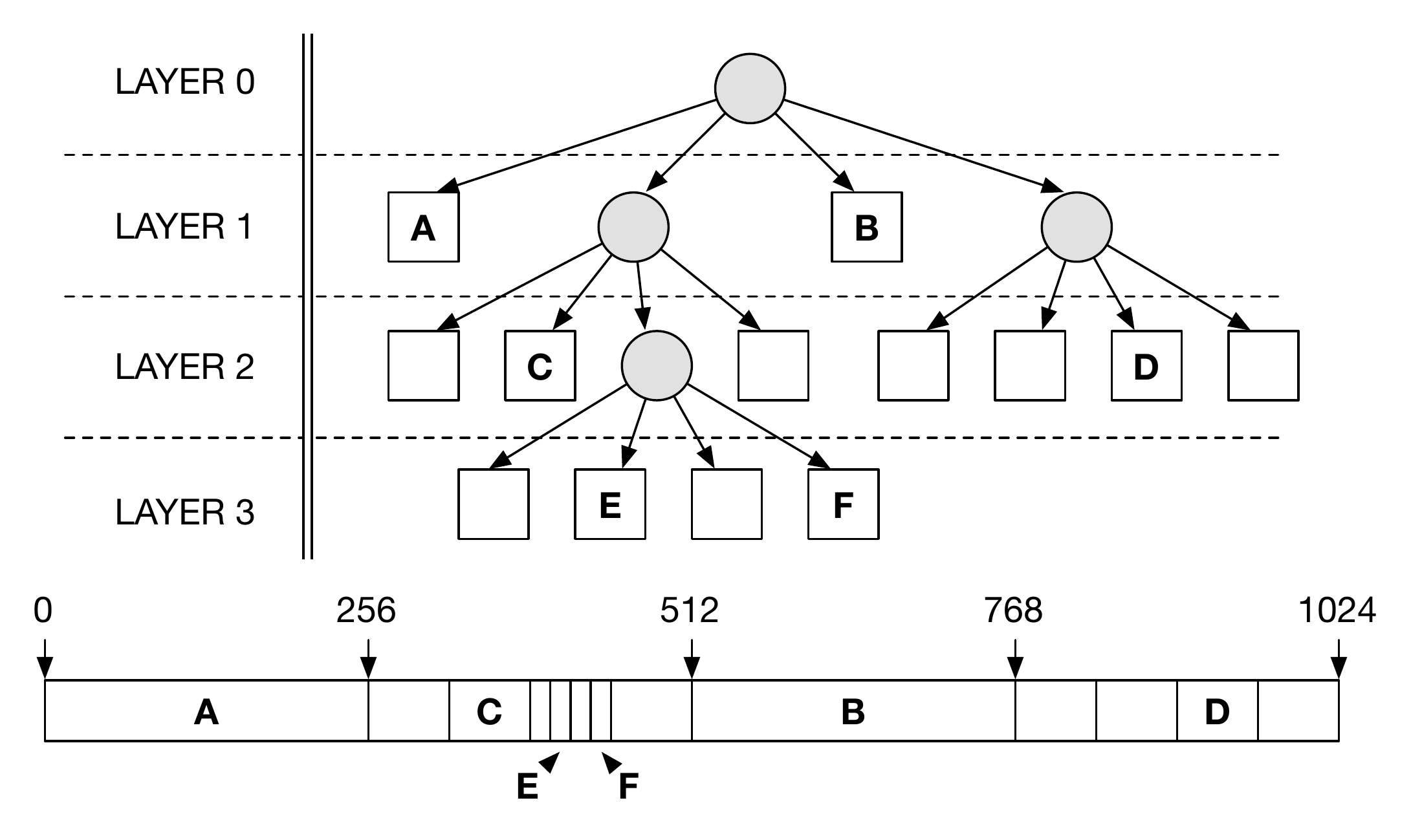}
    } 
    \caption{
        \tree data structure on a target column with value range from 0 to 1024.
        The node fanout is set to 4.
        The boxes represent the leaf nodes and the circles represent the internal nodes.
        The ruler bar shows how \tree partitions the range of the target column.
    }
    \label{figure:tree}
\end{figure}

\tree is a $k$-ary tree structure that maps the values in the target column $M$
to that in the host column $N$.
Its construction algorithm recursively divides $M$'s value range into 
$k$ uniform sub-ranges until every entry pair $(m, n)$ from $M$ and $N$ covered by 
the corresponding sub-range 
can be well estimated by a simple linear regression-based data mapping.
As a tree-based data structure, \tree uses leaf nodes to maintain the detailed data mappings, 
with its internal nodes providing fast navigation to these leaf nodes.
\cref{figure:tree} shows a \tree structure constructed on a target column whose value
range is from 0 to 1024.

{\bf Leaf node.}
A leaf node in \tree is associated with a sub-range $r$
of the target column $M$.
We define that a range $r$ has two elements: a lower bound $lb$ and an upper bound $ub$. 
Given a set of column entries $M^r$ from $M$ covered by $r$ 
(i.e., $\forall m \in M^r, r.lb\leq m\leq r.ub$), the leaf node 
attempts to provide an approximate linear mapping from $M^r$ to 
its corresponding set of column entries $N^r$ in the host column $N$.
Such a mapping is represented using a linear function 
$n=\beta m+\alpha \pm \epsilon$, where $m$ and $n$ represent column values from $M^r$ and $N^r$, $\beta$ and $\alpha$ respectively denote the function's 
slope and intercept, 
and $\epsilon$ denotes the confidence interval.

\tree computes $\beta$ and $\alpha$ using the standard linear regression formula~\cite{linearregression}
listed below:
\begin{footnotesize}
	\begin{align}
& \alpha=\overline{N^r}-\beta\overline{M^r} \qquad \beta=\frac{cov(M^r, N^r)}{var(M^r)} \notag
\end{align}
\end{footnotesize}
\noindent where 
$\overline{N^r}$ and $\overline{M^r}$ respectively denote the average values of elements in $N^r$ and $M^r$, 
$var(M^r)$ is the variance of elements in $M^r$, and $cov(M^r, N^r)$ presents the covariance of the corresponding elements in $M^r$ and $N^r$.
Based on the above equations, both $\alpha$ and $\beta$ can be computed with \textit{one scan} of the data in $M^r$ and $N^r$.

Please note that, instead of adopting gradient descent to iteratively converge to local optimal, 
we directly compute the solution using ordinary least square method, which is computationally easy and and works well for univariate cases.

Different from the slope and intercept, the confidence interval $\epsilon$ can be computed 
based on a user-defined parameter, called $error\_bound$, 
as will be elaborated in \cref{subsec:tuning}. 

The function $n=\beta m+\alpha \pm \epsilon$ captures an approximate linear correlation between columns $M$ and $N$ under the sub-range $r$ in $M$.
Given an $m$ in $M^r$, it bounds the corresponding $n$ to be in the range $(\beta m+\alpha-\epsilon, \beta m+\alpha+\epsilon)$. However, not all the 
entry pairs $(m, n)$ from $M^r$ and $N^r$ are necessarily covered by the computed linear function.
We call these entry pairs as \textit{outliers}.
The leaf node maintains all these outliers in an \textit{outlier buffer}, 
which is implemented as a hash table mapping from $m$ to the corresponding tuple's identifier, 
which can be either a primary key 
or a tuple location, as we will elaborate in \cref{sec:design}.

{\bf Internal node.}
An internal node in \tree functions as a navigator that routes the queries to their targeted leaf nodes.
Similar to the leaf nodes, each internal node is associated to a range in the target column $M$.
However, instead of maintaining any mapping to the host column, an internal node only 
maintains a fixed number of pointers pointing to its child nodes, 
each of which can be either a leaf node or another internal node.
To perform a lookup, an internal node can easily navigate the query to the corresponding child 
node whose range covers the input value.

\subsection{Construction}

\begin{algorithm}[t]
  \begin{small}
    \SetAlgoLined
    \SetKwFunction{push}{Push}
    \SetKwFunction{pop}{Pop}
    \SetKwFunction{isnotempty}{IsNotEmpty}
    \SetKwFunction{getvalue}{GetValue}
    \SetKwFunction{compute}{Compute}
    \SetKwFunction{validate}{Validate}
    \SetKwFunction{split}{Split}
    \SetKwFunction{getKey}{GetKey}
    \SetKwFunction{getValue}{GetValue}
    \SetKwFunction{materialize}{Materialize}
    \SetKwFunction{extractTable}{ProjectTable}
    \SetKwFunction{computerange}{GetHostRange}
    \SetKwFunction{splitNode}{SplitNode}
    \SetKwFunction{splitTable}{SplitTable}
    \SetKwFunction{size}{Size}
    \SetKwFunction{add}{Add}
    \SetKwFunction{deleteTmpTable}{DeleteTmpTable}
    \SetKwProg{Fn}{Function}{}{}
    \KwData{base table $T$, target column ID $cid_M$, host column ID $cid_N$, value range $R$}
    \KwResult{\tree's root node $root$}
    Node root($R$)\;
    TmpTable fullTmpTable = \extractTable($T$, $cid_M$, $cid_N$)\;
    FIFOQueue queue\;
    queue.\push(Pair(root, fullTmpTable))\;
    \While{\upshape queue.\isnotempty()}{
        Pair pair = queue.\pop()\;
        Node node = pair.\getKey()\;
        TmpTable tmpTable = pair.\getValue()\;
        \compute(tmpTable, node)\;
        \If{\upshape !\validate(tmpTable, node)} {
        	Node[] subNodes = \splitNode(node, $\mathit{node\_fanout}$)\;
        	TmpTable[] subTables = \splitTable(tmpTable, subNodes)\;
        	\ForEach{\upshape i \textbf{in} (0 to $\mathit{node\_fanout}-1$)}{
        		queue.\push(Pair(subNodes[i], subTables[i]))\;
        	}
        }
    \deleteTmpTable(tmpTable)\;
    }
    \KwRet root\;
   
    \Fn{\validate{tmpTable, node}}
    {
        \ForEach{\upshape entry \textbf{in} tmpTable}
        {
            \If{\upshape entry.host $\notin$ node.\computerange(entry.target)}
            {node.outliers.\add(entry.target, entry.tid)\;}
            \If{\upshape node.outliers.\size() > $outlier\_ratio$*tmpTable.\size()}
            {
                \KwRet false\;
            }
        }
       \KwRet true\;
    }
    \caption{Index construction in \tree}
    \label{alg:indexconstruction}
  \end{small}
\end{algorithm}

An RDBMS can efficiently construct a \tree upon the user's request. 
\cref{alg:indexconstruction} details the construction steps. 
The construction algorithm takes as input the base table $T$, 
the target column ID $cid_M$, the host column ID $cid_N$, and the target column's full range $R$.
The range $R$ contains the minimum and maximum values in the target column 
and can be easily obtained from the RDBMS's optimizer statistics. 
The algorithm also requires a set of \tree's pre-defined parameters for computation. 

This construction algorithm utilizes a FIFO (first-in-first-out) queue to build the \tree in a top-down fashion.
Each element in the FIFO queue is a pair that contains a \tree node and the node's corresponding temporary table.
The temporary table for a \tree node is a sub-table of $T$, which selects rows with target column in the node's range, 
and projects only the target and host columns along with each tuple's identifier.

\tree's construction starts by creating a root node with its range set to the whole range $R$, and pushing the root node 
along with its projected temporary table into the FIFO queue.
It then does the following steps iteratively until the FIFO queue is empty: (1) retrieve the ($tmpTable$, $node$) pair from the FIFO queue;
(2) compute $node$'s $\beta$, $\alpha$, and $\epsilon$ and determine whether 
the generated linear mapping can well cover its corresponding entry pairs;
(3) if (2) returns false, then split $node$ and $tmpTable$ respectively 
into multiple child nodes and sub-tables, 
then push the corresponding pairs of child node and sub-table back to the FIFO queue.

The \texttt{Compute} function scans a node's temporary table to compute the parameters for the linear function. The  \texttt{Validate} function scans the temporary table again, and 
validates whether each pair of target and host column values can be covered by the linear function.
Any unqualified entry is inserted to the node's outlier buffer.
A node's linear function is determined to be not \textit{good enough} if the outlier buffer exceeds a pre-defined ${outlier\_ratio}$, 
which is a ratio of the outlier buffer size to the total number of tuples covered by the node.
In this case, step (3) is triggered, which drops all the generated content in $node$ 
and splits it into a fixed number (equals to the pre-defined $\mathit{node\_fanout}$ parameter) of equal-range child nodes.
The users can limit the maximum depth of the tree structure by setting the 
parameter ${max\_height}$.

The user-defined parameters for \tree can directly control the confidence 
interval of the linear functions as well as the outlier buffer size,
consequently affecting the performance. 
We discuss the parameters in \cref{subsec:tuning}.
We further elaborate several optimization
strategies in \cref{sec:appendix-discussion}.

\rev{We now perform a complexity analysis for \mbox{\cref{alg:indexconstruction}}. 
\mbox{\tree} uses \mbox{\texttt{Compute}} function to scan the tuples covered by every tree node to derive a linear regression model. 
If the generated \mbox{\tree} is always a balanced full tree, 
then running linear regressions for all the tree nodes at the same height 
takes a full scan of all tuples. 
A \mbox{\tree} with height equal to $h$ will perform $h$ full scans in total. 
As $h$ is bounded by the parameter ${max\_height}$, 
we can conclude that the average and worst case complexities are \mbox{$O(N)$}.}


\subsection{Lookup}
\label{subsec:indexinternal}

\begin{algorithm}[t]
  \begin{small}
    \SetAlgoLined
    \SetKwFunction{push}{Push}
    \SetKwFunction{pop}{Pop}
    \SetKwFunction{add}{Add}
    \SetKwFunction{lookup}{Lookup}
    \SetKwFunction{intersect}{Intersect}
    \SetKwFunction{union}{Union}
    \SetKwFunction{computerange}{GetHostRange}
    \SetKwFunction{isoverlapping}{IsOverlapping}
    \SetKwFunction{isnotempty}{IsNotEmpty}
    \SetKwFunction{isleaf}{IsLeaf}
    \KwData{root node $root$, predicate $P$}
    \KwResult{range set $RS$, tuple identity set $IS$}
    Set<Range> RS\;
    Set<TupleID> IS\;
    FIFOQueue queue\;
    queue.\push(root)\;
    \While{\upshape queue.\isnotempty()}{
        node = queue.\pop()\;
        \eIf{\upshape node.\isleaf()}{
            Range r = \intersect(node.range, $P$)\;
            RS.\add(node.\computerange(r))\;
            IS.\add(node.outliers.\lookup(r))\;
        }
    {
        \ForEach{\upshape child \textbf{in} node.children}{
          \If{\upshape child.\isoverlapping($P$)}
            {
                queue.\push(child)\;
            }
        }
    }
    }
    RS = \union(RS)\;
    \KwRet $RS$ and $IS$\;
    \caption{Index lookup in \tree}
    \label{alg:indexlookup}
  \end{small}
\end{algorithm}

\tree allows users to perform both point and range lookups on the target column $M$
to get the corresponding results on the host column $N$. 
Instead of returning results that exactly match the query predicates, 
\tree' lookup algorithm returns approximate results.
\system will perform additional lookups on the host indexes and further validate 
the results and generate exact matchings.

\cref{alg:indexlookup} lists the details of \tree's lookup algorithm. 
The algorithm takes as input \tree's root node $root$ and a query predicate $P$ on the target column $M$.
It generates as output a set of value ranges $RS$ on the host column $N$ as well as a set of tuple identifiers $IS$. 
Without losing generality, we consider $P$ to be a value range on $M$ with two 
elements: lower bound $lb$ and upper bound $ub$.
A point query predicate has its lower bound equals to its higher bound.
The lookup starts from $root$ and runs a breadth-first search using a FIFO queue. 
The \tree iterates every single node in the queue and performs a lookup if the node 
is a leaf node. On confronting an internal node,
\tree retrieves its child nodes and checks whether each child's range overlaps with $P$. Any overlapping child node will be pushed to the FIFO queue.
The lookup algorithm continues iterating until the queue is empty.

\tree performs a lookup on a leaf node by taking the following
steps.
First, it computes an intersection between the query predicate $P$ and the node's value range. 
The intersection result is a value range $r$.
Using this range, the node can then use its linear function to compute the estimated range on $N$ that covers the exact matching.
The estimated range will be either
$(\beta \times r.lb+\alpha-\epsilon, \beta \times r.ub+\alpha+\epsilon)$
or 
$(\beta \times r.ub+\alpha-\epsilon, \beta \times r.lb+\alpha+\epsilon)$,
depending on the sign (positive or negative) of the slope $\beta$.
Not all the matchings are covered by the linear function.
Hence, the leaf node further retrieves a set of tuple identifiers
from its outlier buffer.
These identifiers can be used to directly fetch the corresponding tuples
from the RDBMS without looking up the host index.
Before terminating the algorithm, \tree computes a union among all the 
elements in $RS$. This is because the returned ranges generated by 
different leaf nodes can overlap.
Computing the union can help reduce the number of elements in $RS$.

\subsection{Maintenance}

\begin{algorithm}[t]
  \begin{small}
    \SetAlgoLined
    \SetKwFunction{tupleinsert}{Insert}
    \SetKwFunction{tupledelete}{Delete}
    \SetKwFunction{isleaf}{IsLeaf}
    \SetKwFunction{push}{Push}
    \SetKwFunction{remove}{Remove}
    \SetKwFunction{getvalue}{GetValue}
    \SetKwFunction{computerange}{GetHostRange}
    \SetKwFunction{traverse}{Traverse}
    \SetKwProg{Fn}{Function}{}{}
    \KwData{root node $root$, target column value $m$, host column value $n$, tuple ID $tid$}
    \Fn{\tupleinsert{$root$, $m$, $n$, $tid$}}{
    Node node = \traverse($root$)\;
    \If{\upshape $n$ $\notin$ node.\computerange($m$)}{
        node.outliers.\add($m$, $tid$)\;
    }
    }
    \Fn{\tupledelete{$root$, $m$, $n$, $tid$}}{
    Node node = \traverse($root$)\;
        node.outliers.\remove($m$, $tid$)\;
    }
    \Fn{\traverse{node, $m$}}{
    \eIf{\upshape node.\isleaf()}
    {
        \KwRet node\;
    }
    {
    \ForEach{\upshape child \textbf{in} node.children}
    {
        \If{\upshape $m$ $\in$ child.range}{}
        {\KwRet \traverse{child}\;}
    }
    }
    }
    \caption{Index insertion and deletion in \tree}
    \label{alg:indexinsertdelete}
  \end{small}
\end{algorithm}

At system runtime, \tree can dynamically support all of the commonly used database operations, including
insertions, deletions, and updates. 
This makes \tree a drastic departure from existing machine learning-based solutions, which rely on a long-running 
training phase to reconstruct the index structures from scratch.
\tree also reorganizes the \tree structure at runtime to ensure the best query performance. 
\cref{alg:indexinsertdelete} demonstrates how \tree processes
insertions and deletions.

{\bf Insertion.}
Tuple insertions in \tree are performed swiftly with little runtime change to its internal structures.
Given the to-be-inserted tuple's target column value $m$, host column value $n$,
and tuple identifier $tid$, \tree starts the insertion by locating the leaf
node containing the range covering $m$.
After that, \tree checks whether the node's corresponding range of the host column can cover $n$ (using the leaf node's linear function). 
If not, then \tree inserts this tuple's $m$ and $tid$ into the outlier buffer.
Otherwise, the insertion algorithm directly terminates.
The outlier buffer size of certain leaf nodes may grow to be too large, consequently
degrading \tree's lookup performance.
In this case, 
\tree invokes structure reorganization to further split these nodes, as we shall discuss shortly.

{\bf Deletion.}
The tuple deletion algorithm in \tree shares a similar routine as the insertion.
However, \tree does not perform any computation after locating the leaf node.
Instead, it directly checks the outlier buffer and removes the corresponding entry if exists.
Frequent tuple deletion from the index can cause suboptimal space utilization problem,
meaning that \tree can potentially use less number of leaf nodes to accurately capture
the column correlations. \tree also relies on the structure reorganization to handle 
this issue.



{\bf Reorganization.}
As we have mentioned above,
\tree reorganizes its internal structure on demand to optimize
the index efficiency in terms of both lookup performance and space utilization. 
\tree detects reorganization opportunities based on two criteria.
First, the outlier buffer size of a certain leaf node reaches a threshold ratio
compared to the total number of tuples covered in the corresponding range;
second, the number of deleted tuples covered by the leaf node's corresponding 
range reaches a threshold compared to the total number of tuples.
For the first case, \tree directly splits the leaf node into multiple equal-range
child nodes, as described in \cref{alg:indexconstruction}.
For the second case, \tree checks the node's neighbors to determine 
whether merging is beneficial.

\rev{\mbox{\tree} uses a dedicated background thread to execute the 
reorganization procedure, but it offloads the detection of candidate nodes  
to each insert/delete operation.
Specifically, \mbox{\tree} maintains a FIFO queue to record nodes 
where merge or split can be made. 
Once an insert operation finishes its procedure and 
detects that the outlier buffer size of its visited
leaf node has reached the threshold, it then adds the pointer to the leaf node 
into the FIFO queue. Delete operations proceed in a similar manner, 
but they add into the FIFO queue the pointer to the parent of 
the visited leaf node. 
Every entry 
in the queue is attached with a flag to identify whether this node is 
a candidate for split or merge.}

\rev{To perform structure reorganization, the background thread scans the 
target column to obtain all the tuples that fall into the affected value range.
It then computes the linear functions and populates the outlier buffers before
installing the new node(s) into the tree structure. To reduce the latency 
of the reorganization procedure, \mbox{\tree} periodically performs 
batch structure reorganization, meaning that the background thread 
can reorganize several candidate nodes in one scan.
On confronting drastic workload change, \mbox{\tree} can reorganize entire subtree 
at once (as we shall see in \mbox{\cref{subsection:exp:maintenance}}).
\mbox{\tree} also supports online structure reorganization, 
which enables concurrent lookup/insert/delete operations with little 
interference.
\mbox{\tree} ensures the structure consistency by leveraging 
a very simple yet efficient synchronization protocol. 
Due to the space limit, we provide 
more details in \mbox{\cref{sec:appendix-maintenance}}.}

\subsection{Parameters}
\label{subsec:tuning}
\tree requires the users to pre-define some parameters before the index construction, including 
$\mathit{node\_fanout}$, $max\_height$, and $outlier\_ratio$. There is another important parameter, 
called $error\_bound$, which is used to determine the confidence interval $\epsilon$ of each leaf node.

The $error\_bound$ parameter represents the expected number of host column $N$ values covered by the range returned from searching a \tree node for
a point query on the target column $M$. So, by setting this parameter, \tree roughly measures the number of false positives for a point query.
For a given leaf node with range $r$ on column $M$, its linear function returns an estimated range 
$(\beta \times r.lb+\alpha-\epsilon, \beta \times r.ub+\alpha+\epsilon)$ on column $N$ (We assume 
slope $\beta$ to be positive. Our discussion be easily generalized to cases with negative slopes.).
For any point $m\in r$ on column $M$, the linear function returns a range $(\beta \times m+\alpha-\epsilon, \beta \times m+\alpha+\epsilon)$ on column $N$.
Let $n$ be the number tuples covered by the leaf node. Assuming that values on column $N$ are uniformly distributed, 
then the expected number of values of $N$ for a point query (i.e. $error\_bound$) can be estimated as
$error\_bound = \frac{2 \epsilon}{\beta(r.ub-r.lb)+2 \epsilon}\times n \approx \frac{2 \epsilon}{\beta(r.ub-r.lb)}\times n$.
So, now given a desired $error\_bound$ parameter value, we can derive $\epsilon \approx \frac{\beta(r.ub-r.lb)\times error\_bound}{2n}$.


In theory, $error\_bound$ should be set carefully, since larger $error\_bound$ generates
larger $\epsilon$, which subsequently results in larger returned ranges for
the upcoming host index lookup. 
Too small $error\_bound$ can also cause performance degradation, 
since more tuples covered by the 
corresponding range may be identified as outliers, consequently causing node 
splitting, yielding much deeper tree structure.
Fortunately, in practice, the configuration of $error\_bound$ does not 
impact the performance too much.
This is because database users are more likely to issue range queries on secondary key
columns, in which case the amount of false positives brought by large $\epsilon$ 
is negligible compared to the amount of the tuples covered by the range query predicate.
Hence, \system adopting \tree can enjoy a very competitive end-to-end performance 
compared to conventional complete-tree indexes, even with larger $error\_bound$.

\section{\system}
\label{sec:design}

\tree lookup returns only approximate results. To obtain the real matching
for the input queries, \system further needs to remove all the false positive
results.
In this section, we first discuss tuple identifier schemes in existing 
RDBMSs, and show how \system can work with these different schemes and generate accurate query results.

\subsection{Tuple Identifiers}

A secondary index built on a certain (set of) column(s) provides a mapping from the columns' key values
to the corresponding tuples' identifiers. \rev{Tuple identifiers can be implemented in two different ways depending on the RDBMS's performance
requirement}~\cite{wu2017empirical,gvk+14}.
\system's indexing mechanism is general enough to work with both schemes.


An RDBMS adopting \textit{logical pointers} stores the primary key of the corresponding
tuple in each secondary index's leaf nodes.
The rationale behind logical pointers is that any update to tuple locations will not influence the secondary indexes.
However, the drawback of this mechanism is that the RDBMS has to perform an additional lookup
on the primary index every time a secondary index lookup happens. Popular RDBMSs like MySQL 
adopt this mechanism.


An RDBMS adopting the other identifier mechanism, called \textit{physical pointers}, 
directly stores tuple locations (in the format of "blockID+offset") in each secondary index's leaf nodes.
While avoiding traversing the primary index during a secondary index lookup, 
an RDBMS using physical pointers has to update each index's corresponding leaf node
once any tuple location is changed.
Several DBMSs such as PostgreSQL employ this scheme.

\subsection{Lookup in \system}
\begin{figure}[t!]
    \centering
    \subfloat{
        \includegraphics[width=\columnwidth]
            {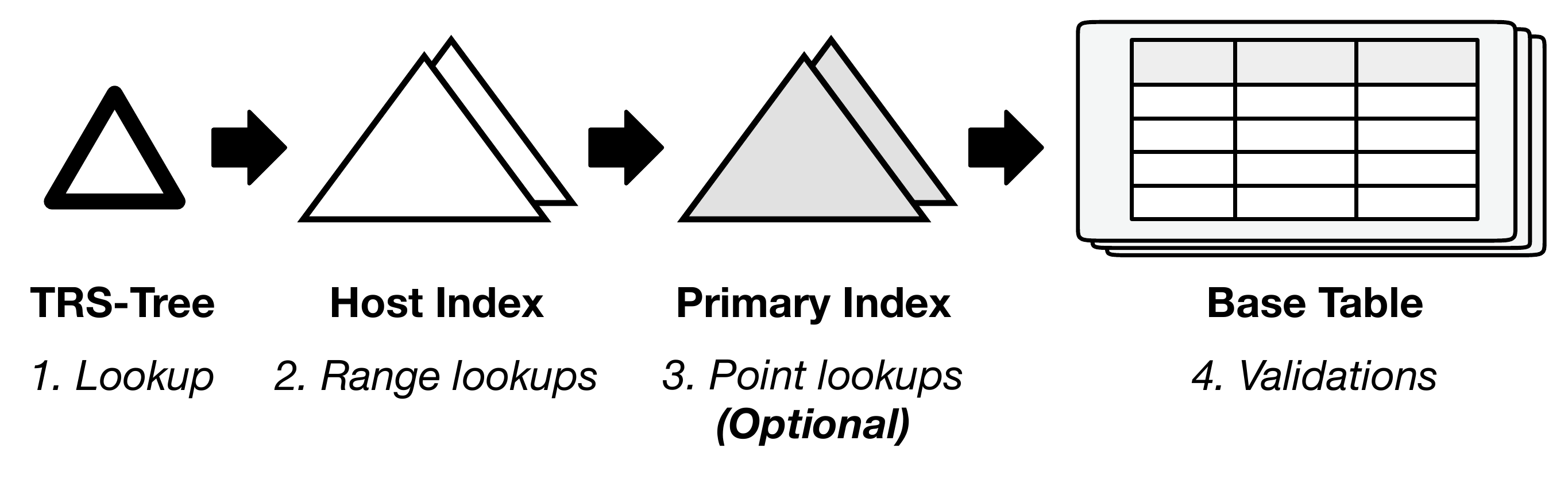}
    } 
    \caption{
        The workflow of \system's lookup mechanism.
    }
    \label{figure:pointers}
\end{figure}

\system can generate accurate lookup results for both tuple identifier schemes.
\cref{figure:pointers} shows the entire workflow of \system's lookup mechanism.
We list the key steps as follows:

  \textbf{Step 1.} \textit{\tree lookup} -- This step performs a lookup on \tree as described in \cref{subsec:indexinternal}. 
  The results are a set of ranges on the host column and a set of tuple identifiers.
  
  \textbf{Step 2.} \textit{Host index lookup} -- 
  This step performs lookups on the host index with the returned host column ranges as inputs.
  The result is a set of tuple identifiers, which is further unioned with the set of identifiers returned from Step 1.
  
  \textbf{Step 3.} \textit{Primary index lookup (optional)} -- 
  This step occurs only if the RDBMS adopts logical pointers as tuple identifiers.
  It looks up the primary index with the 
  returned set of tuple identifiers as inputs. 
  The result is a set of tuple locations.
  
  \textbf{Step 4.} \textit{Base table validation} -- 
  This step fetches the actual tuples using tuple locations and 
  validates whether each tuple satisfies the input predicates. 
  This step returns all the qualified results to the input query.

Please note that a primary index can also serve as the host index, in which case the lookup
procedure shall be the same.

Compared to conventional secondary index methods, \system can bring in additional 
overhead due to the extra host index lookup phase as well as the base table validation phase.
The overhead is exacerbated when using logical pointers as the tuple identifier scheme,
as it involves unnecessary primary index lookups for unqualified matchings. However, we argue that such overhead is insignificant when performing 
range queries, which are prevalent for secondary indexes.
This is because the number of false positives for range queries is quite small when compared
to that of the qualified tuples.
Moreover, as a \tree greatly reduces the space consumption,
it brings huge benefit to modern main-memory RDBMSs, where 
memory space is precious.

\section{Discussion}
\label{sec:discussion}

\rev{\mbox{\textbf{Tradeoff between space and computation.}}
Compared to conventional indexing mechanisms, \mbox{\tree} 
achieves great space saving by sacrificing access performance. 
While the actual space used by \mbox{\tree} is dependent on the 
correlation quality (i.e., how correlate the two columns are), 
we can indeed strike a balance by tuning the $error\_bound$ parameter. 
Let us consider a scenario with $max\_height$
set to 1, where \mbox{\tree} becomes a single-node, 
single-layer structure. Now we tune the $error\_bound$ parameter and analyze 
how \mbox{\tree} behaves.
In an extreme case where $error\_bound$ is set to 0, \mbox{\system} shall 
identify every single data that 
cannot be perfectly covered by the generated linear function as outlier. 
In this case, \mbox{\system} can consume more memory but 
achieves optimal lookup performance. 
The increase of $error\_bound$ can effectively drop the memory consumption 
in the expense of reducing lookup performance. 
This is because \mbox{\tree} enlarges the returned bound for the lookups,
and consequently introducing more false positives.
However, as we shall see in our experiments, 
\mbox{\tree}'s performance is actually not quite sensitive 
to the value of $error\_bound$ parameter, as long as it is set to small enough.
The key reason is that \mbox{\tree} navigates lookups based on simple linear function computation, which is much cheaper than chasing pointers and retrieving every single elements from the standard index structure.}

\textbf{Fault tolerance.}
The RDBMS must periodically persist
\system's \tree into the underlying storage for fault tolerance. 
Depending on the scenario, \tree can 
function either like a disk-based index and persist leaf nodes on disks or 
like a pure in-memory index which relies on write-ahead logging and checkpointing for persistence.
We leave the detailed discussion 
as a future work.



Due to the space limit, please refer to \mbox{\cref{sec:appendix-discussion}}
for detailed discussion on correlation recovery, optimization, complex machine learning models, 
and several other issues.
\section{Evaluation}
\label{sec:evaluation}

%

\subsection{Experiment Setups}

We implemented \system in two different RDBMSs: \database, a main-memory 
RDBMS prototype built internally in IBM -- Almaden,
and PostgreSQL~\cite{postgresql}, a disk-based RDBMS that is widely used in 
backing modern database applications.
We performed all the experiments on a commodity machine running Ubuntu 16.04
with one 6-core Intel i7-8700 processor clocked at 3.20 GHz. The machine has a 16 GB DRAM
and one PCIe attached 1 TB NVMe SSD.

\rev{We use three different applications to evaluate \mbox{\system}: 
\mbox{\textsc{Synthetic}}, \mbox{\textsc{Stock}}, and \mbox{\textsc{Sensor}}.
We provide the detailed descriptions of these applications 
in \mbox{\cref{sec:appendix-applications}}}.

Throughout this section, we compare two mechanisms:

\textbf{$\diamond$ \system}: the correlation-based secondary indexing mechanism proposed in this paper.

\textbf{$\diamond$ Baseline}: the standard B+-tree-based secondary indexing mechanism used in conventional RDBMSs.

\rev{The B+-tree in \mbox{\database} is fully maintained in memory. 
It is highly optimized for 
modern CPUs with many advanced techniques such as cache-conscious layout and SIMD 
instructions (for numerical keys) applied. The node size is set to 256 bytes.
PostgreSQL instead implements a page-based B+-tree backed by buffer pool. In our experiments, 
we have reconfigured the buffer pool size to ensure that the B+-tree is fully cached in memory.}

Without any explicit declaration, 
we set \tree's parameters, including $\mathit{node\_fanout}$, $max\_height$, $outlier\_ratio$, and $error\_bound$, 
to 8, 10, 0.1, and 2, respectively. \tree achieves a good space-computation tradeoff with this configuration, as we shall see later.

\subsection{Real-World Applications}

\begin{figure}[t!]
    \centering
     \fbox{
     \includegraphics[width=0.9\columnwidth]
         {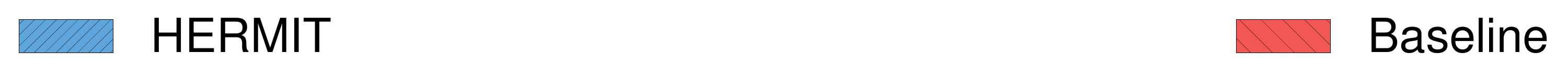}
     }
    \subfloat[Logical Pointer]{
        \includegraphics[width=0.48\columnwidth]
            {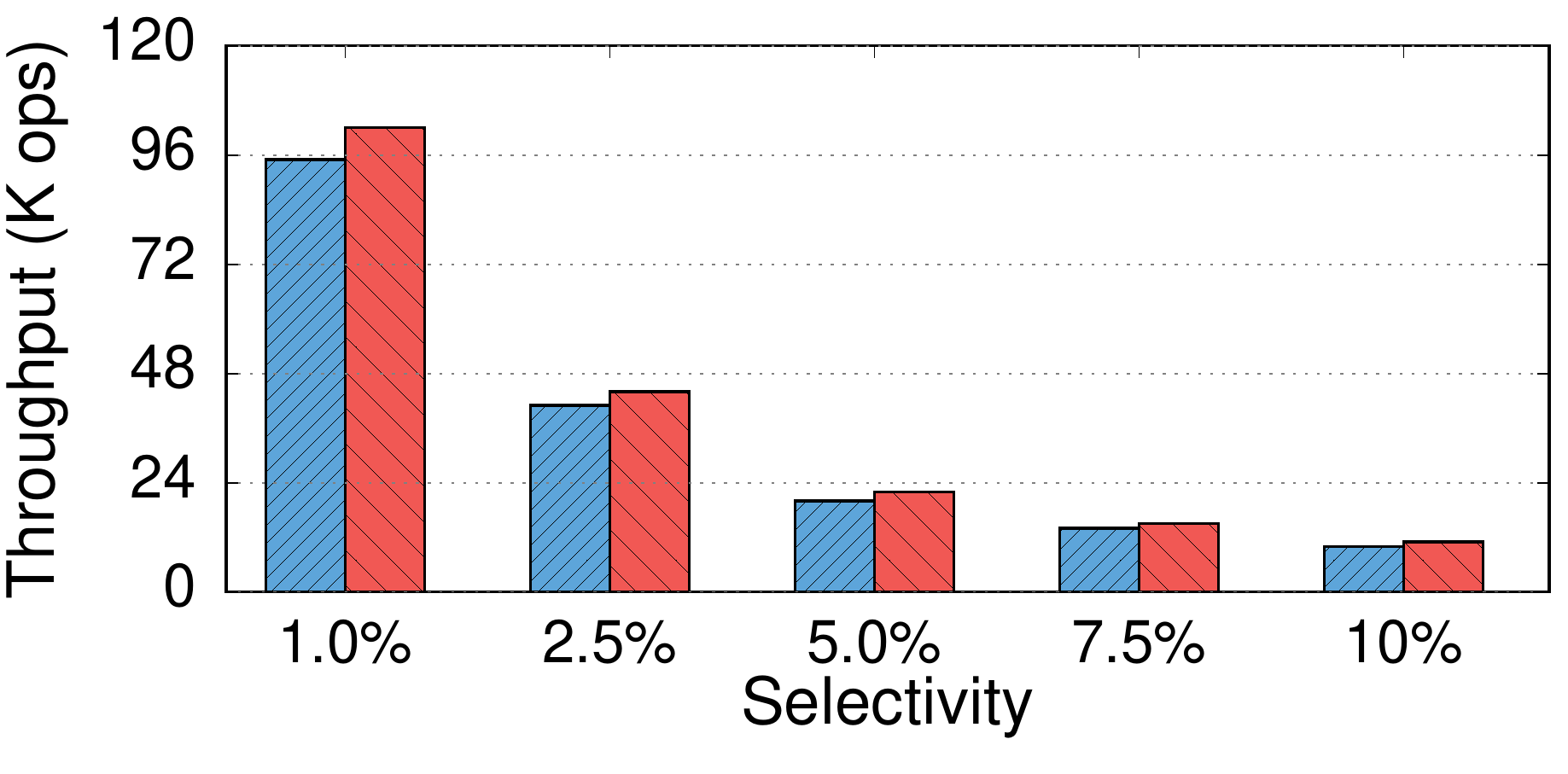}
        \label{figure:experiments:stock-selectivity-logical}
    }
    \subfloat[Physical Pointer]{
        \includegraphics[width=0.48\columnwidth]
            {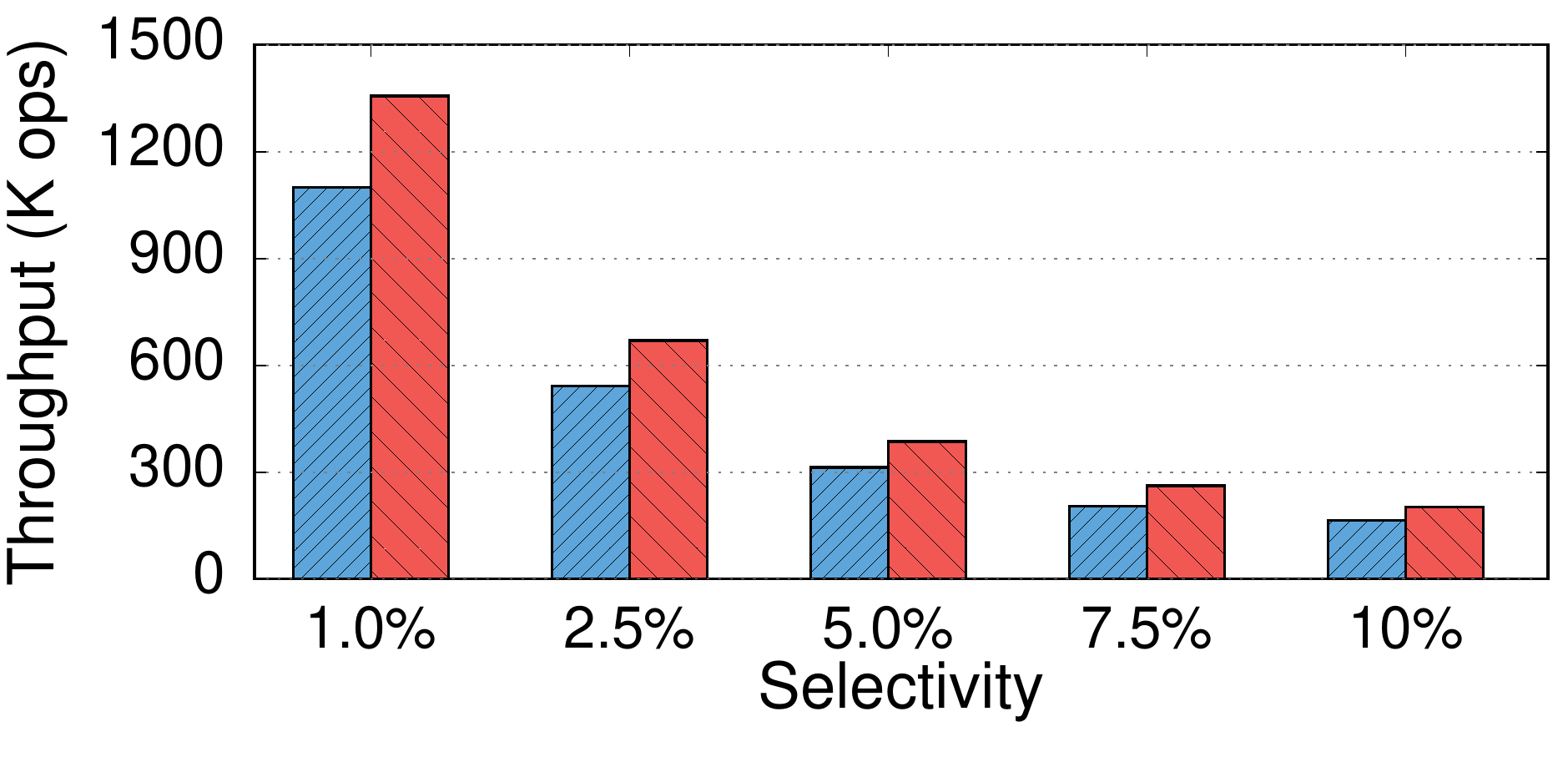}
        \label{figures:experiments:stock-selectivity-physical}
    }
    \caption{
        Range lookup throughput with different selectivities (\textsc{Stock}).
    }
    \label{figures:experiments:stock-selectivity}
\end{figure}

\begin{figure}[t!]
    \centering
    \subfloat[Memory Consumption]{
        \includegraphics[width=0.48\columnwidth]
            {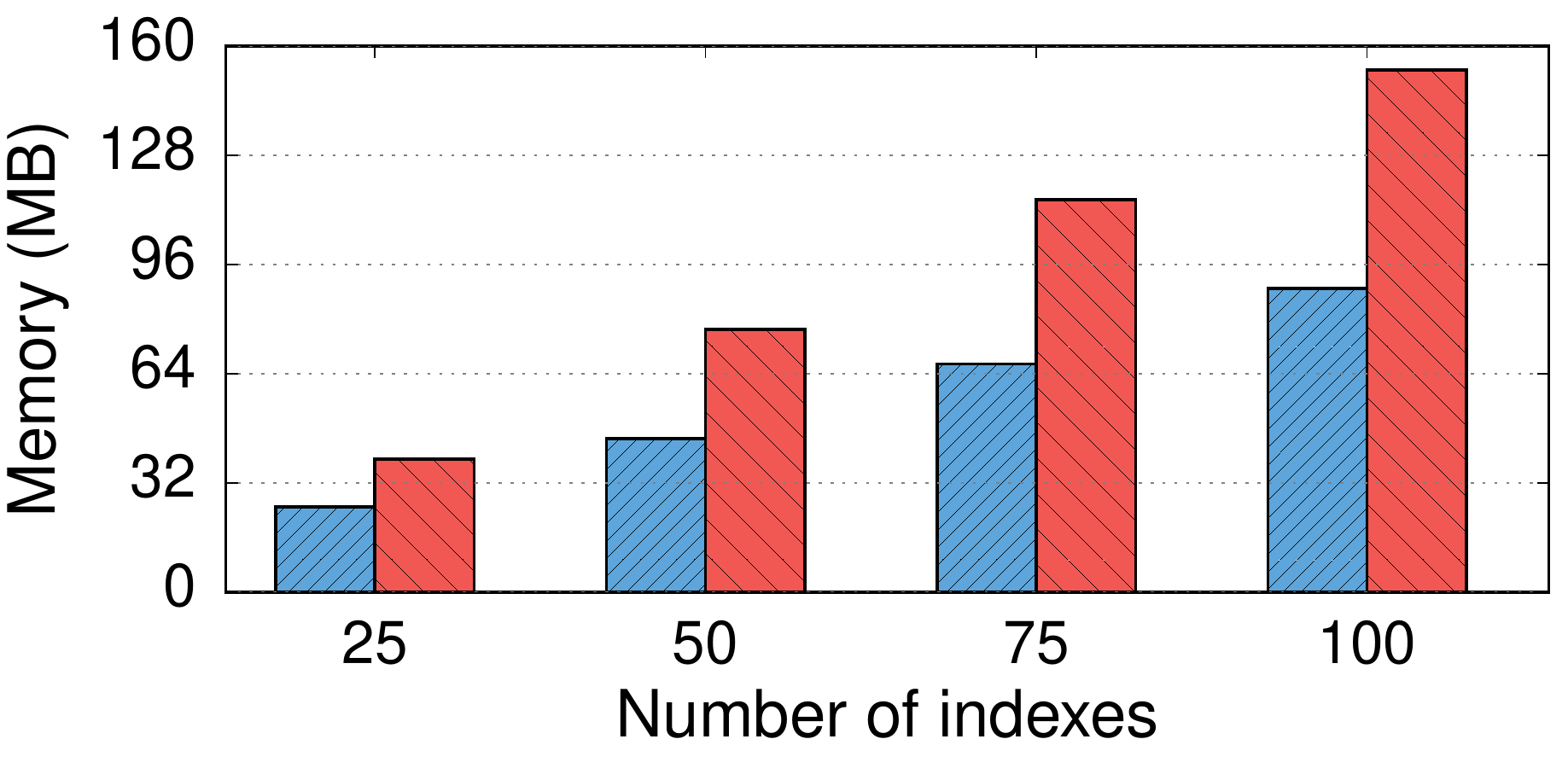}
        \label{figures:experiments:stock-memory-indexcount}
    }
    \subfloat[Space Breakdown]{
        \includegraphics[width=0.48\columnwidth]
            {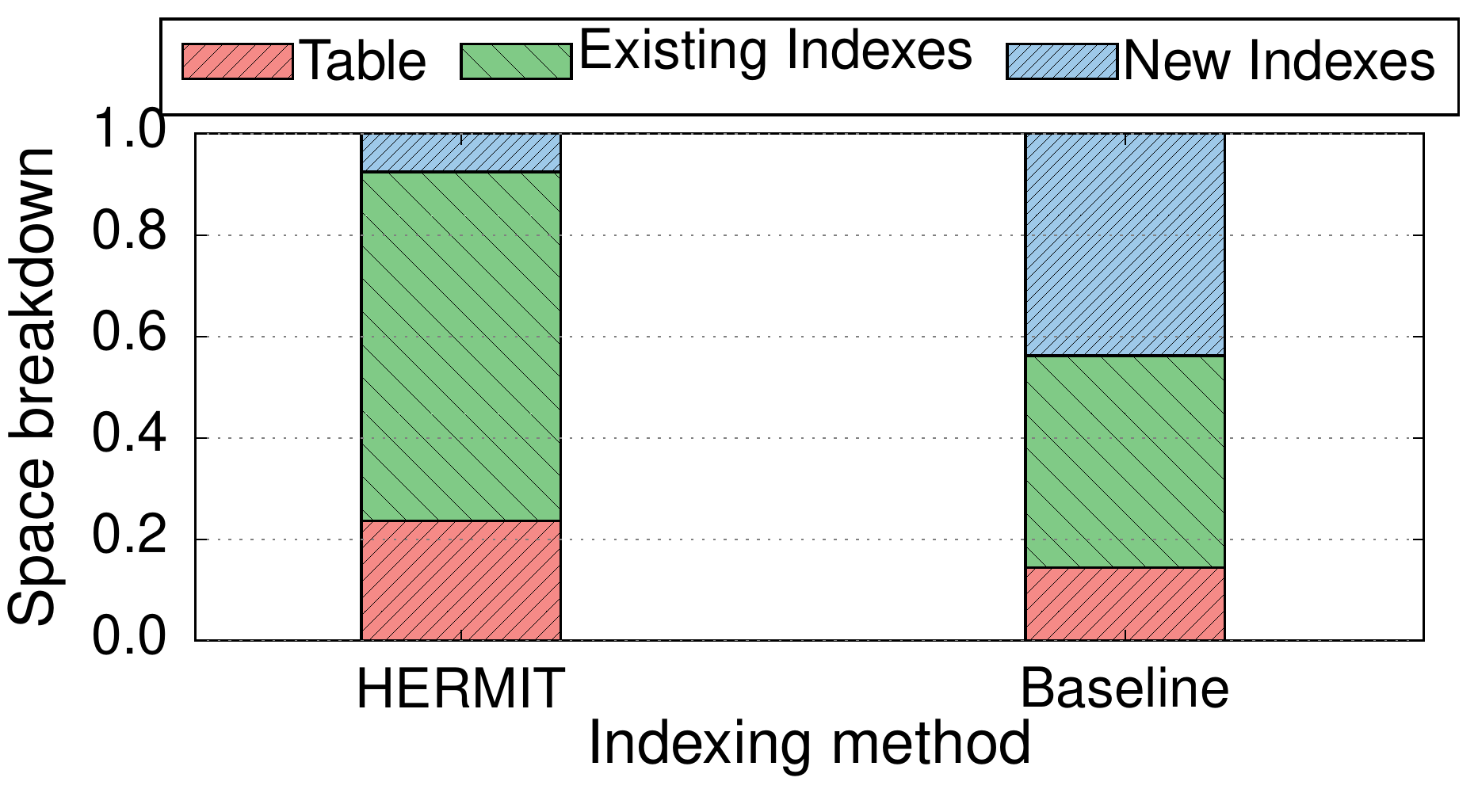}
        \label{figures:experiments:stock-memory-breakdown}
    }
    \caption{
        Memory consumption with different numbers of indexes (\textsc{Stock}).
    }
    \label{figures:experiments:stock-point}
\end{figure}

\begin{figure}[t!]
    \centering
    \subfloat[Logical Pointer]{
        \includegraphics[width=0.48\columnwidth]
            {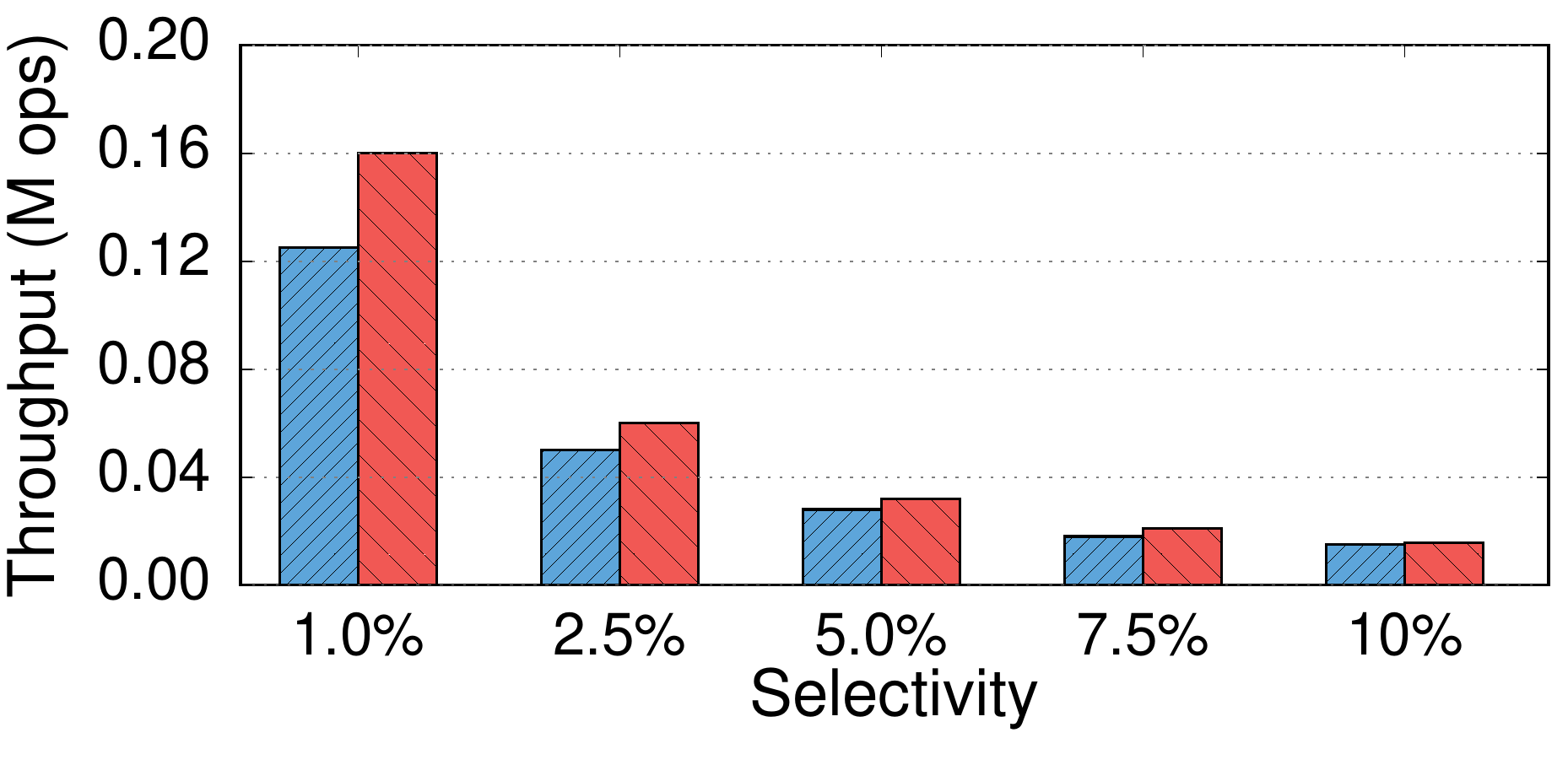}
        \label{figures:experiments:sensor-selectivity-logical}
    }
    \subfloat[Physical Pointer]{
        \includegraphics[width=0.48\columnwidth]
            {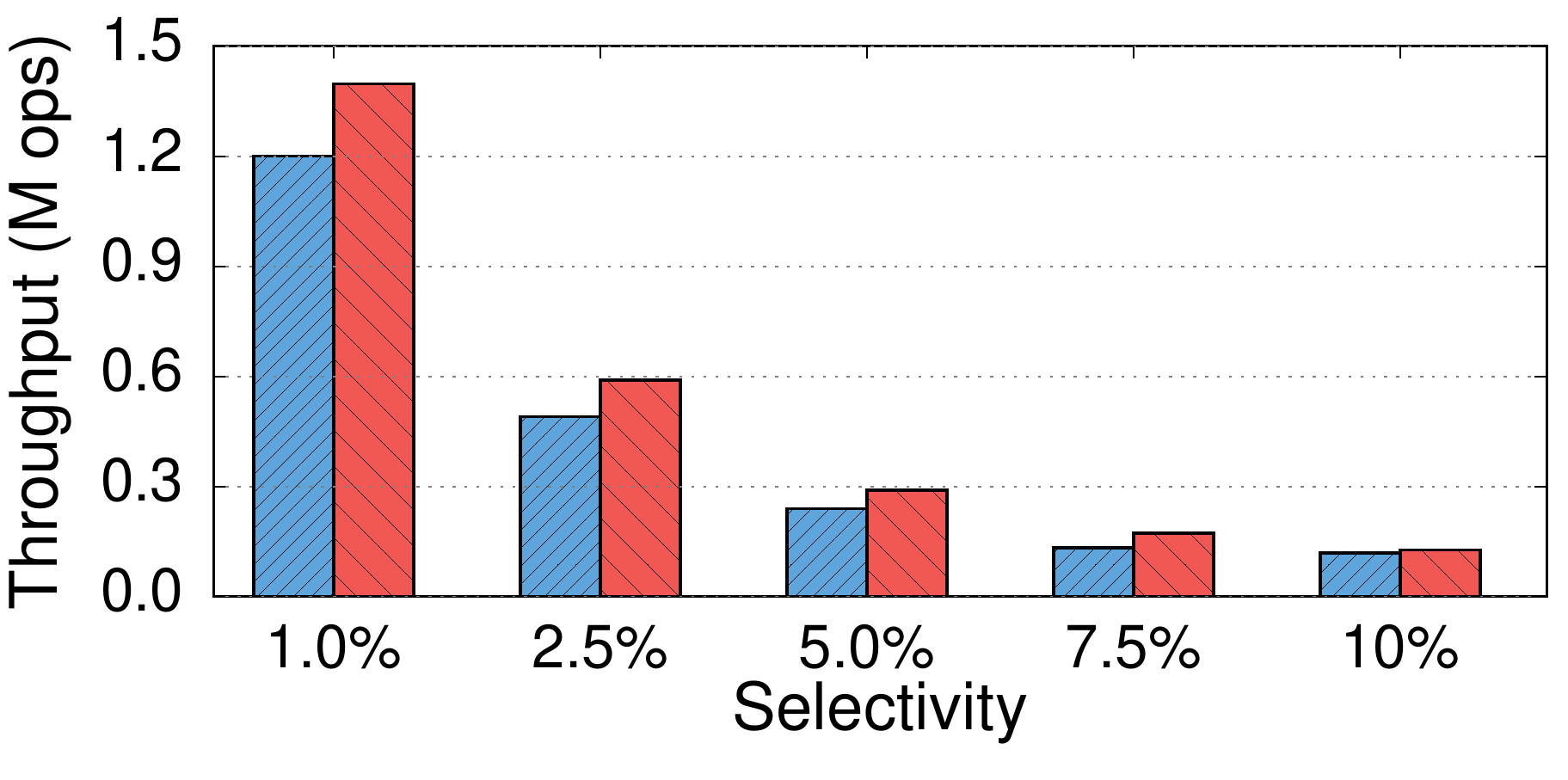}
        \label{figures:experiments:sensor-selectivity-physical}
    }
    \caption{
        Range lookup throughput with different selectivities (\textsc{Sensor}).
    }
    \label{figures:experiments:sensor-selectivity}
\end{figure}

\begin{figure}[t!]
    \centering
    \subfloat[Memory Consumption]{
        \includegraphics[width=0.48\columnwidth]
            {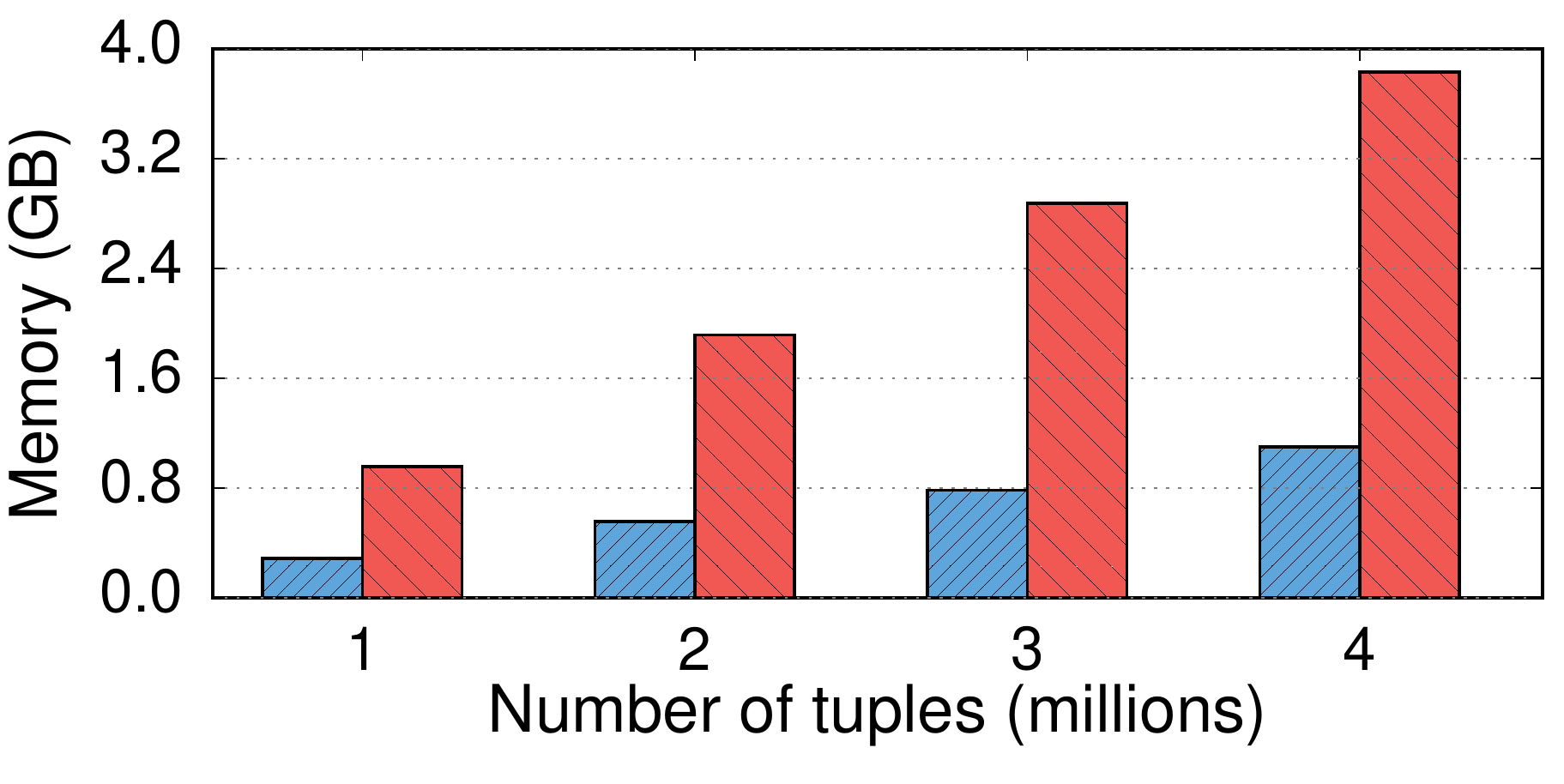}
        \label{figure:experiments:sensor-memory-tuplecount}
    }
    \subfloat[Space Breakdown]{
        \includegraphics[width=0.48\columnwidth]
            {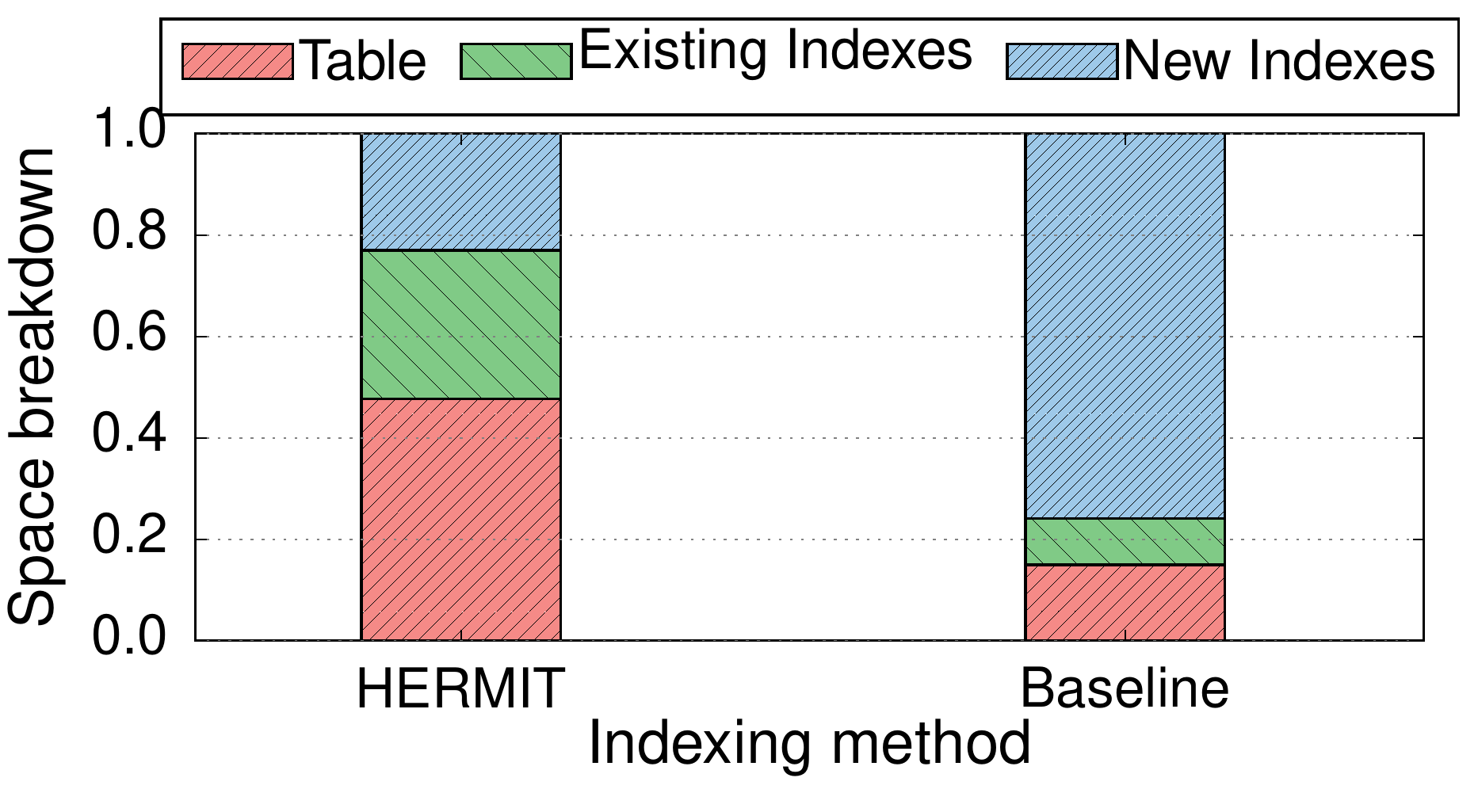}
        \label{figures:experiments:sensor-memory-breakdown}
    }
    \caption{
        Memory consumption with different numbers of tuples (\textsc{Sensor}).
    }
    \label{figures:experiments:sensor-memory}
\end{figure}


\rev{In this subsection, we evaluate \mbox{\system}'s performance using 
real-world applications in \mbox{\database}}.

The first experiment uses the \textsc{Stock} application to evaluate \system's range
query performance. It is simple for \system's \tree to model the correlations 
between a stock's daily highest and lowest prices, as they form a near-linear correlation. 
\rev{One thing worth noticing is that there does not exist any strict bound between 
the two prices, and stock price can increase/drop by over 50\% in a single day
(see PG\&E stock (\mbox{\texttt{NYSE: PCG}})). \mbox{\system} shall identify and maintain 
these readings as outliers}.
\cref{figures:experiments:stock-selectivity} shows the range lookup throughput 
with different selectivities.
As we can see, \system yields a very competitive performance to the baseline 
solution, which requires building a complete secondary index on every single column.
While \system suffers slightly from the overhead caused by false positive removal,
its influence to the overall throughput is reduced with the increase of the selectivity. 

We then measure \system's memory consumption by changing the number of stocks stored 
in the table. 
\cref{figures:experiments:stock-memory-indexcount} shows the result. 
When setting the number of indexes to 25, it means we store 25 stocks in the table, 
as we build one index for each stock's lowest price column.
The result indicates that \system's \tree takes little memory space, and 
the RDBMS adopting \system consumes only half of the 
memory compared to adopting the baseline solution, which creates one index
for each column.
\rev{In fact, \mbox{\system} in this case spends a great fraction of memory for storing 
outliers, and this guarantees a small false positive ratio, as we shall see later.}
\cref{figures:experiments:stock-memory-breakdown} provides a space breakdown to confirm this finding.

Next, we test \system's performance using the \textsc{Sensor} application.
Supporting fast data retrieval in this application is challenging, as each sensor reading
column has a non-linear correlation with the average reading columns.
\cref{figures:experiments:sensor-selectivity} shows the throughput with different
range lookup selectivities. When setting the selectivity to 1.0\%, \system yields a 
throughput that is around 22\% lower than the baseline solution. However, the performance 
gap diminishes with the growth of the selectivity. This is because 
\system generates approximate results, and the time ratio of filtering out 
false positives decreases with the increase of the result size.

We then use the same application to measure \system's memory consumption.
As \cref{figure:experiments:sensor-memory-tuplecount} shows, the space consumed by the 
baseline solution grows much faster than that consumed by \system.
According to the analytics in \cref{figures:experiments:sensor-memory-breakdown},
the baseline solution spends most of the memory maintaining newly created secondary indexes.
In contrast, \system's \tree takes much less space, and most of the space is used 
for storing outliers.

\subsection{Lookup}
In this subsection, we evaluate \system's range and point lookup performance 
using the \textsc{Synthetic} application, 
with both \textsc{Linear} and \textsc{Sigmoid} correlation functions.
\rev{We perform all the experiments in \mbox{\database}}.

We use both \system and the baseline method to index column $col_C$.
Modeling \textsc{Linear} correlation function is trivial using \system' \tree structure. However, 
it can be challenging to model \textsc{Sigmoid} correlation function, which is polynomial.

\begin{figure}[t!]
    \centering
    \subfloat[Logical Pointer]{
        \includegraphics[width=0.47\columnwidth]
            {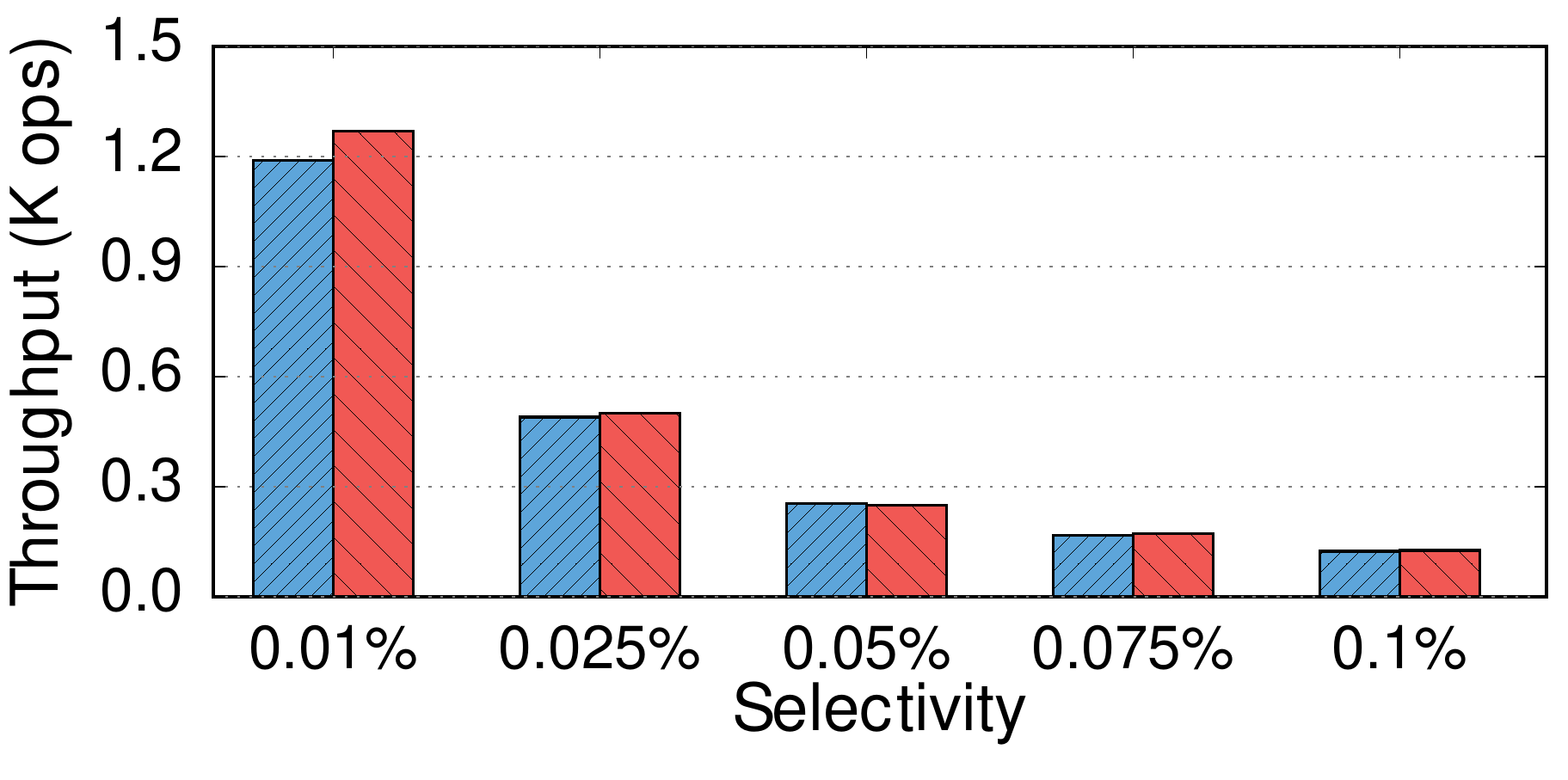}
        \label{figure:experiments:linear-selectivity-logical}
    }
    \subfloat[Physical Pointer]{
        \includegraphics[width=0.48\columnwidth]
            {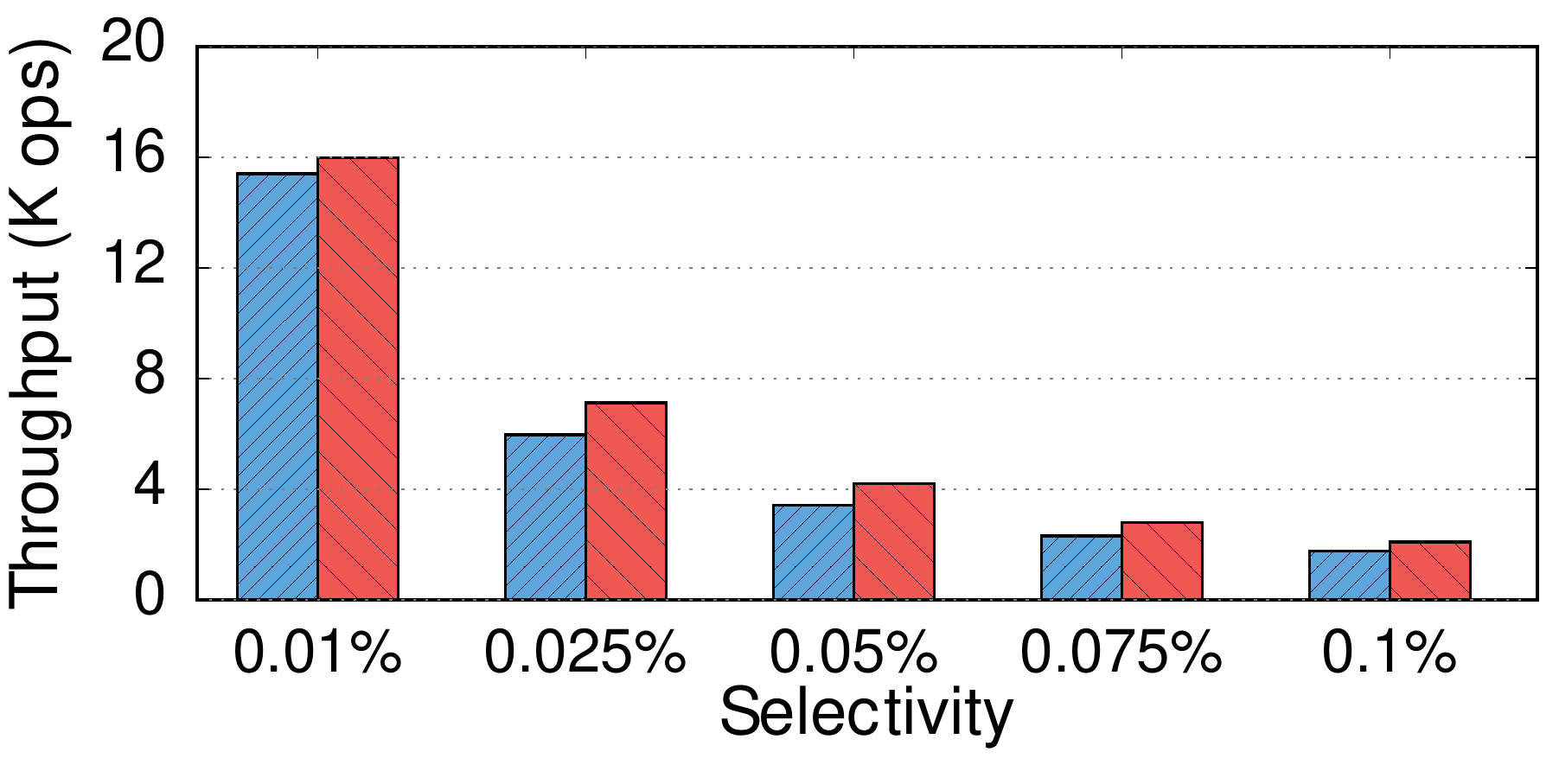}
        \label{figures:experiments:linear-selectivity-physical}
    }
    \caption{
        Range lookup throughput with different selectivities (\textsc{Synthetic} -- \textsc{Linear}).
    }
    \label{figures:experiments:linear-selectivity}
\end{figure}

\begin{figure}[t!]
    \centering
    \subfloat[Logical Pointer]{
        \includegraphics[width=0.48\columnwidth]
            {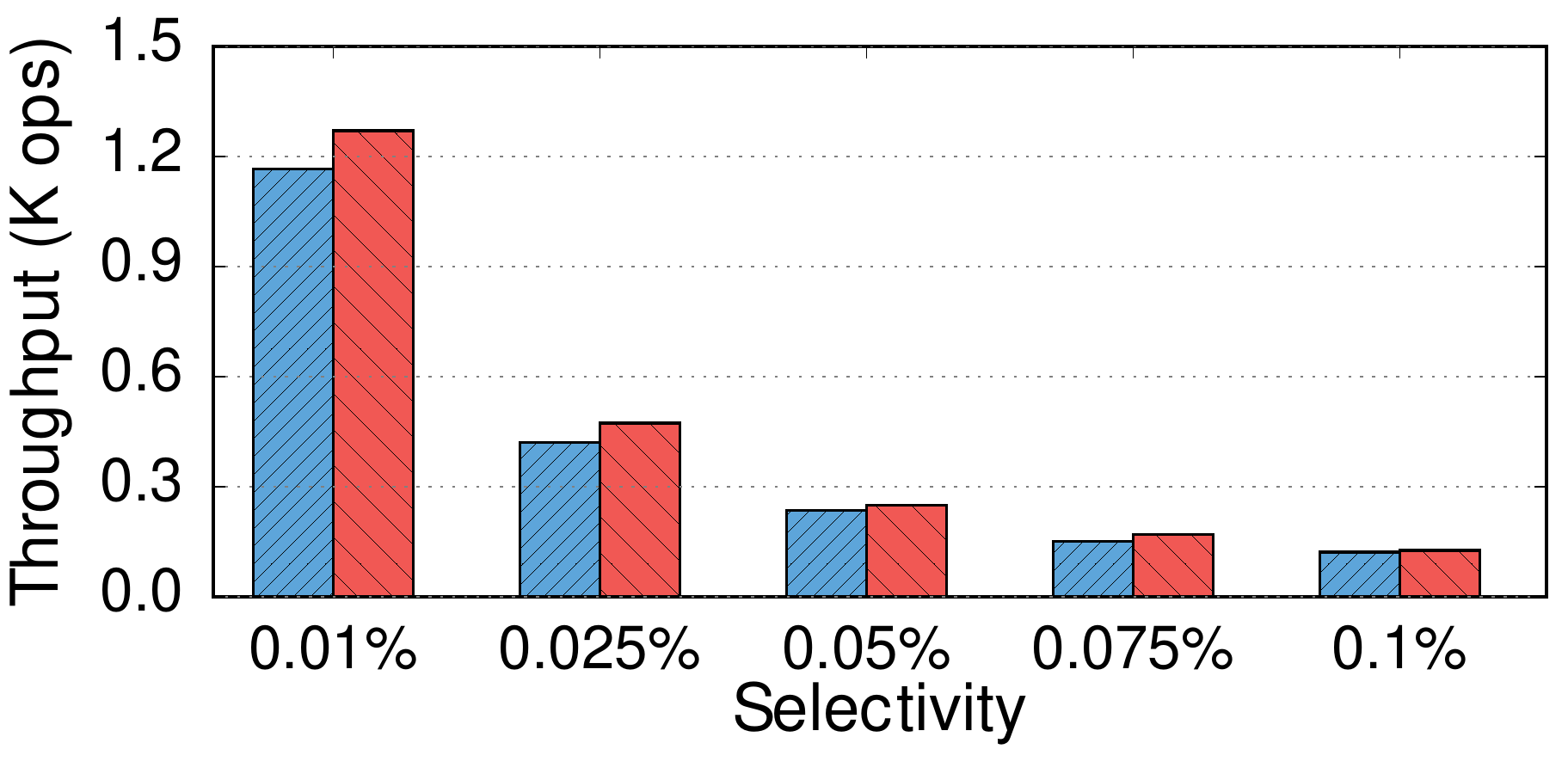}
        \label{figure:experiments:sigmoid-selectivity-logical}
    }
    \subfloat[Physical Pointer]{
        \includegraphics[width=0.48\columnwidth]
            {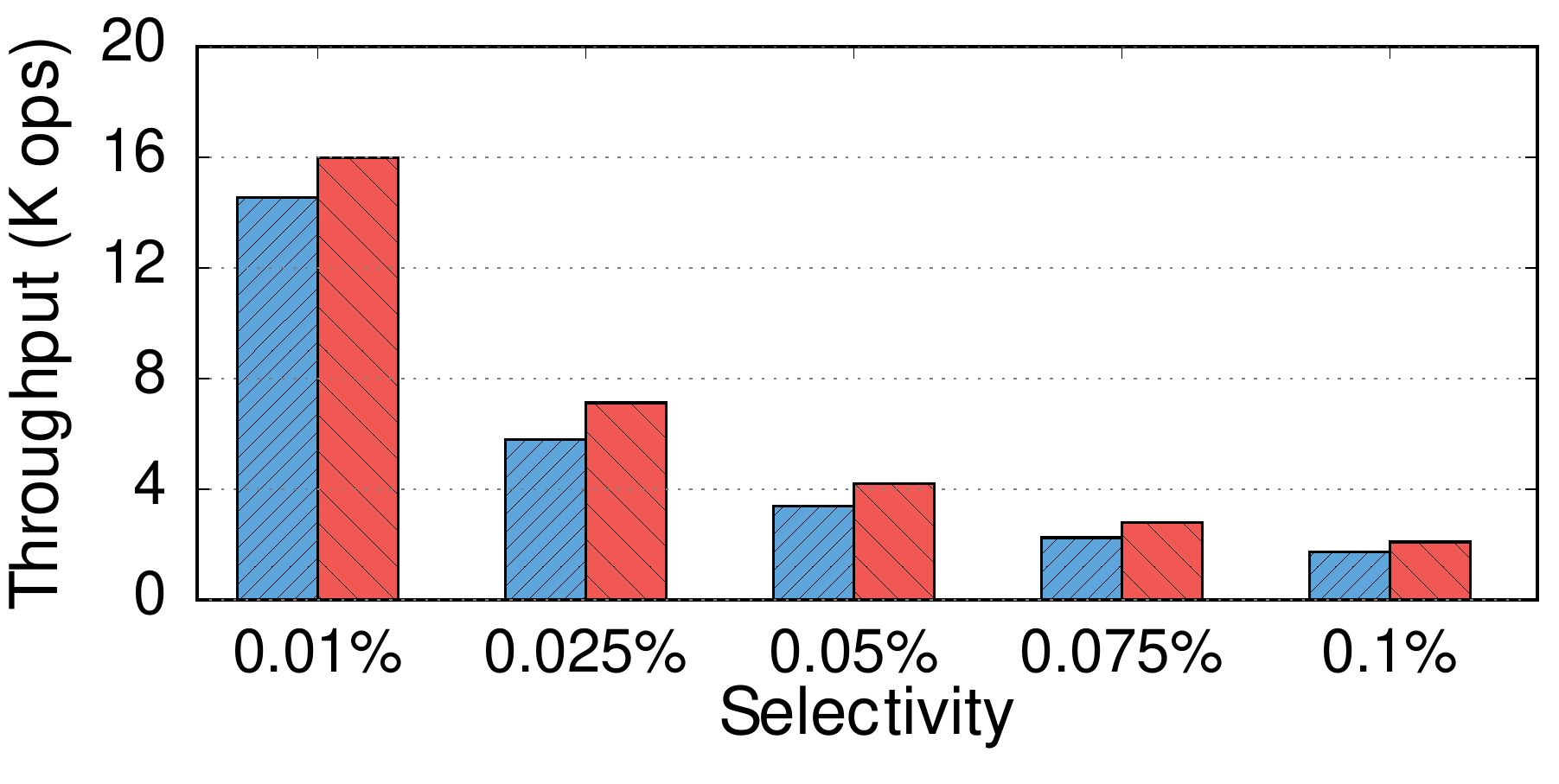}
        \label{figures:experiments:sigmoid-selectivity-physical}
    }
    \caption{
        Range lookup throughput with different selectivities (\textsc{Synthetic} -- \textsc{Sigmoid}).
    }
    \label{figures:experiments:sigmoid-selectivity}
\end{figure}

\cref{figures:experiments:linear-selectivity} depicts the performance of 
\system and the baseline mechanism with \textsc{Linear} correlation function.
We set the number of tuples to 20 million and measure the throughput changes of the range lookup queries
with different query selectivities.
We also adopt different tuple identifier methods to show their impacts on the performance.
The result indicates that \system's performance is very close to that achieved by the baseline. 
Using logical pointers, \system and the baseline respectively 
proceed 1.19 and 1.27 K operations per second (K ops) with selectivity set to 0.01\%.
This gap is reduced with the increase of selectivity.
The experiments with physical pointers also demonstrate the same results.
One of the reasons is that \system's \tree only needs to use a single leaf node to model the correlation function,
yielding optimal performance. 
We then use \textsc{Sigmoid} function to test whether \tree can efficiently model complex correlations.
The results in \cref{figures:experiments:sigmoid-selectivity} show that the performance gap is little changed.
Observing this, we decide to perform a detailed analysis to understand where the time goes.

\begin{figure}[t!]
    \centering
     \fbox{
     \includegraphics[width=0.9\columnwidth]
         {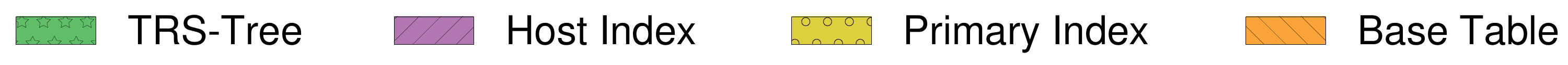}
     }
    \subfloat[Logical Pointer]{
        \includegraphics[width=0.48\columnwidth]
            {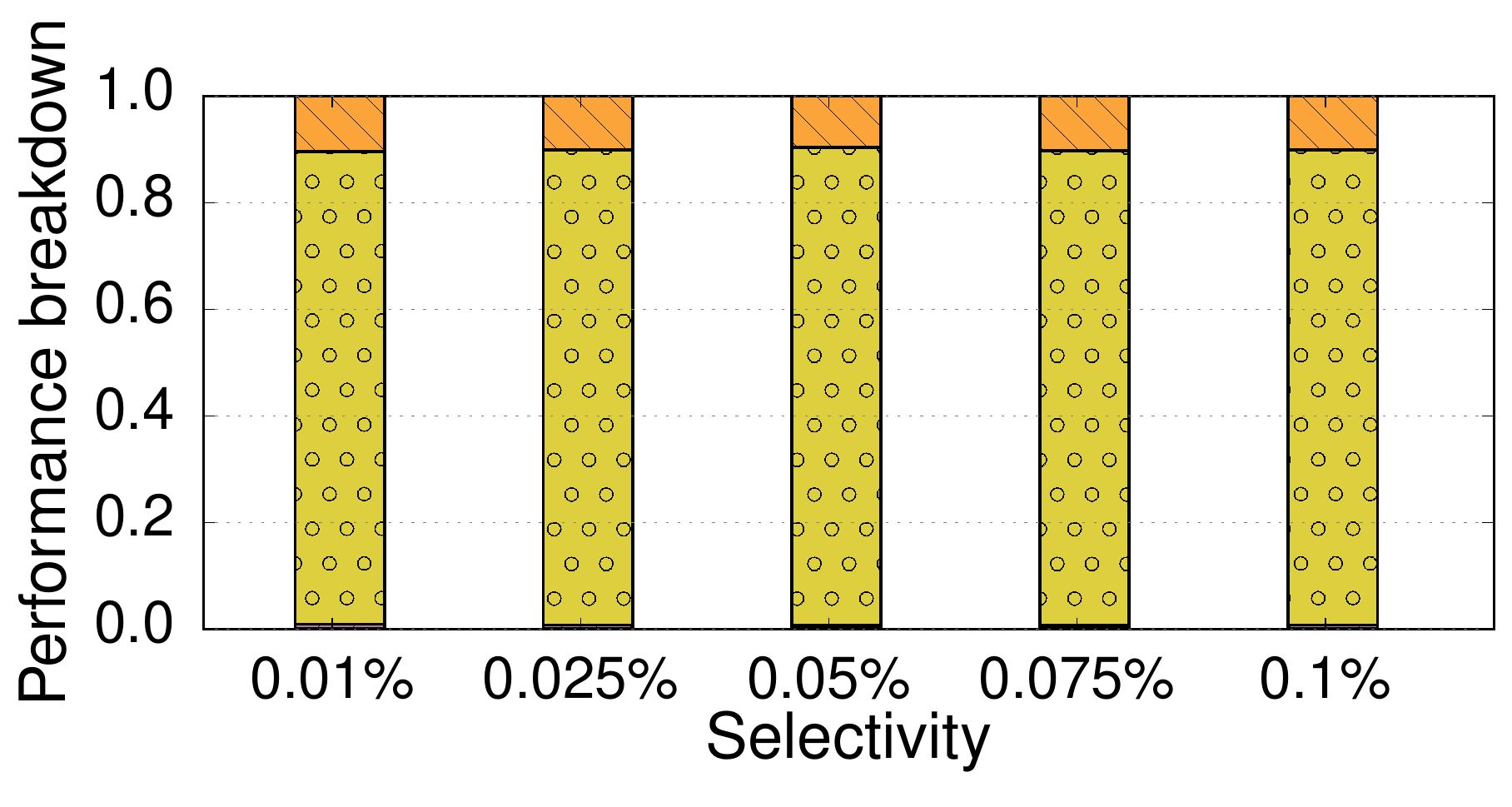}
        \label{figure:experiments:sigmoid-selectivity-breakdown-logical}
    }
    \subfloat[Physical Pointer]{
        \includegraphics[width=0.48\columnwidth]
            {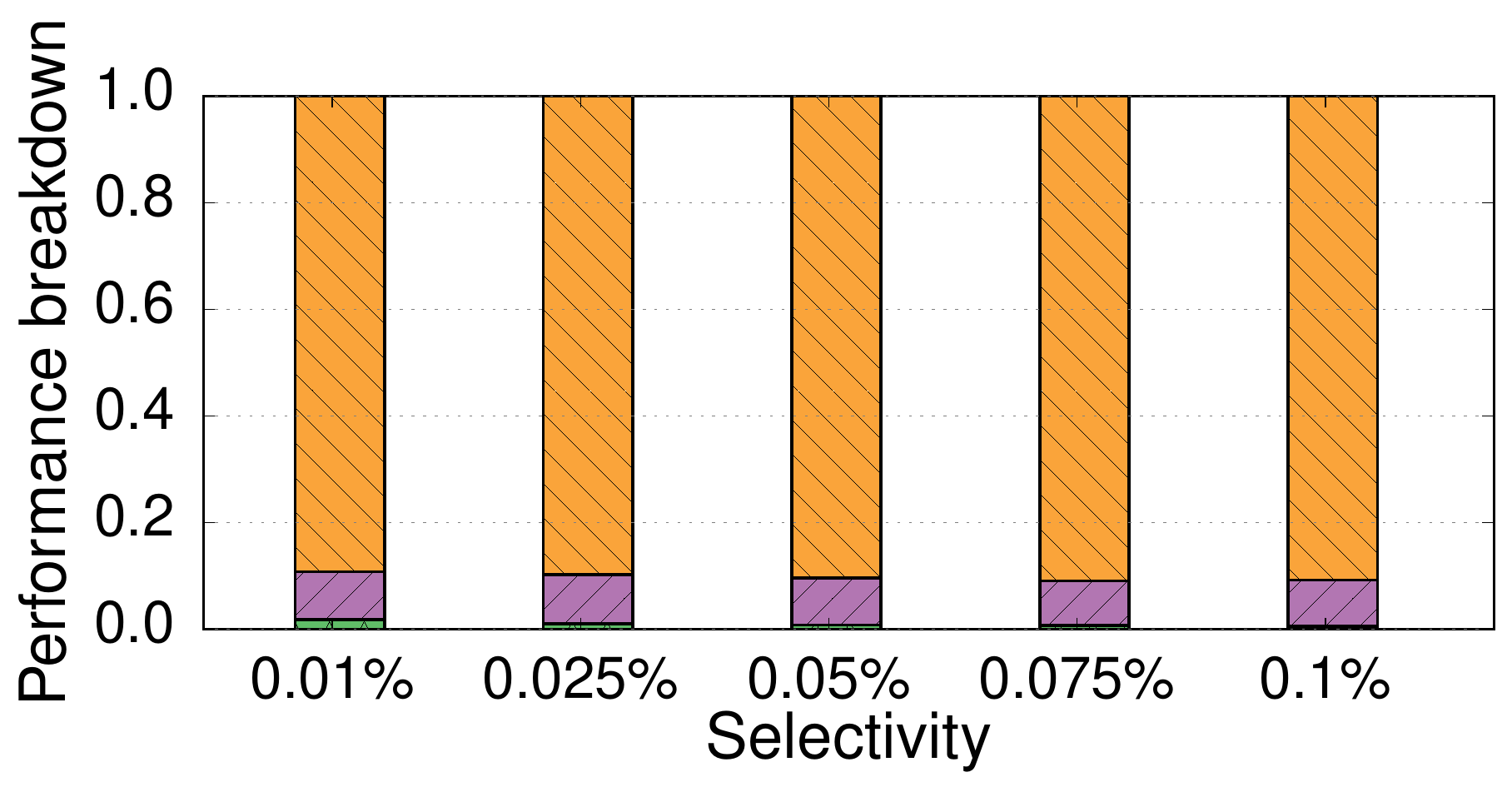}
        \label{figures:experiments:sigmoid-selectivity-breakdown-physical}
    }
    \caption{
        \system's range lookup performance breakdown with different selectivities 
        (\textsc{Synthetic} -- \textsc{Sigmoid}).
    }
    \label{figures:experiments:sigmoid-selectivity-breakdown}
\end{figure}

\begin{figure}[t!]
    \centering
     \fbox{
     \includegraphics[width=0.9\columnwidth]
         {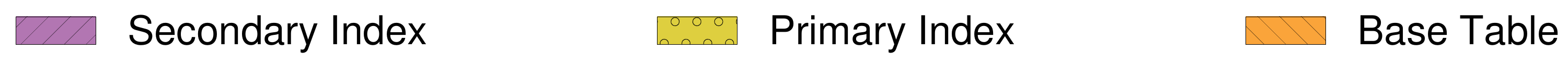}
     }
    \subfloat[Logical Pointer]{
        \includegraphics[width=0.48\columnwidth]
            {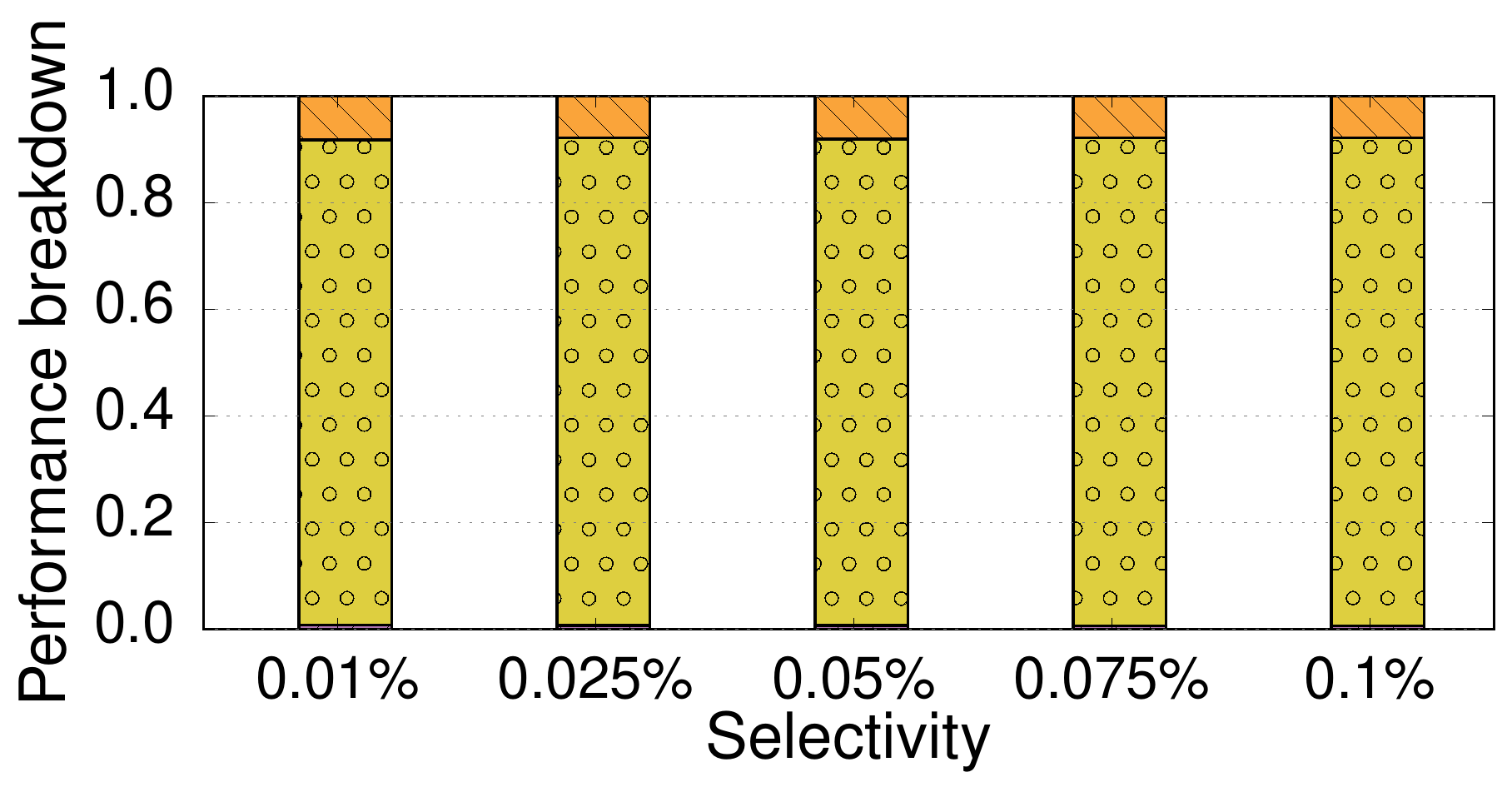}
        \label{figure:experiments:sigmoid-selectivity-breakdown-logical-baseline}
    }
    \subfloat[Physical Pointer]{
        \includegraphics[width=0.48\columnwidth]
            {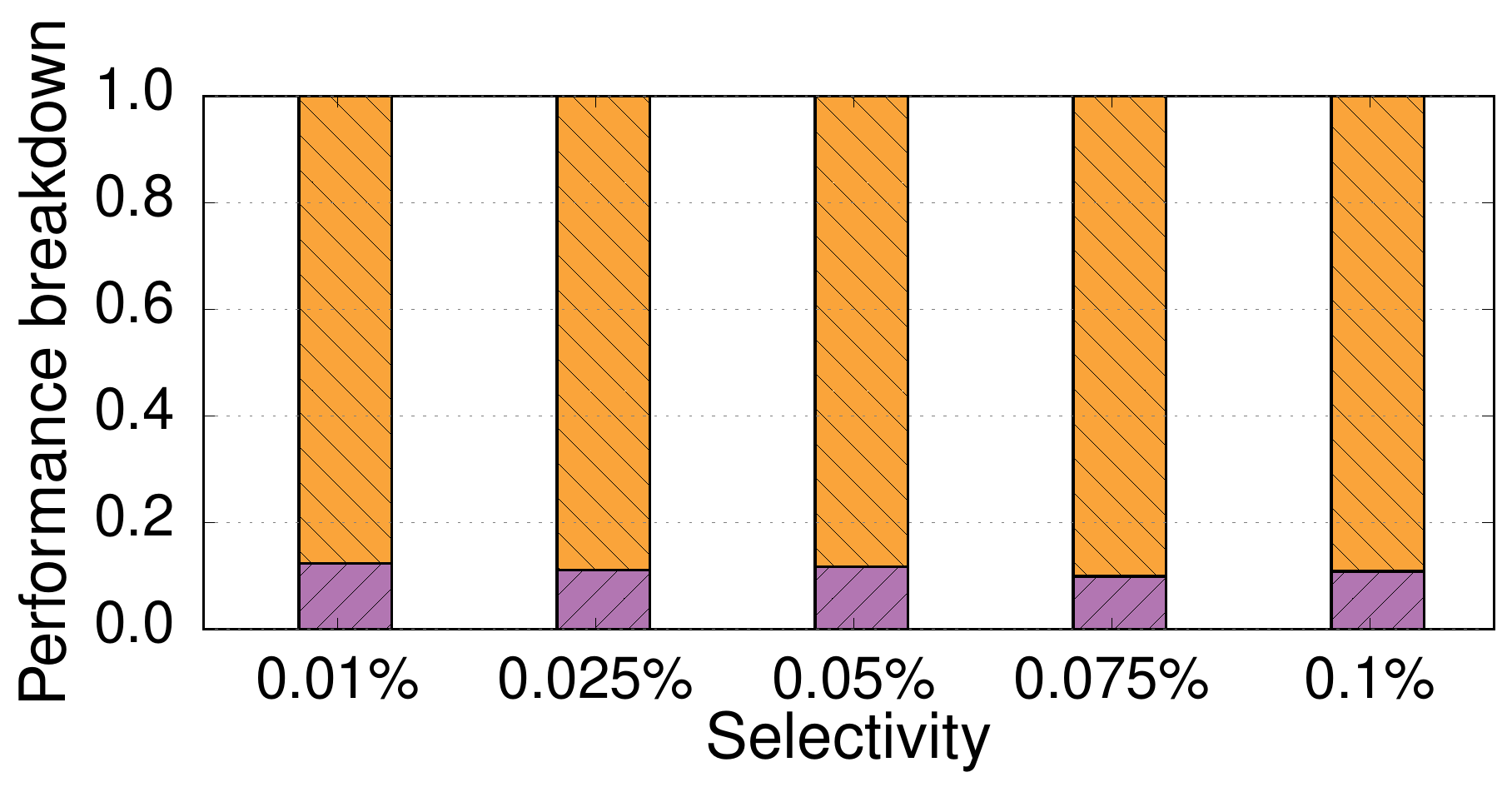}
        \label{figures:experiments:sigmoid-selectivity-breakdown-physical-baseline}
    }
    \caption{
        Baseline's range lookup performance breakdown with different selectivities 
        (\textsc{Synthetic} -- \textsc{Sigmoid}).
    }
    \label{figures:experiments:sigmoid-selectivity-breakdown-baseline}
\end{figure}

\cref{figures:experiments:sigmoid-selectivity-breakdown} and 
\cref{figures:experiments:sigmoid-selectivity-breakdown-baseline}
respectively show the performance breakdown of \system and the baseline method. The time includes both CPU and memory IO.
Recall that \system performs lookups through the following steps: 
\tree lookup, host index lookup, primary index lookup (optional), and base table validation.
In contrast, the baseline method only needs to perform secondary index lookup, 
primary index lookup (optional), and base table access.
With logical pointers as tuple identifiers, both methods spend over 90\% of 
their time on the primary index lookup.
This is inevitable because an RDBMS using logical pointers does not directly expose 
the tuple location to any index other than the primary one.
When identifying tuples using physical pointers, the major bottleneck of 
both methods shifted to the base table access. While one-time tuple retrieval from 
the base table using its tuple location seems to be trivial, the total number of the 
fetches is actually equivalent to the total number of returned tuples, which 
can be expensive in range queries.

\begin{figure}[t!]
    \centering
    \subfloat[Logical Pointer]{
        \includegraphics[width=0.48\columnwidth]
            {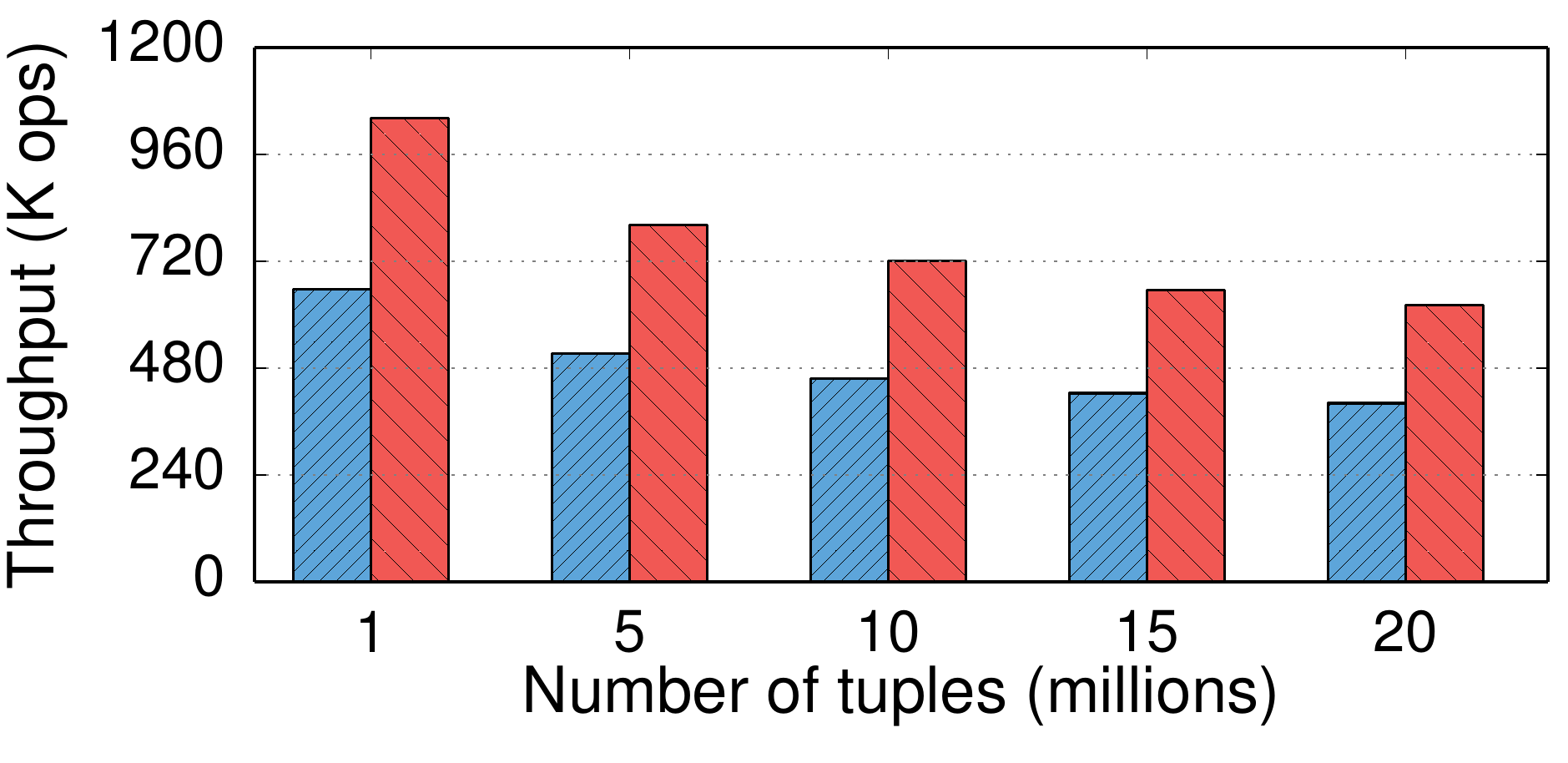}
        \label{figure:experiments:linear-index-logical}
    }
    \subfloat[Physical Pointer]{
        \includegraphics[width=0.48\columnwidth]
            {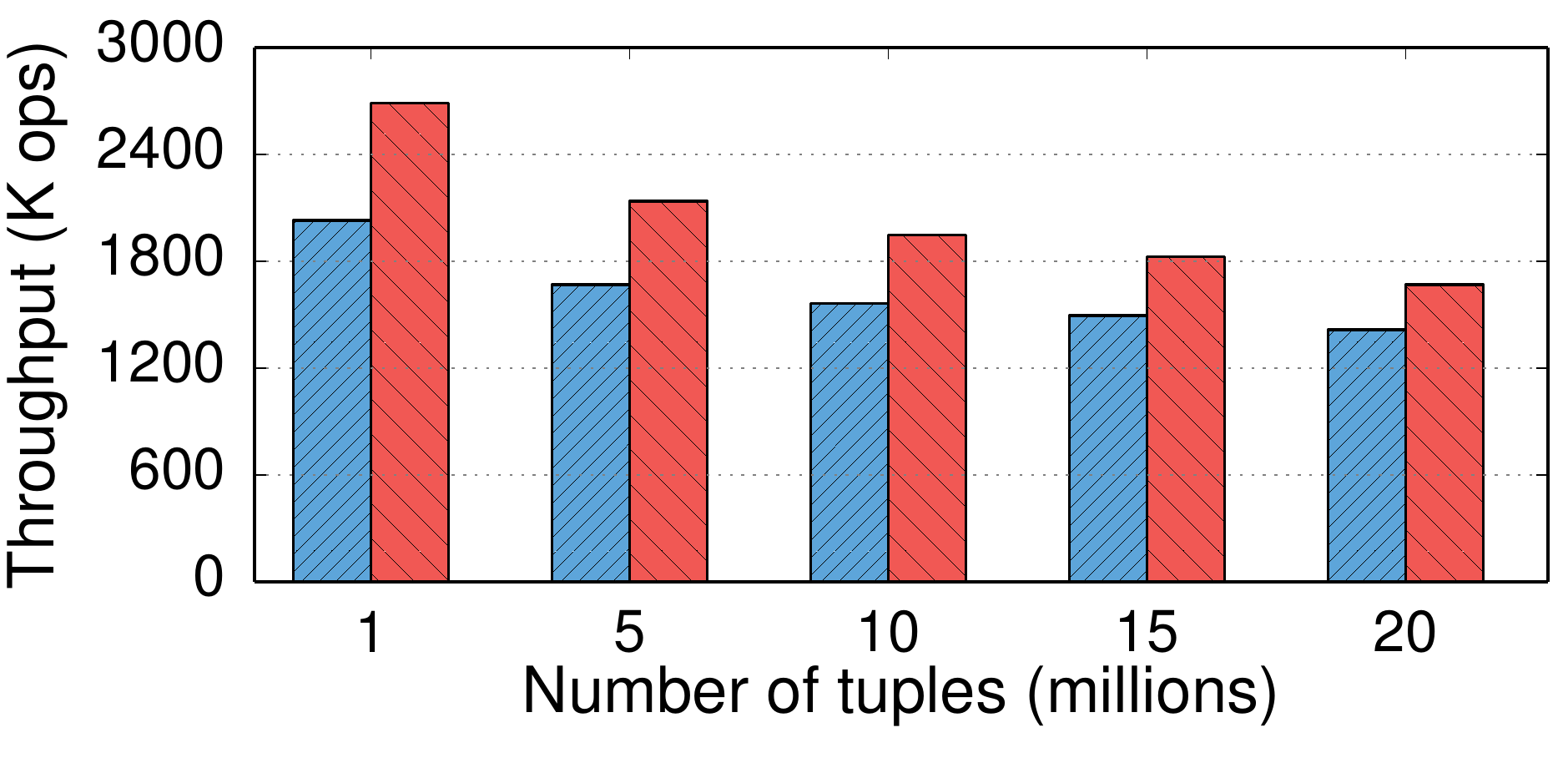}
        \label{figures:experiments:linear-index-physical}
    }
    \caption{
        Point lookup throughput with different numbers of tuples
        (\textsc{Synthetic} -- \textsc{Linear}).
    }
    \label{figures:experiments:linear-index}
\end{figure}

\begin{figure}[t!]
    \centering
    \subfloat[Logical Pointer]{
        \includegraphics[width=0.48\columnwidth]
            {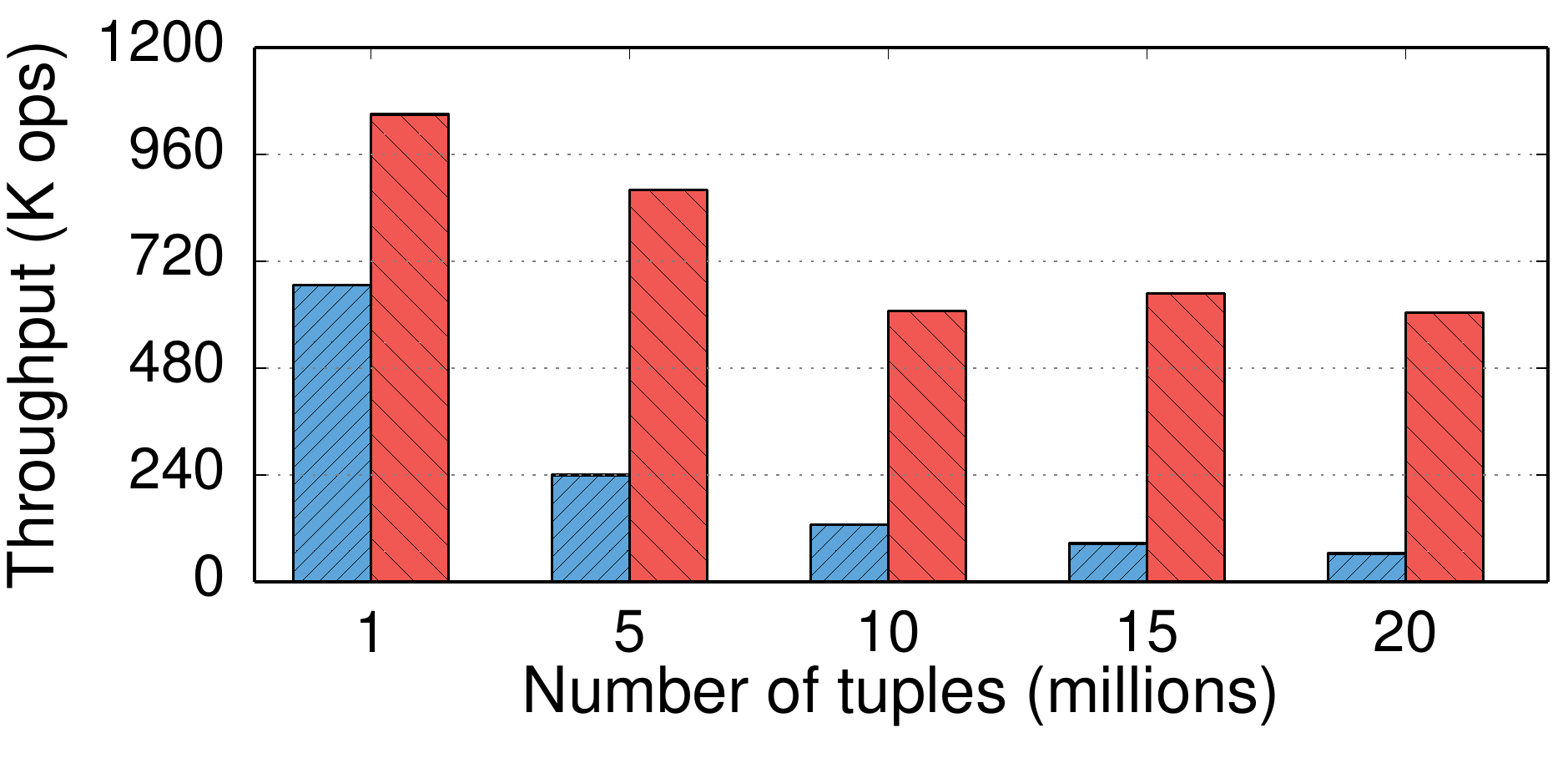}
        \label{figure:experiments:sigmoid-index-logical}
    }
    \subfloat[Physical Pointer]{
        \includegraphics[width=0.48\columnwidth]
            {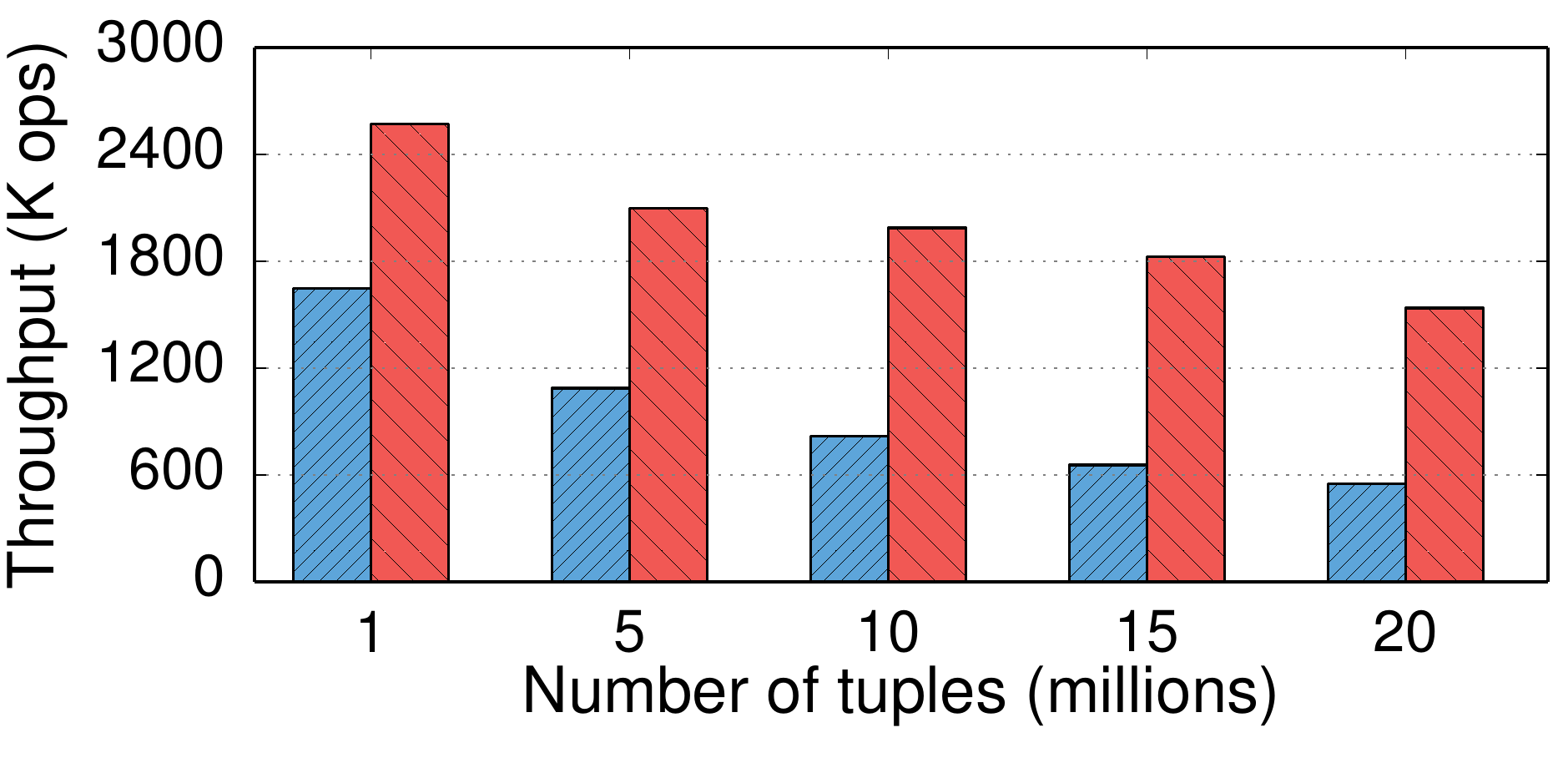}
        \label{figures:experiments:sigmoid-index-physical}
    }
    \caption{
        Point lookup throughput with different numbers of tuples
        (\textsc{Synthetic} -- \textsc{Sigmoid}).
    }
    \label{figures:experiments:sigmoid-index}
\end{figure}

Despite the high efficiency for range queries, \system suffers from some performance degradation 
on point lookups, due to its introduction of false positives.
We now measure the point lookup throughput by increasing the number of tuples in the database.
\cref{figures:experiments:linear-index} shows the result with \textsc{Linear} correlation function.
Using logical pointers for tuple identifiers, 
\system's throughput is 35\% lower than that achieved by the baseline method, 
when the number of tuples is set to 20 millions.
This is because \system's \tree lookup results in not only multiple unnecessary lookups on host 
and primary indexes, but also additional validation phase on the base table.
We also observe that such a performance gap is reduced to 15\% when switching the identifier method 
to physical pointers.
The key reason is that the absence of expensive primary index lookups helped reduce \system's performance
overhead.
\cref{figures:experiments:sigmoid-index} shows the point lookup performance when 
using \textsc{Sigmoid} correlation function.
\system's performance degrades with more tuples.
The key reason is that the increasing tuple count makes 
correlation function more difficult to model, hence \system's \tree can generate
more false positives for its subsequent processes, eventually degrading the 
performance.

\begin{figure}[t!]
    \centering
     \fbox{
     \includegraphics[width=0.9\columnwidth]
         {figures/experiments/time_breakdown_legend.pdf}
     }
    \subfloat[Logical Pointer]{
        \includegraphics[width=0.48\columnwidth]
            {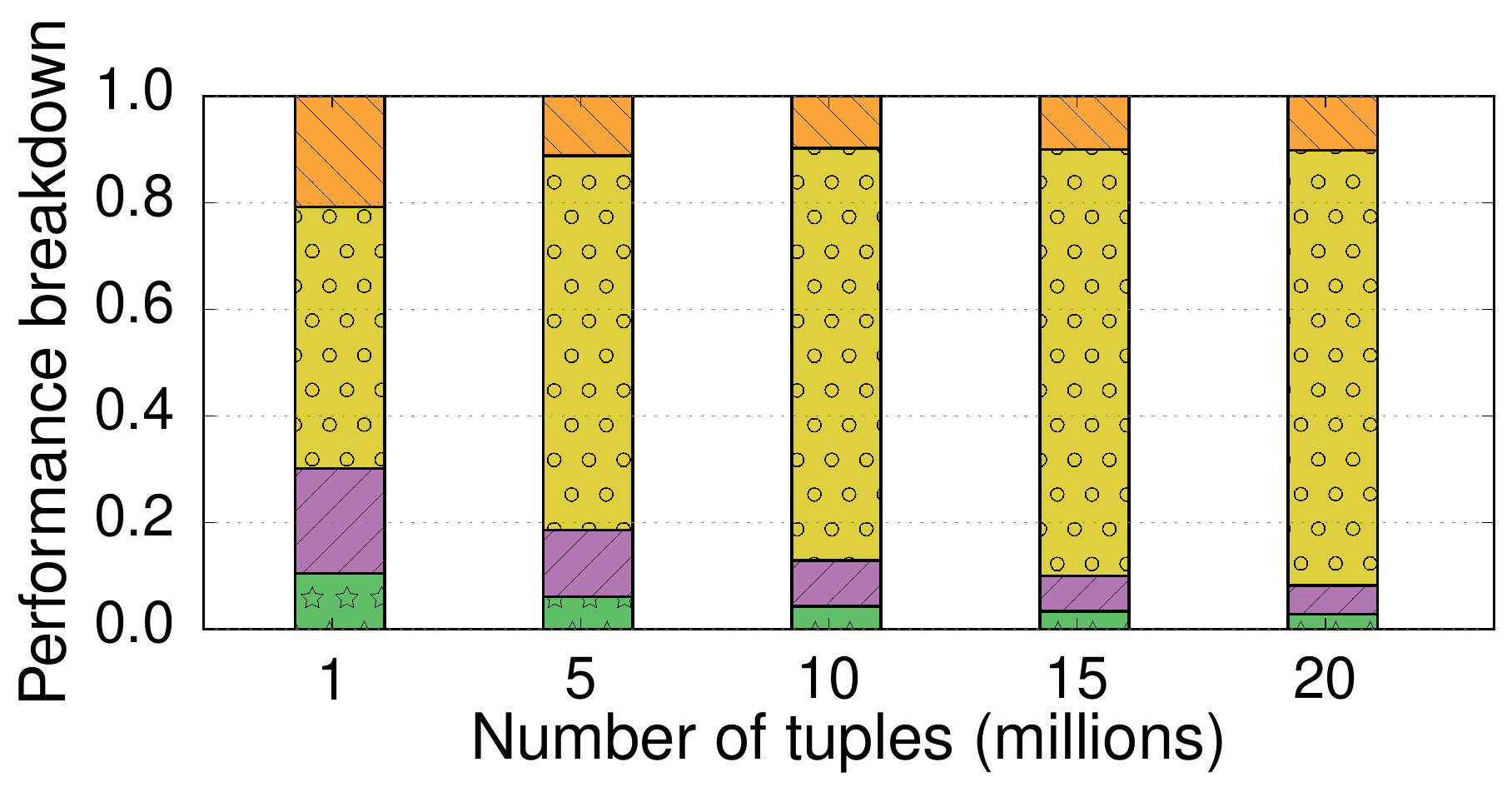}
        \label{figure:experiments:sigmoid-point-breakdown-logical}
    }
    \subfloat[Physical Pointer]{
        \includegraphics[width=0.48\columnwidth]
            {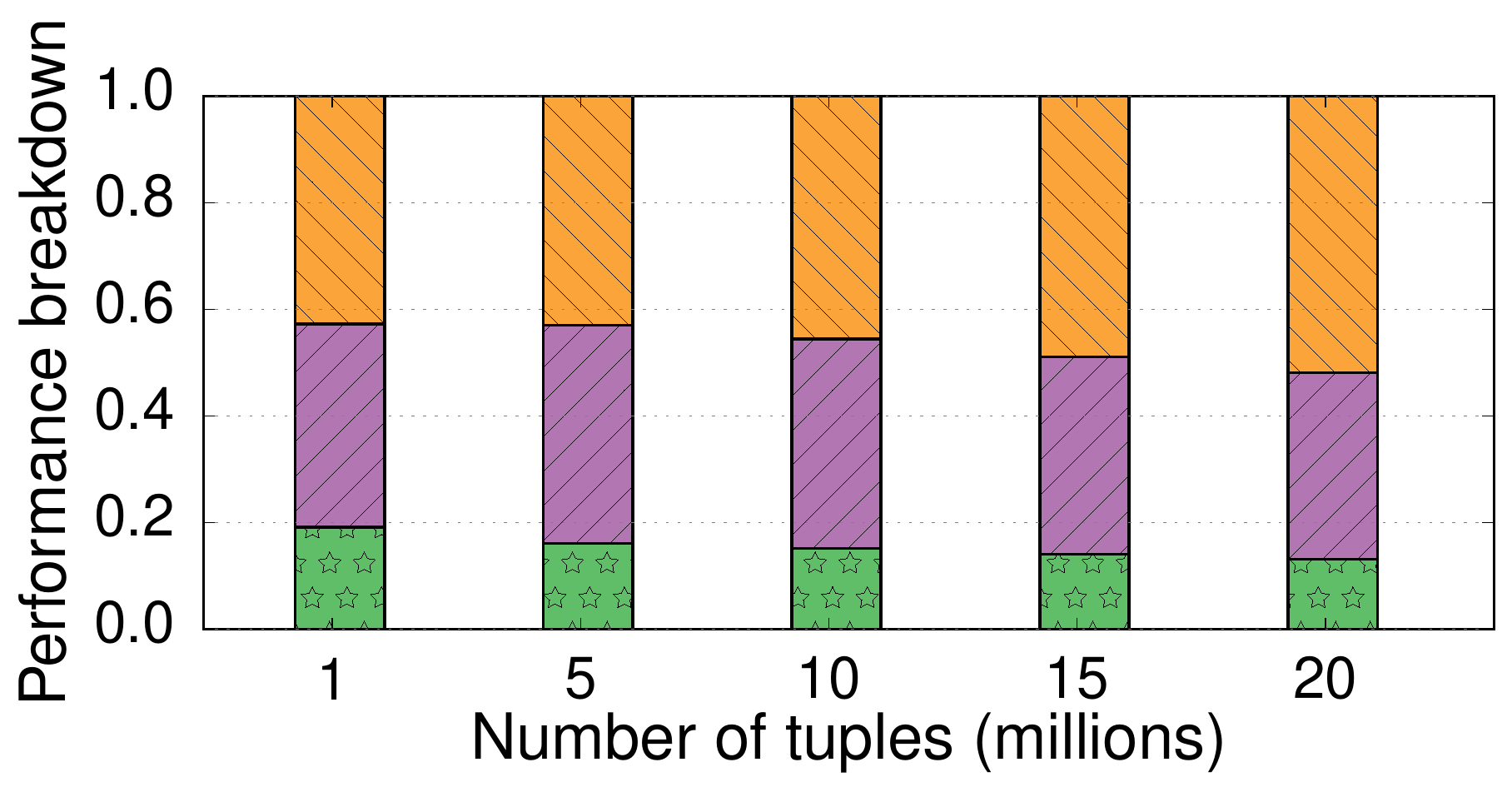}
        \label{figures:experiments:sigmoid-point-breakdown-physical}
    }
    \caption{
        \system's point lookup performance breakdown with different 
        tuple counts (\textsc{Synthetic} -- \textsc{Sigmoid}).
    }
    \label{figures:experiments:sigmoid-point-breakdown}
\end{figure}

\begin{figure}[t!]
    \centering
     \fbox{
     \includegraphics[width=0.9\columnwidth]
         {figures/experiments/time_breakdown_legend1.pdf}
     }
    \subfloat[Logical Pointer]{
        \includegraphics[width=0.48\columnwidth]
            {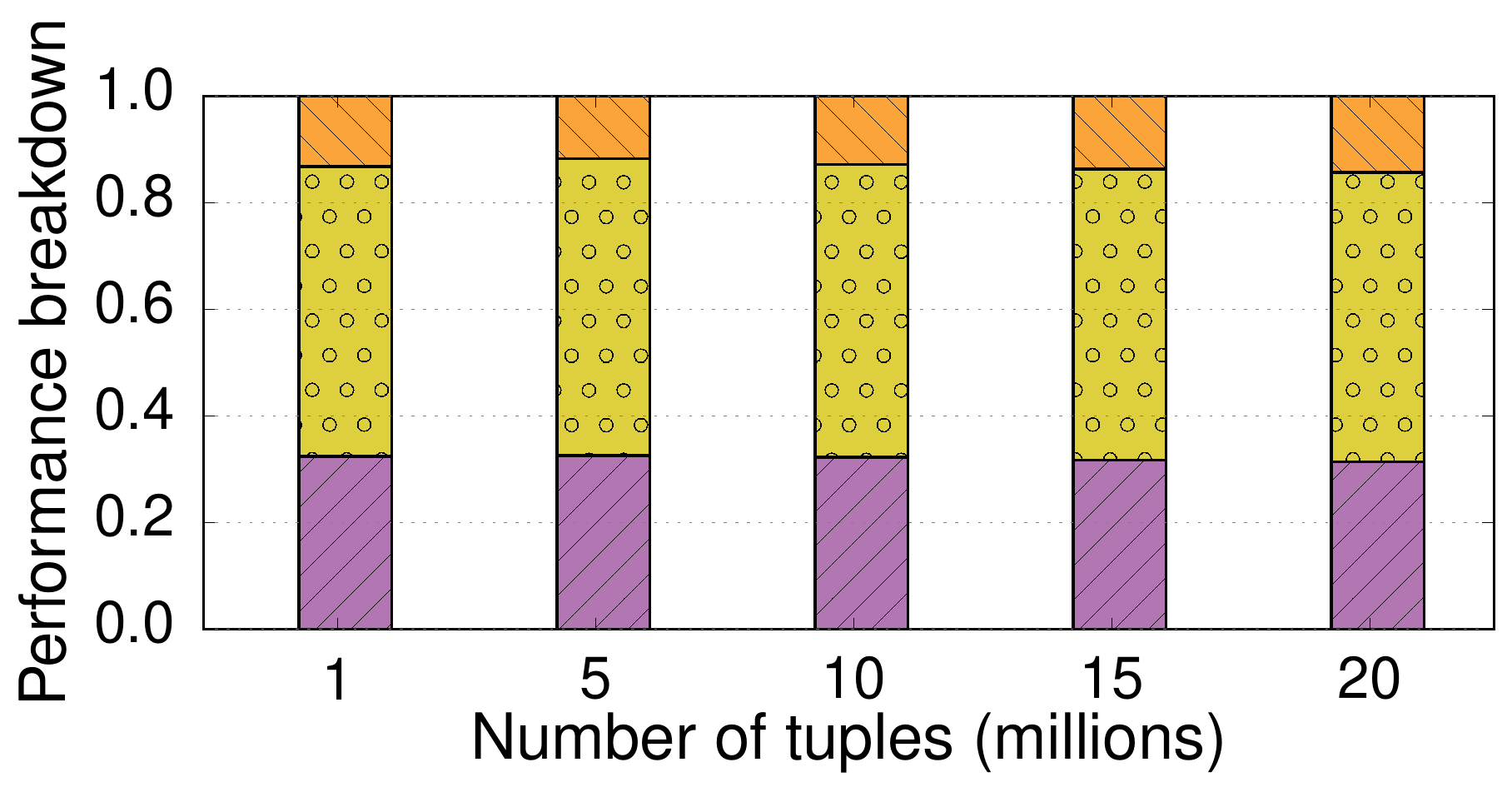}
        \label{figure:experiments:sigmoid-point-breakdown-logical-baseline}
    }
    \subfloat[Physical Pointer]{
        \includegraphics[width=0.48\columnwidth]
            {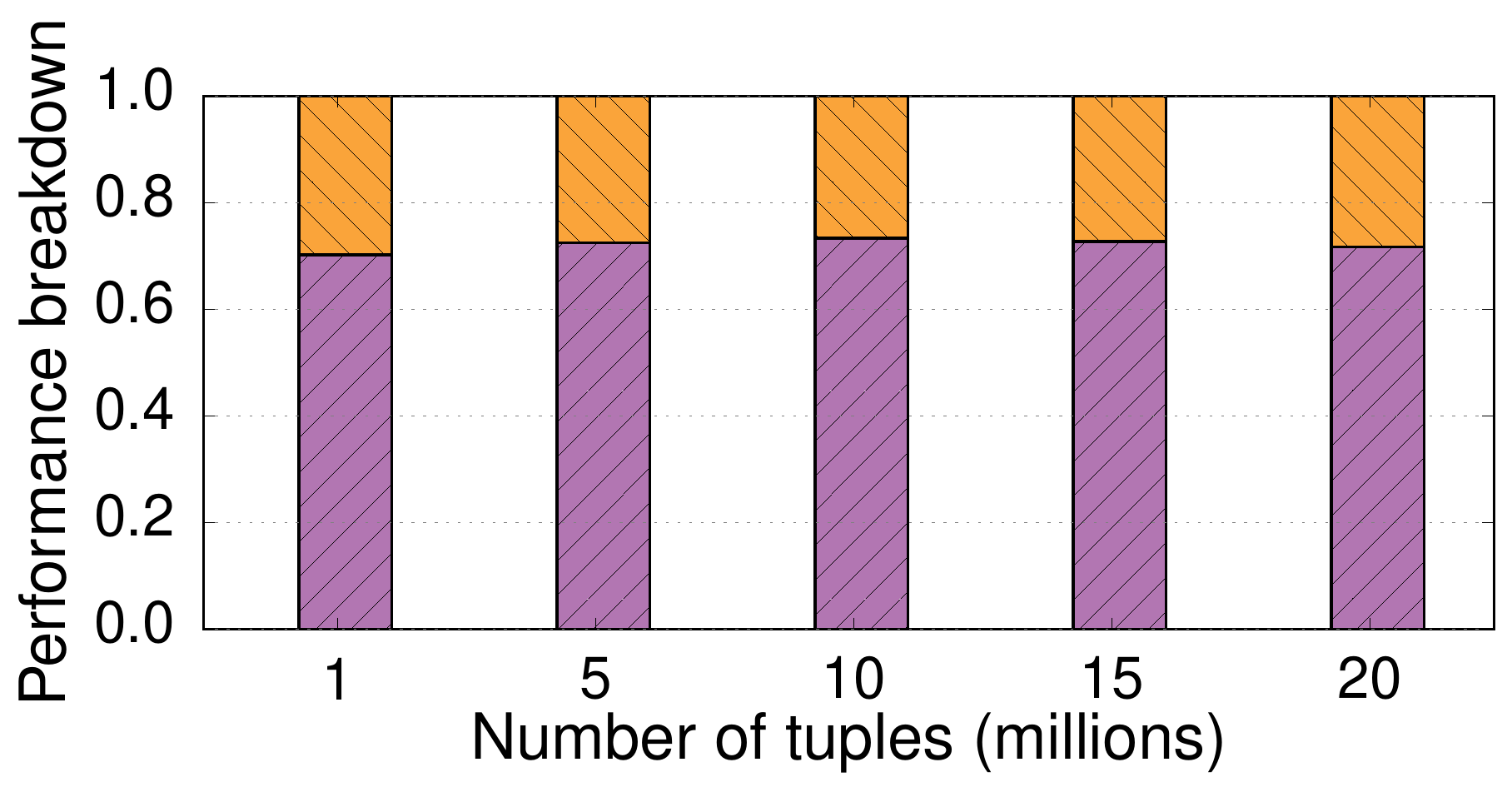}
        \label{figures:experiments:sigmoid-point-breakdown-physical-baseline}
    }
    \caption{
        Baseline's point lookup performance breakdown with different 
        tuple counts (\textsc{Synthetic} -- \textsc{Sigmoid}).
    }
    \label{figures:experiments:sigmoid-point-breakdown-baseline}
\end{figure}

We further perform a performance breakdown to better understand the point queries.
\cref{figures:experiments:sigmoid-point-breakdown} and \cref{figures:experiments:sigmoid-point-breakdown-baseline}
show the results.
There are two points worth noticing.
First, using logical pointers, \system spends an increasing amount of time on primary index lookup
when the tuple count increases.
As explained above, this is because the larger tuple count indicates more complex correlation relations, 
and \system has to waste more time on retrieving unqualified tuples from the primary index.
Second, compared to the baseline method, \system spends a larger portion of time on the base table.
This is because \system has to validate every single tuple fetched from the base table to 
filter out false positives.

\begin{figure}[t!]
    \centering
     \fbox{
     \includegraphics[width=0.9\columnwidth]
         {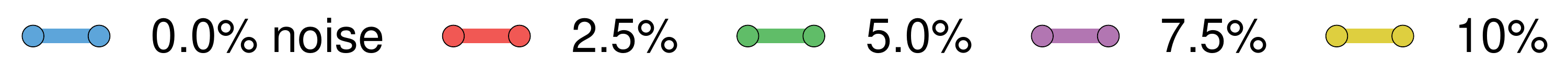}
     }
    \subfloat[\textsc{Linear} Correlation]{
        \includegraphics[width=0.48\columnwidth]
            {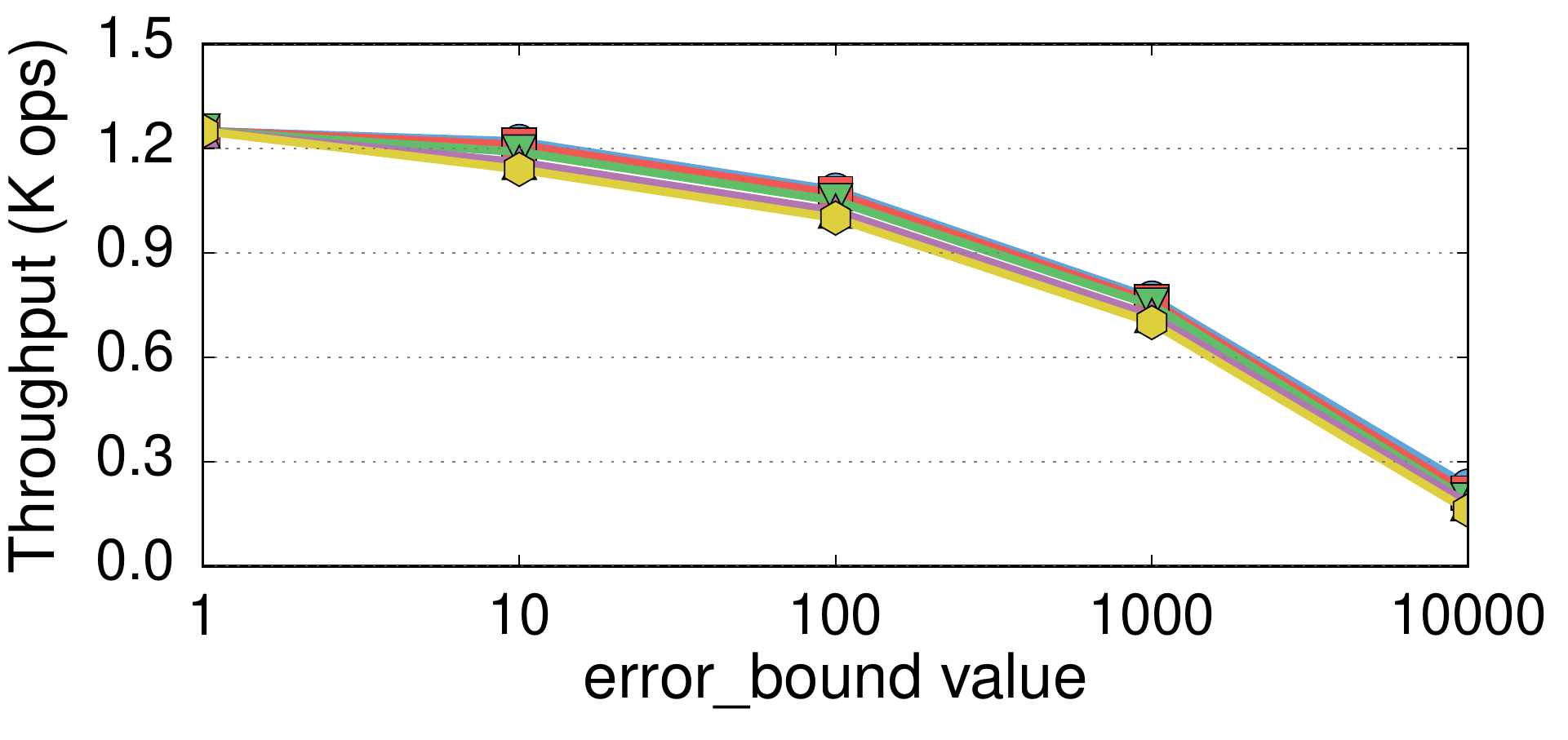}
        \label{figure:experiments:tradeoff-throughput-linear-range}
    }
    \subfloat[\textsc{Sigmoid} Correlation]{
        \includegraphics[width=0.48\columnwidth]
            {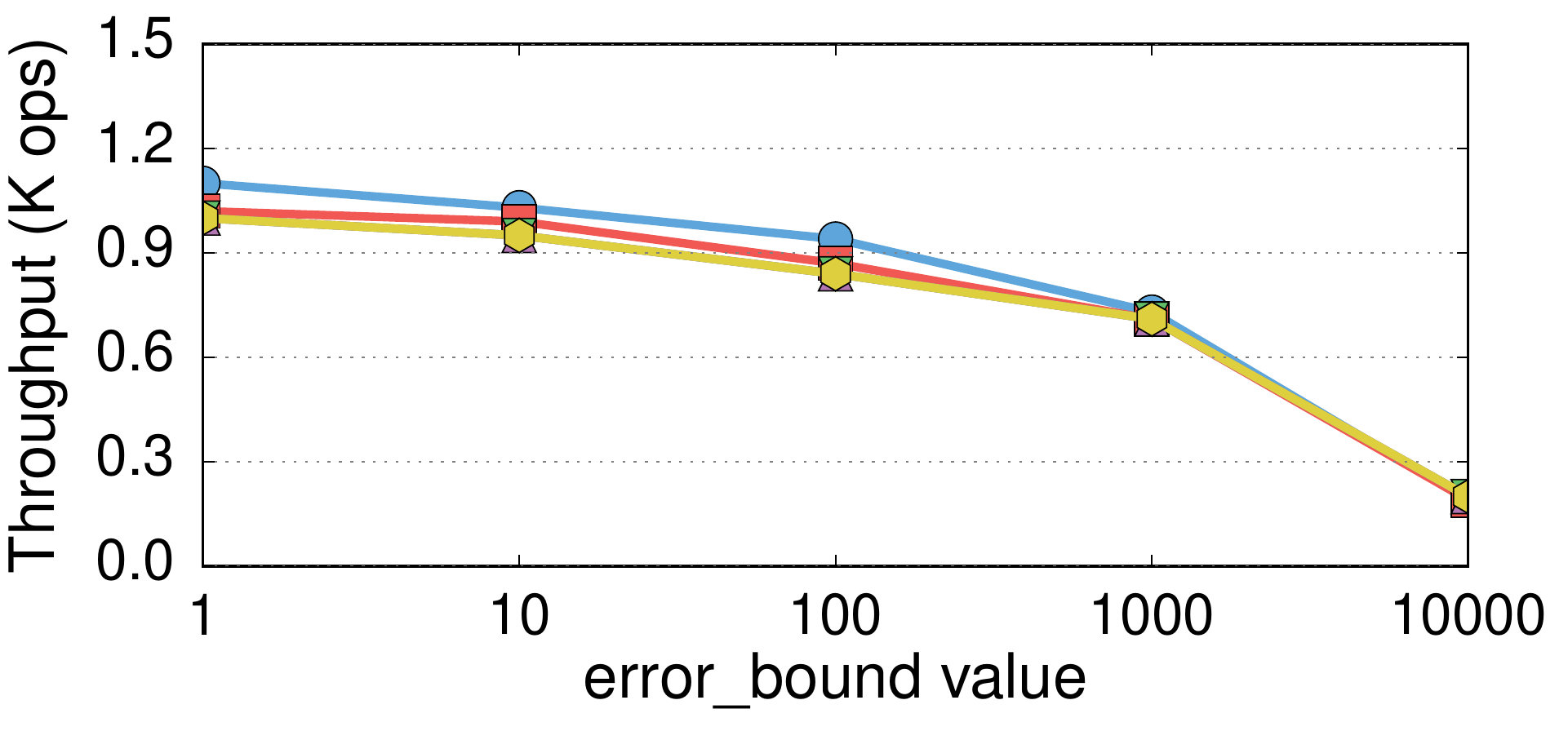}
        \label{figure:experiments:tradeoff-throughput-sigmoid-range}
    }
    \caption{
        Range lookup throughput with different percentages of injected noises and 
        $error\_bound$ values (\textsc{Synthetic}).
    }
    \label{figures:experiments:tradeoff-throughput-range}
\end{figure}

\begin{figure}[t!]
    \centering
    \subfloat[\textsc{Linear} Correlation]{
        \includegraphics[width=0.48\columnwidth]
            {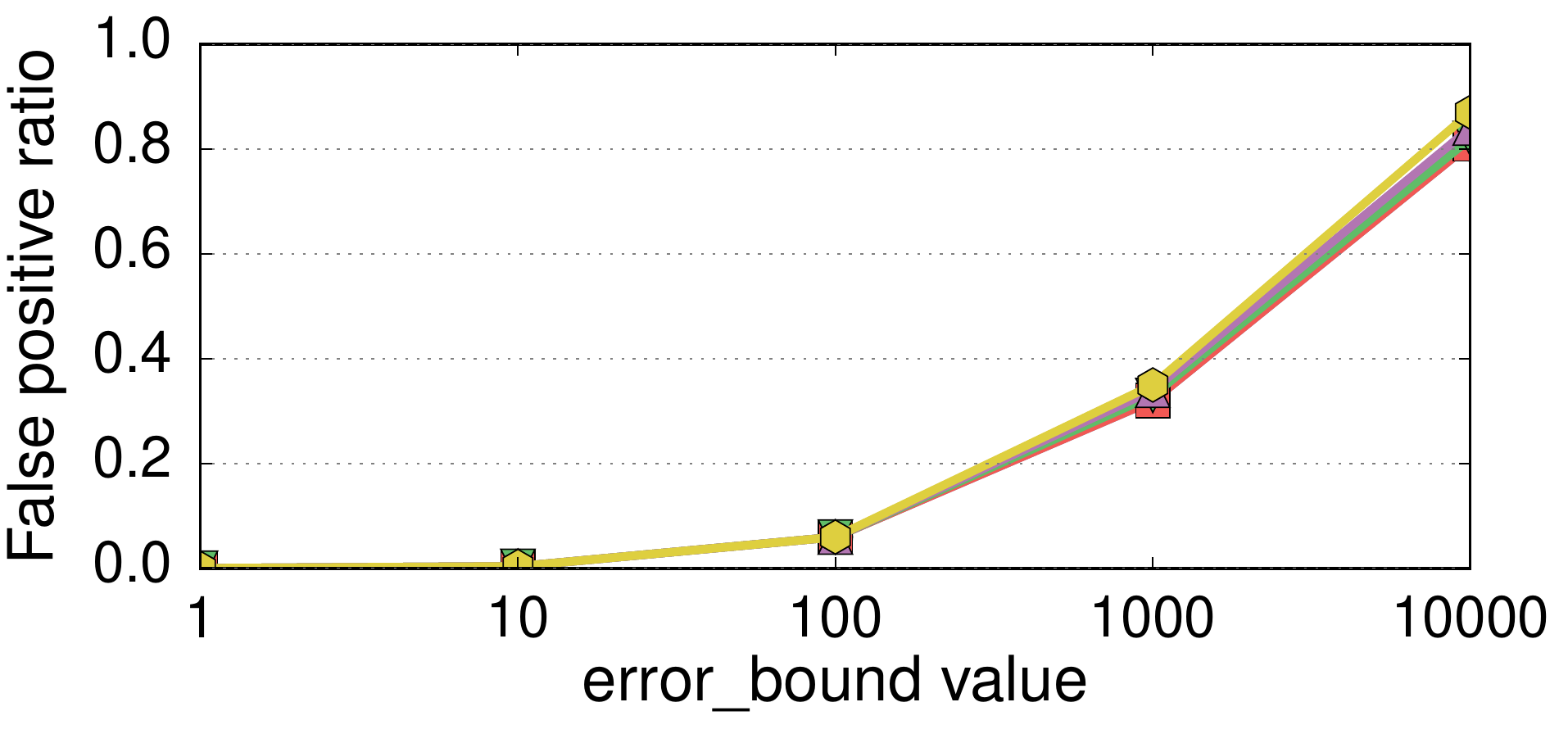}
        \label{figure:experiments:tradeoff-fpr-linear-range}
    }
    \subfloat[\textsc{Sigmoid} Correlation]{
        \includegraphics[width=0.48\columnwidth]
            {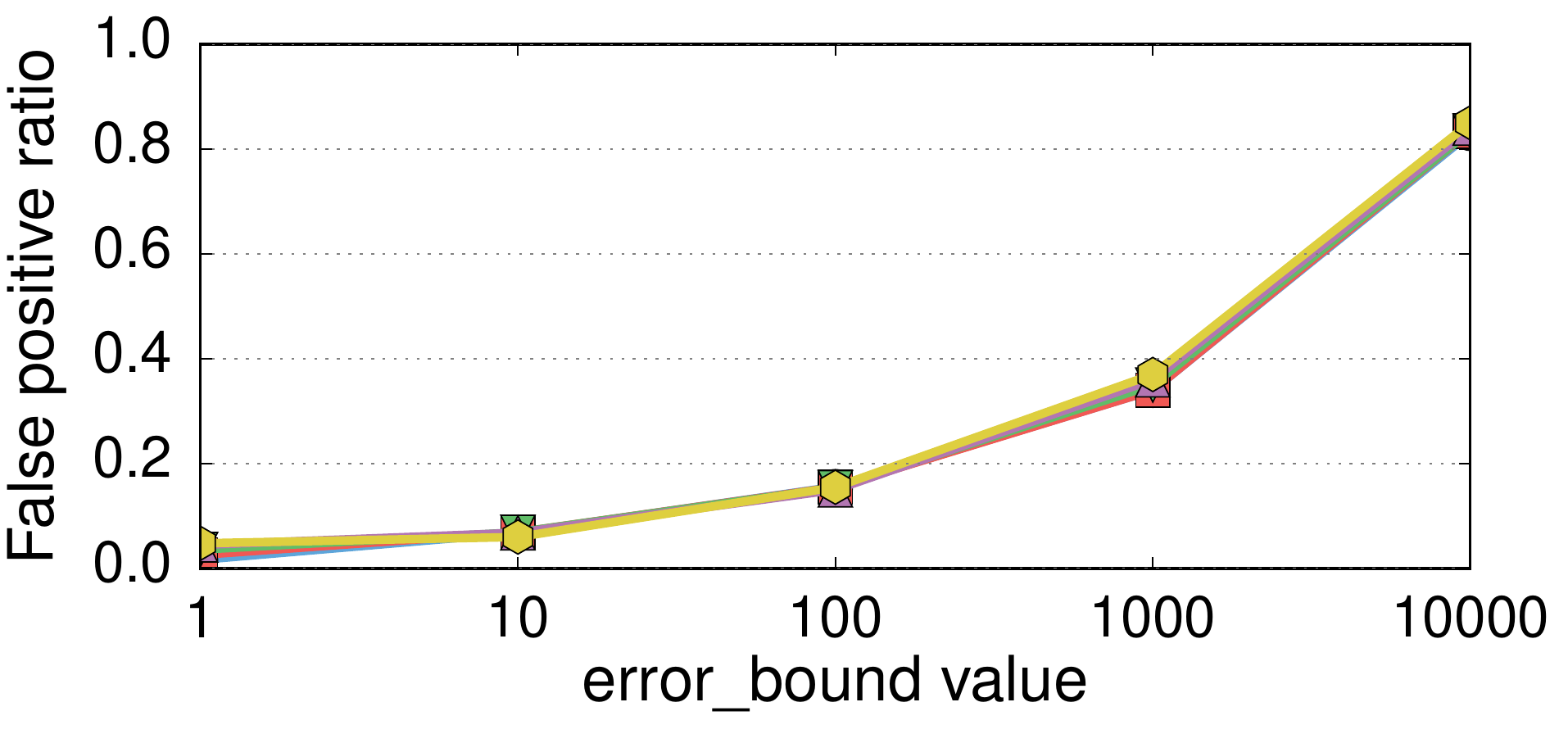}
        \label{figure:experiments:tradeoff-fpr-sigmoid-range}
    }
    \caption{
        Range lookup false positive ratio with different percentages of injected noises and 
        $error\_bound$ values (\textsc{Synthetic}).
    }
    \label{figures:experiments:tradeoff-fpr-range}
\end{figure}

\rev{
\mbox{\system}'s performance can be affected by the correlation quality as well as the 
user-defined parameters. Now we control the percentages of the injected noise 
as well as the value of $error\_bound$ to see how \mbox{\system} behaves.
\mbox{\cref{figures:experiments:tradeoff-throughput-range}} shows 
the range lookup (selectivity set to 0.01\%) throughput 
with different percentages of injected noise and different $error\_bound$ values.
We use logical pointers in the experiment. We observe that \mbox{\system}'s performance drops
drastically with the increase of $error\_bound$. This is because 
larger $error\_bound$ indicates more false positives, and \mbox{\system} has to perform
redundant secondary index lookups and rely on the validation phase to remove 
unqualified tuples. 
This is confirmed by \mbox{\cref{figures:experiments:tradeoff-fpr-range}}, which shows 
that the false positive rate reaches up to 80\% when $error\_bound$ is set to 10,000.
An interesting finding is that \mbox{\system}'s performance remains stable with the increase 
of noise percentage. The key reason is that \mbox{\system} is capable to capture any outlier 
that fall beyond its generated linear function, and it can effectively find 
these outliers from the corresponding outlier buffers.
} 

\begin{figure}[t!]
    \centering
    \subfloat[\textsc{Linear} Correlation]{
        \includegraphics[width=0.48\columnwidth]
            {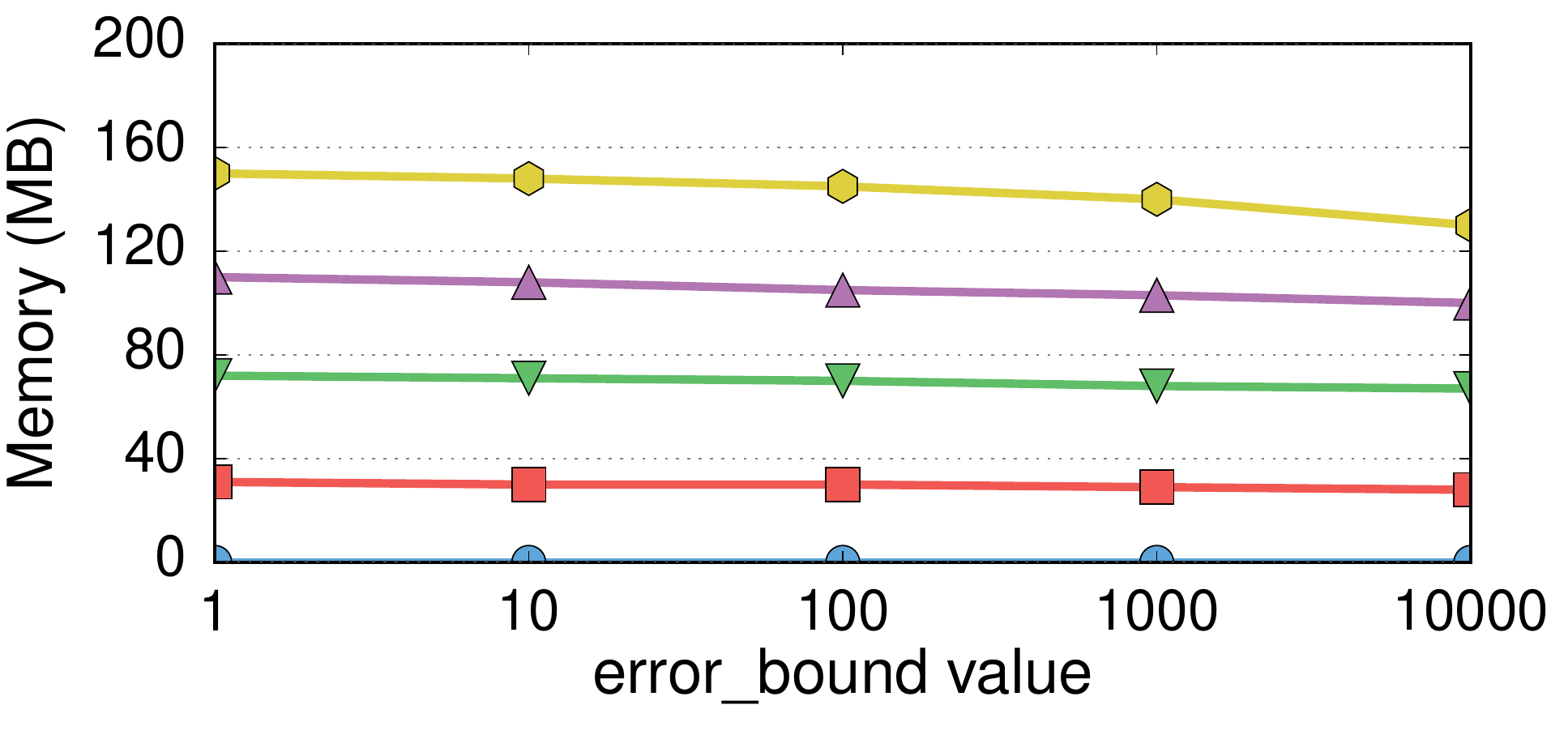}
        \label{figure:experiments:tradeoff-memory-linear-range}
    }
    \subfloat[\textsc{Sigmoid} Correlation]{
        \includegraphics[width=0.48\columnwidth]
            {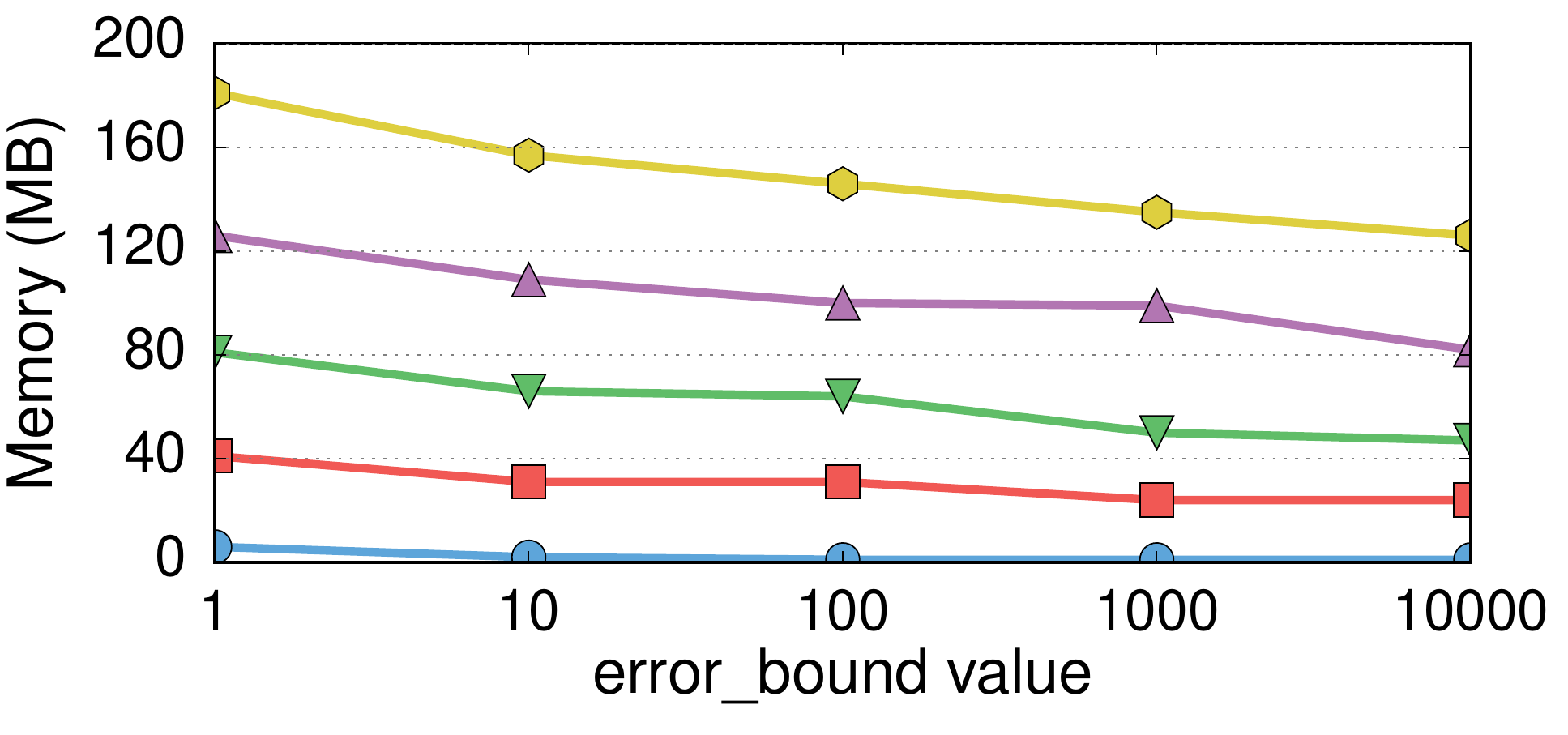}
        \label{figure:experiments:tradeoff-memory-sigmoid-range}
    }
    \caption{
        Memory consumption with different percentages of injected noises and 
        $error\_bound$ values (\textsc{Synthetic}).
    }
    \label{figures:experiments:tradeoff-memory-range}
\end{figure}

\rev{
\mbox{\cref{figures:experiments:tradeoff-memory-range}} further shows how noisy data and 
$error\_bound$ values affect \mbox{\system}'s memory consumption.
Our first finding is that the memory consumption grows linearly with the increase of 
noise percentage. This is because \mbox{\system} stores noisy data in outlier buffers, as 
explained above.
Our second finding is that the memory consumption declines by increasing the $error\_bound$
values. This is because larger $error\_bound$ covers more data, and hence \mbox{\system}'s
\mbox{\tree} can construct less nodes to capture the correlations.
One thing worth mentioning is the memory spent for capturing the \mbox{\textsc{Linear}} and
\mbox{\textsc{Sigmoid}} correlations tend to be the same (close to 120 MB) when $error\_bound$ is set to 10,000.
This matches our expectation, as \mbox{\system}'s \mbox{\tree} spent most of the space for storing 
outliers, and only little space is used for model correlations.
}

\rev{
We observe that memory usage for capturing \mbox{\textsc{Sigmoid}} drops a lot when changing 
$error\_bound$ from 1 to 10, but we do not really see a noticeable decline in throughput. 
This is because \mbox{\tree} can identify too many data as outliers
with $error\_bound$ set to 1. With the increase of $error\_bound$, it efficiently 
captures correlations using linear regression models. It also navigates lookups 
using computation rather than chasing pointers, hence achieving good performance.
}

\subsection{Space Consumption}

\system trades performance for space efficiency. Its goal is to greatly reduce
the storage space while achieving ``good enough'' tuple retrieval speed.
In the last subsection, we showed that \system yields competitive performance
to the conventional secondary index mechanisms when supporting database operations, 
especially range queries. 
Now we measure \system's space efficiency using the \textsc{Synthetic} application.
\rev{We still run all the experiments in \mbox{\database}.}


\begin{figure}[t!]
    \centering
    \subfloat[\textsc{Linear} Correlation]{
        \includegraphics[width=0.48\columnwidth]
            {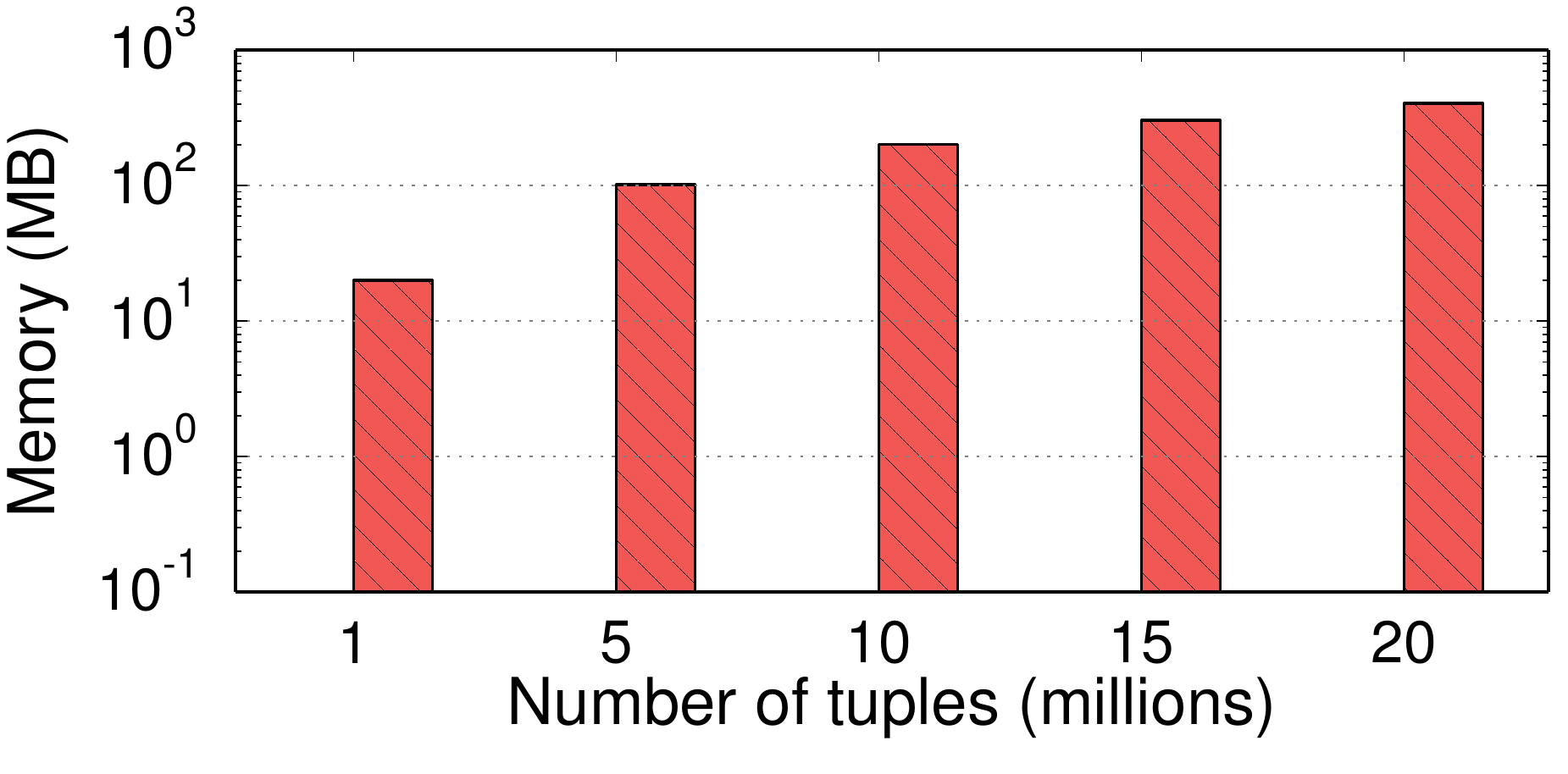}
        \label{figure:experiments:linear-index-memory}
    }
    \subfloat[\textsc{Sigmoid} Correlation]{
        \includegraphics[width=0.48\columnwidth]
            {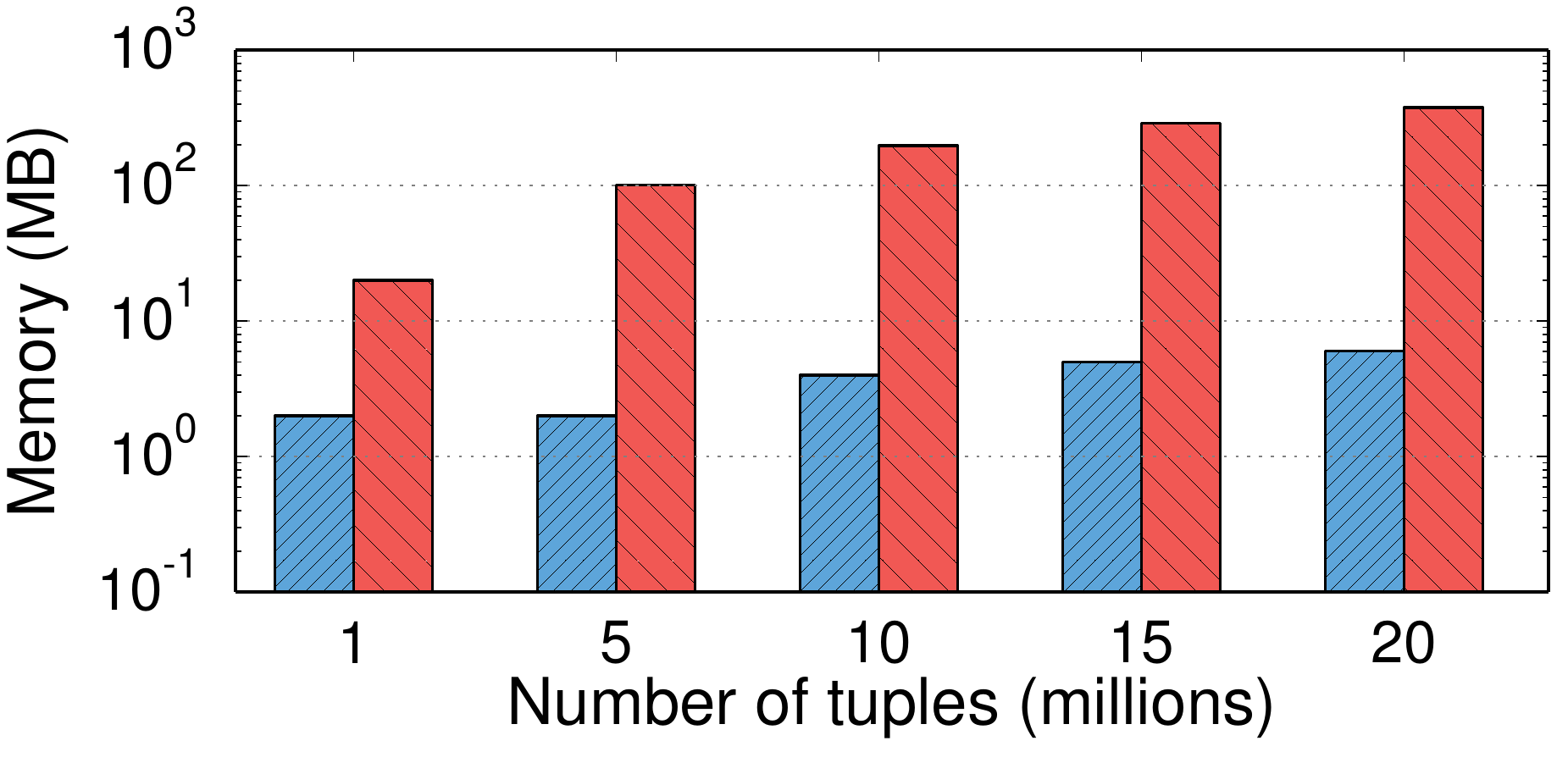}
        \label{figure:experiments:sigmoid-index-memory}
    }
    \caption{
        Index memory consumption with different numbers of tuples (\textsc{Synthetic}).
    }
    \label{figures:experiments:memory}
\end{figure}

\begin{figure}[t!]
    \centering
    \subfloat[Memory Consumption]{
        \includegraphics[width=0.48\columnwidth]
            {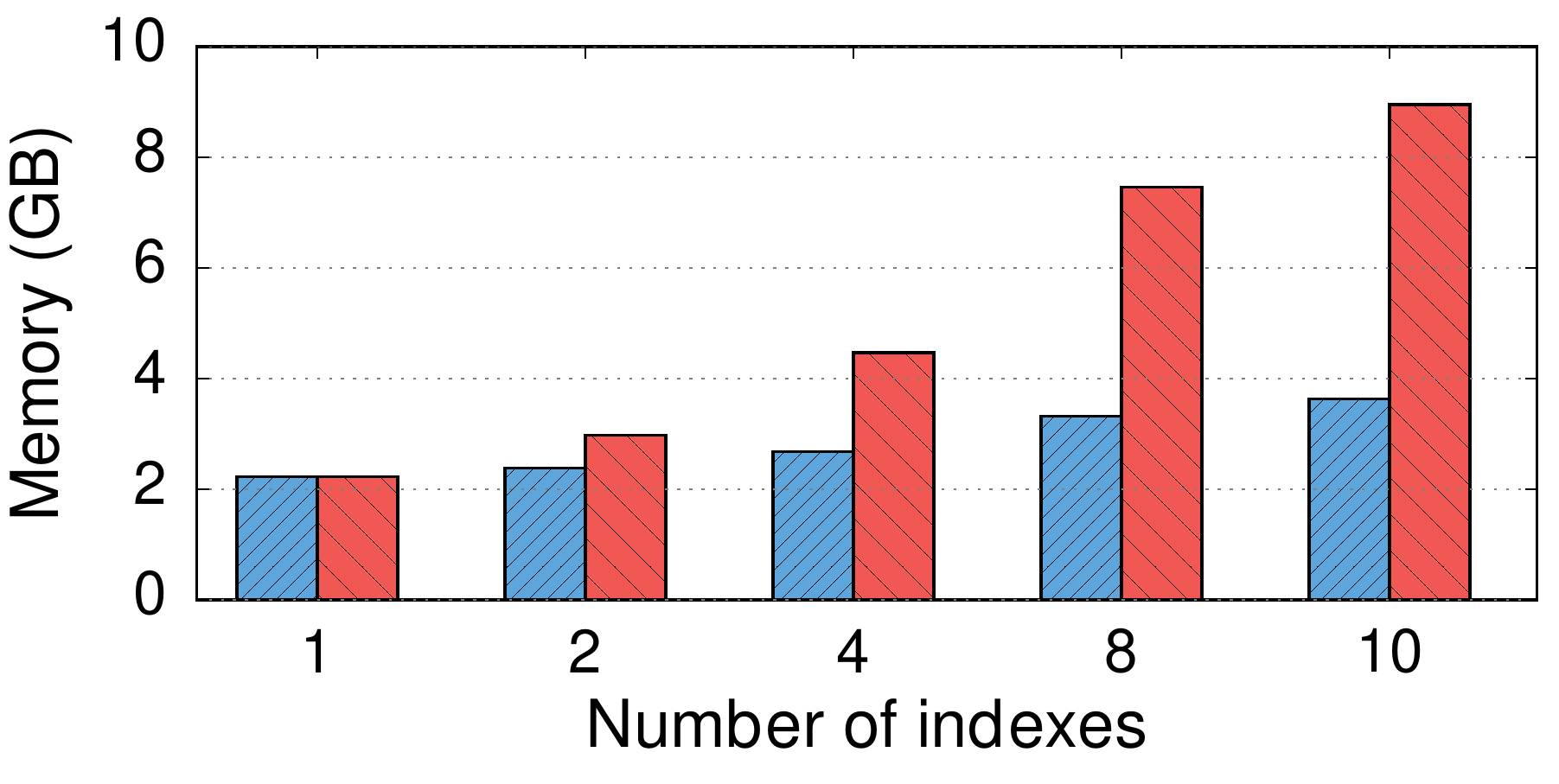}
        \label{figure:experiments:linear-memory-indexcount}
    }
    \subfloat[Space Breakdown]{
        \includegraphics[width=0.48\columnwidth]
            {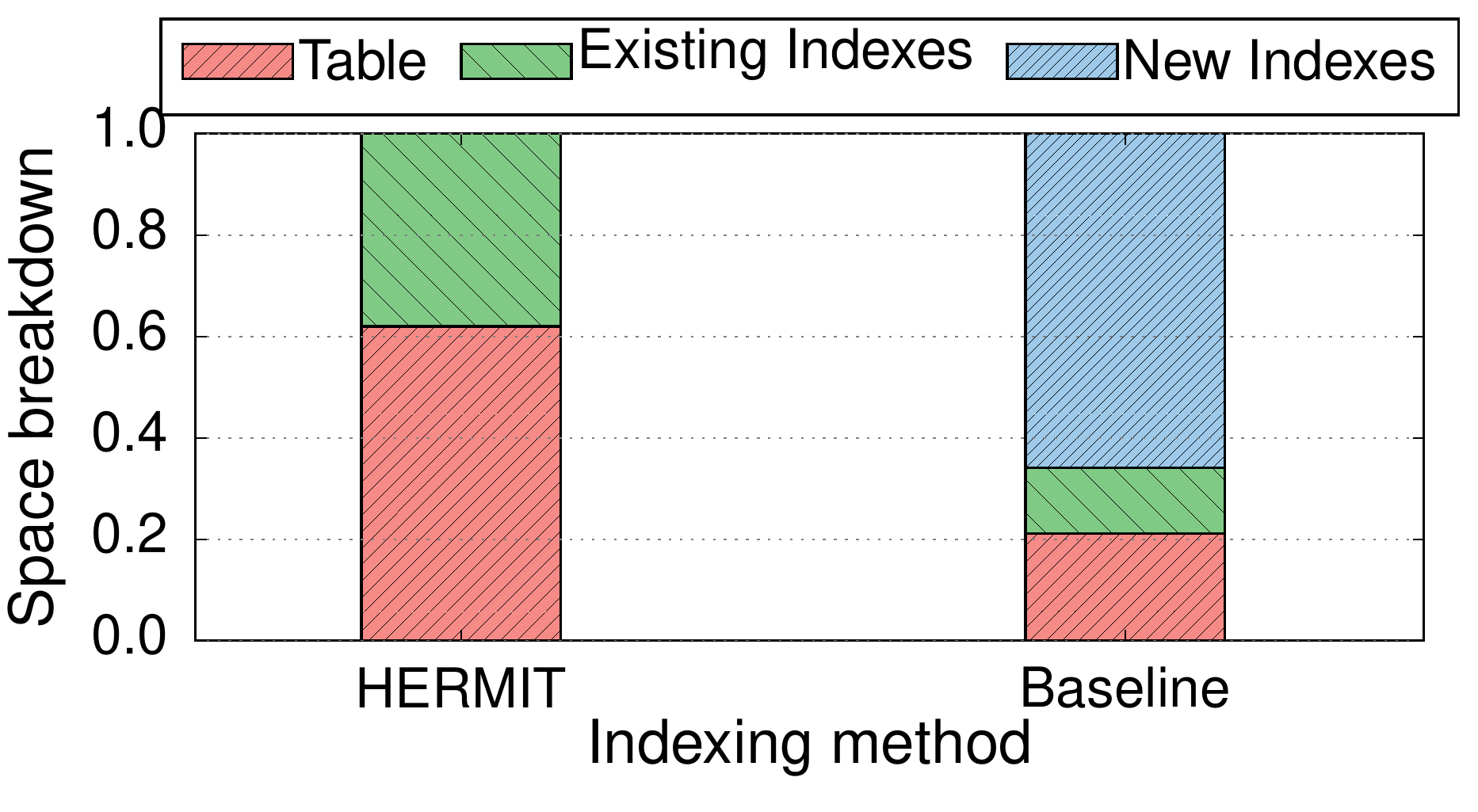}
        \label{figure:experiments:linear-memory-indexcount-breakdown}
    }
    \caption{
        Total memory consumption with different numbers of tuples (\textsc{Synthetic} -- \textsc{linear}).
    }
    \label{figures:experiments:linear-memory}
\end{figure}

The first experiment measures the amount of memory used respectively 
by \tree and conventional secondary index on $col_C$. 
\cref{figures:experiments:memory} shows that, compared to the baseline 
solution, \system takes little space to index the column.
An extreme case is to use \system's \tree for capturing 
\textsc{Linear} correlation function. In this case, \tree only needs to 
use a constant amount of memory (a few bytes) to record the linear function's parameters.
When modeling \textsc{Sigmoid} function, 
\tree consumes more memory (less than 10 MB) as the number of tuples increases,
because \tree needs to construct more leaf nodes to better fit the correlation curve.
However, \system's memory consumption is still negligible compared to the baseline solution, which takes close to 400 MB.


Next, we measure the overall memory consumption caused by \system.
Other than the existing indexes on $col_A$ and $col_B$, we add some 
additional columns and build one index on each of them. All these 
newly added columns are correlated to $col_B$.
\cref{figure:experiments:linear-memory-indexcount} shows that, when adopting 
the baseline solution, the amount of memory consumption grows near linearly
with the increasing number of newly added indexes. 
Specifically, the database used up to
8.5 GB memory when supporting 10 secondary indexes.
In contrast, when adopting \system, the database only consumes 2.4 GB memory,
which is a significant gain in memory utilization 
compared to the baseline solution.
\cref{figure:experiments:linear-memory-indexcount-breakdown} further depicts 
the memory usage breakdown when the number of indexes is set to 10.
The space consumed by \system's \tree is negligible compared to that 
used by the base table and the primary index. However, when adopting 
the baseline solution, the database application has to use over 70\% of 
the memory to maintain the secondary indexes. This result further 
confirms \system's space efficiency.

\begin{figure}[t!]
    \centering
    \subfloat[\textsc{Linear} Correlation]{
        \includegraphics[width=0.48\columnwidth]
            {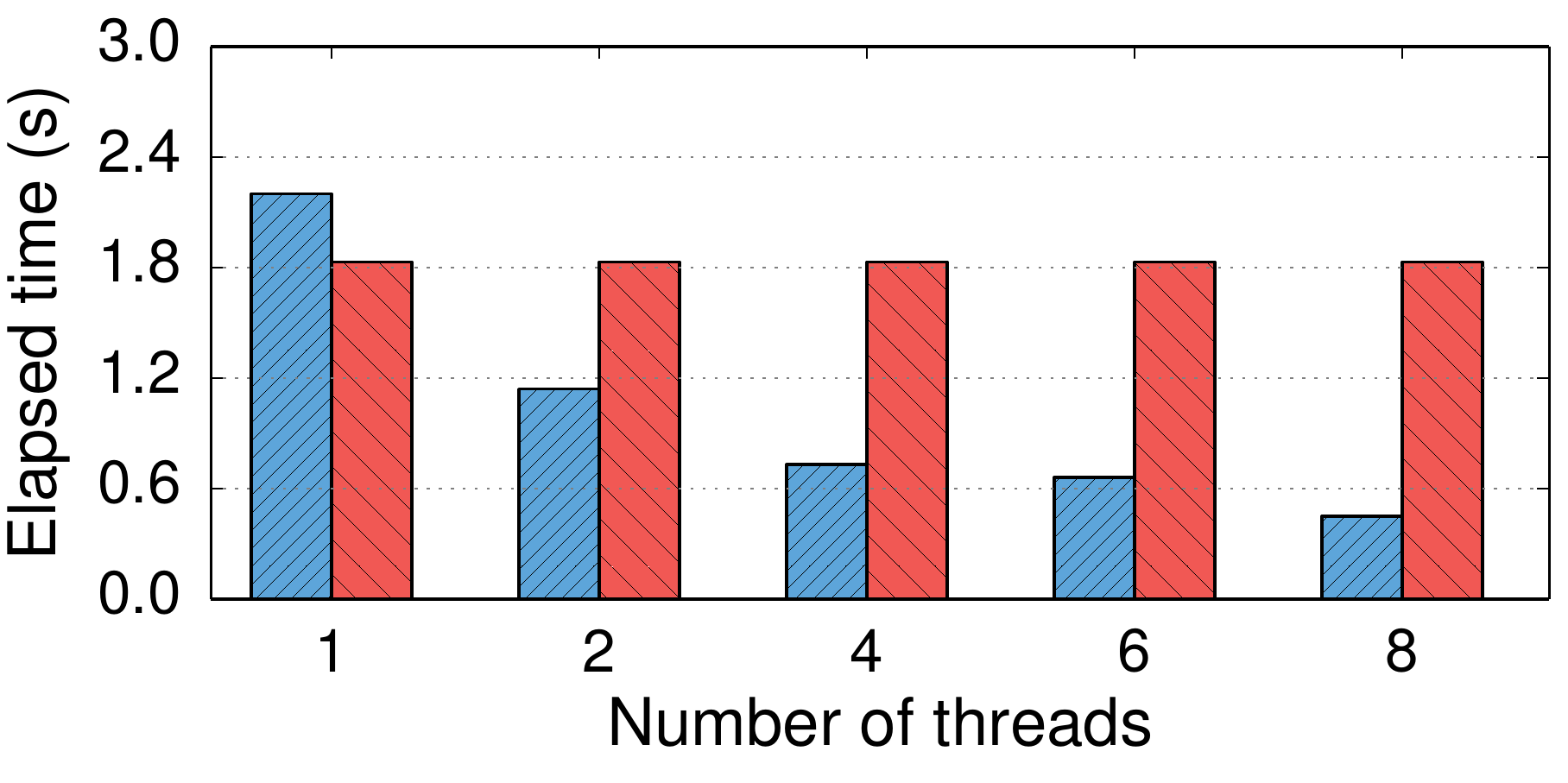}
        \label{figure:experiments:linear-construction}
    }
    \subfloat[\textsc{Sigmoid} Correlation]{
        \includegraphics[width=0.48\columnwidth]
            {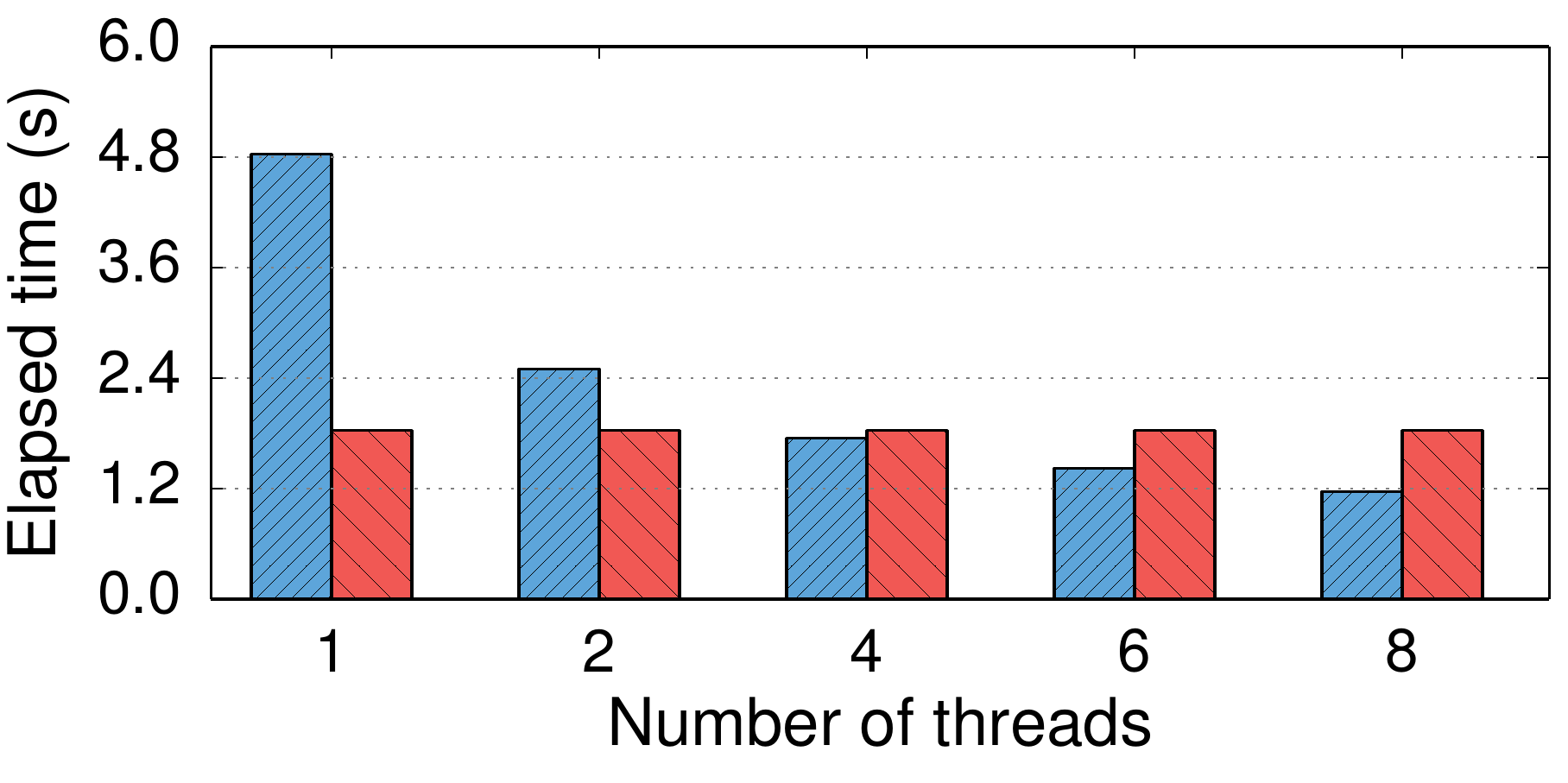}
        \label{figures:experiments:sigmoid-construction}
    }
    \caption{
        Index construction time with different numbers of threads (\textsc{Synthetic}).
    }
    \label{figures:experiments:construction}
\end{figure}
\begin{figure}[t!]
    \centering
    \subfloat[Insertion Throughput]{
        \includegraphics[width=0.48\columnwidth]
            {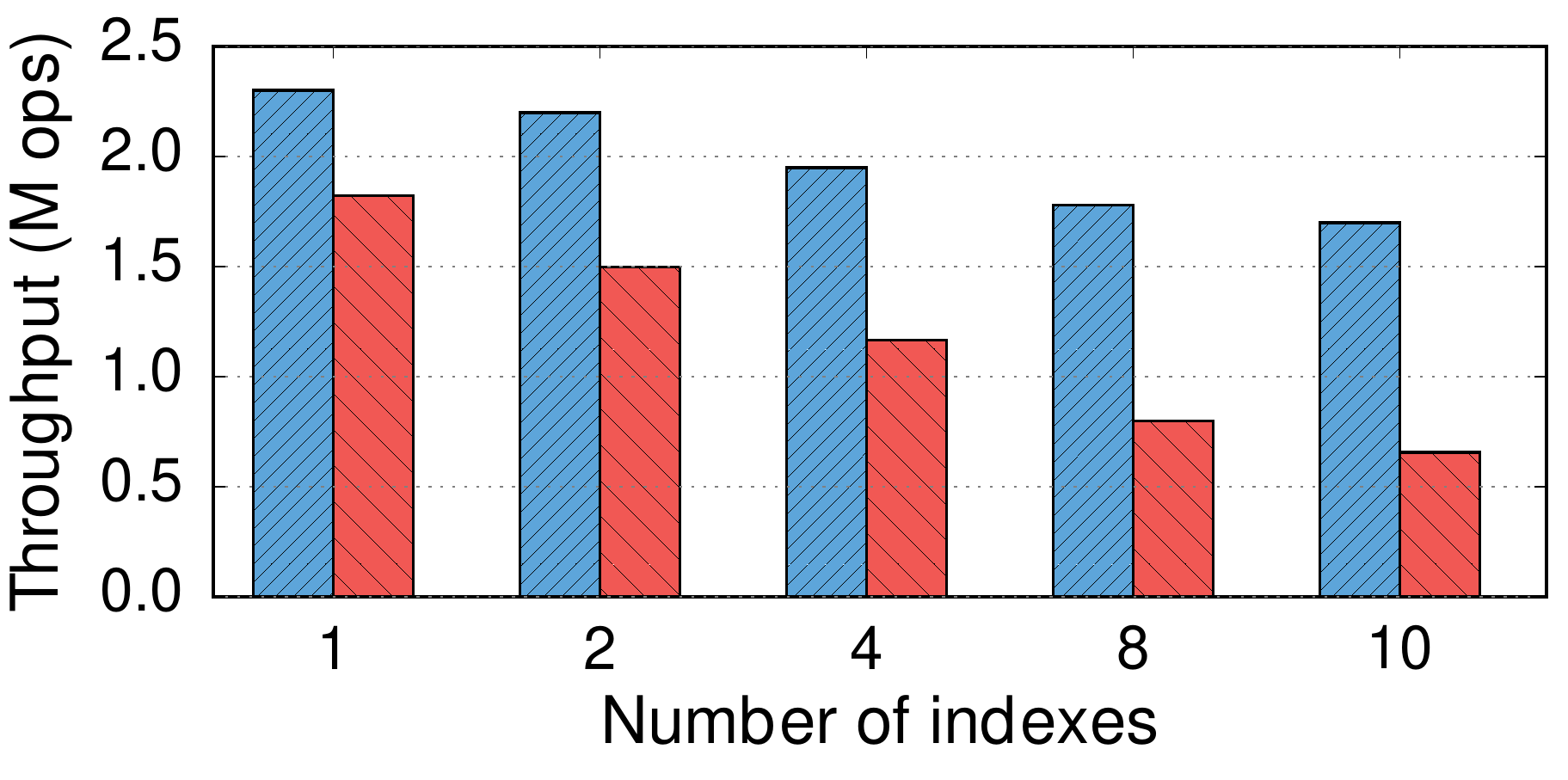}
        \label{figure:experiments:linear-insert-indexcount}
    }
    \subfloat[Performance Breakdown]{
        \includegraphics[width=0.48\columnwidth]
            {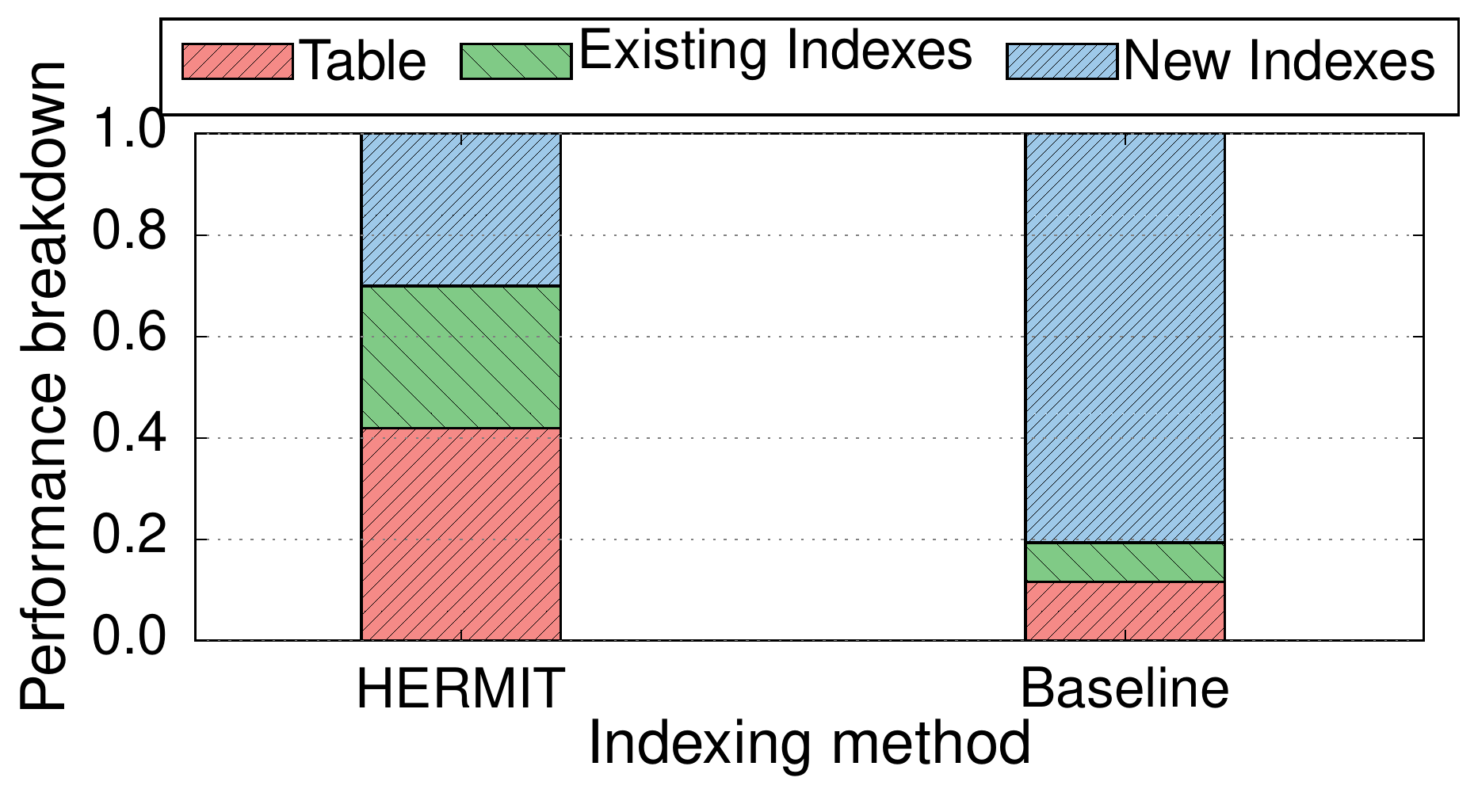}
        \label{figure:experiments:linear-insert-indexcount-breakdown}
    }
    \caption{
        Index insertion throughput with different numbers of indexes 
        (\textsc{Synthetic} -- \textsc{Linear}).
    }
    \label{figures:experiments:linear-insert}
\end{figure}

\subsection{Construction}

The construction of \system's \tree is different from that of the conventional B+-tree-like 
index structures. 
In this experiment, \rev{we measure the time for constructing \mbox{\tree} in 
\mbox{\database} with different numbers of threads}. 
We compare the results with that 
obtained by constructing the B+-tree index.
The B+-tree is built using single-thread bulk loading, 
as it currently does not support multithreading mode. 
We leave the comparison with concurrent B+-tree construction
as a future work.
The results in \cref{figures:experiments:construction} contain two interesting findings. 
First, \tree needs more time to finish the index construction when confronting complex 
correlation functions, such as \textsc{Sigmoid}. This is because \tree needs to perform
multiple rounds of computations to calculate the leaf nodes' linear functions.
Second, \tree's construction time 
drops near linearly with the increase of threads. 
This is because 
\tree constructs its internal structures using a top-down mechanism,
hence it can be easily parallelized.

\subsection{Insertion}

Different from existing machine learning based index structures that require expensive retraining in face of data changes, \system 
can dynamically support operations like insertion, deletion, and updates at runtime.
In this experiment, \rev{we use \mbox{\database} to compare \mbox{\system}'s 
insertion performance with conventional secondary indexes}. 
\cref{figure:experiments:linear-insert-indexcount} depicts the overall insertion throughputs 
with different numbers of indexes. 
Please note that we take into account
the time for updating the primary index and the base table.
These results are obtained with \textsc{Linear} correlation function and logical pointers, 
and we observed the same trend with other configuration combinations. 
As the result shows, when setting the 
number of indexes to 10, \system can process 1.7 million insert operations per second, 
which is 2.6 times higher than that achieved by the conventional secondary index scheme.
The major reason is that \system's \tree only needs to update the leaf nodes' outlier 
buffers when necessary, which is pretty lightweight.
\cref{figure:experiments:linear-insert-indexcount-breakdown} further explains the result.
Using the baseline mechanism, the database application has to spend over 80\% of the time 
for inserting tuples into the secondary indexes. This demonstrates the inefficiency of the 
conventional indexing mechanism for supporting inserts.

\subsection{Maintenance}
\label{subsection:exp:maintenance}

\begin{figure}[t!]
    \centering
    \subfloat[Lookup Throughput]{
        \includegraphics[width=0.48\columnwidth]
            {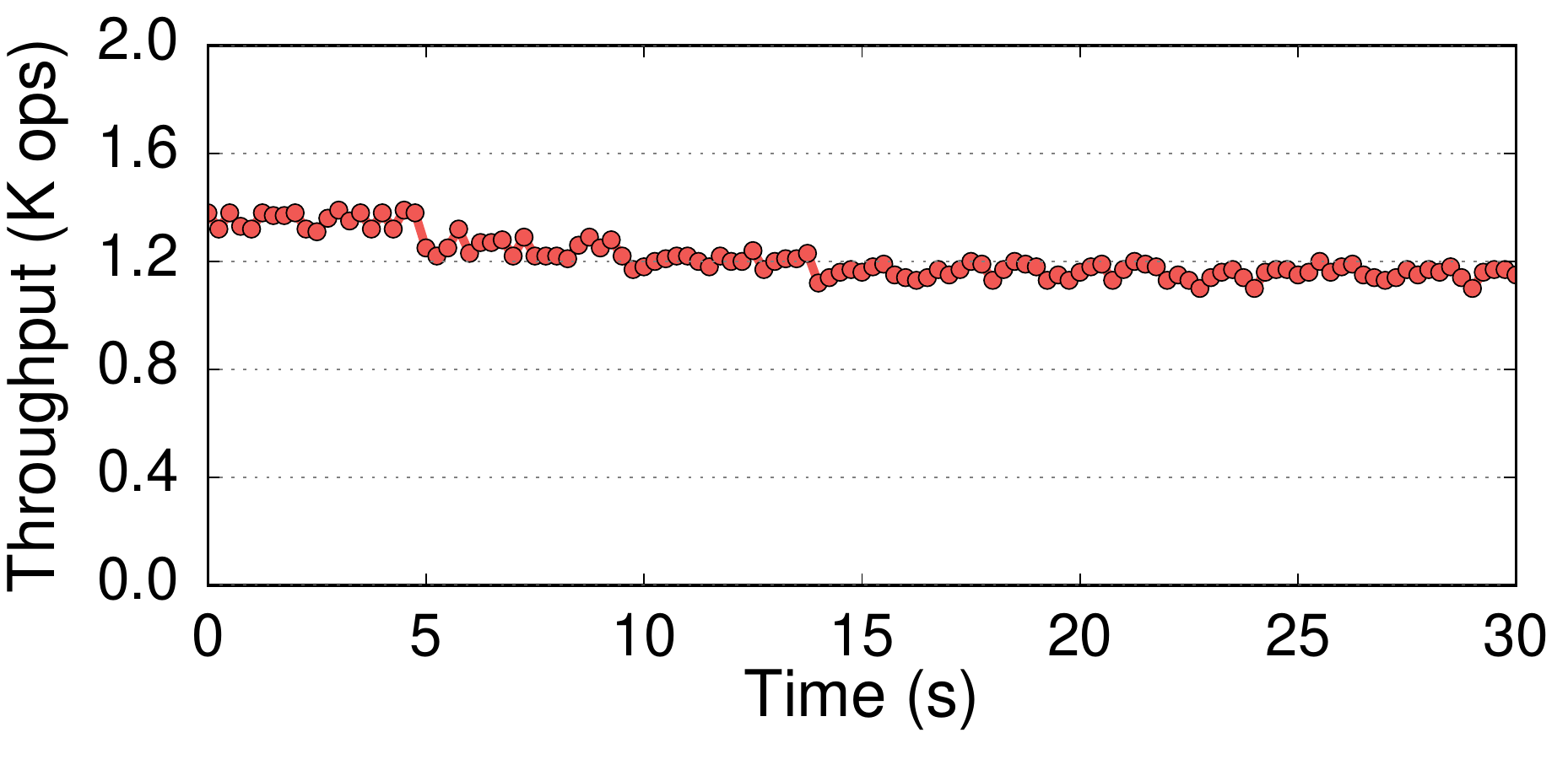}
        \label{figure:experiments:trace-throughput}
    }
    \subfloat[Memory Consumption]{
        \includegraphics[width=0.48\columnwidth]
            {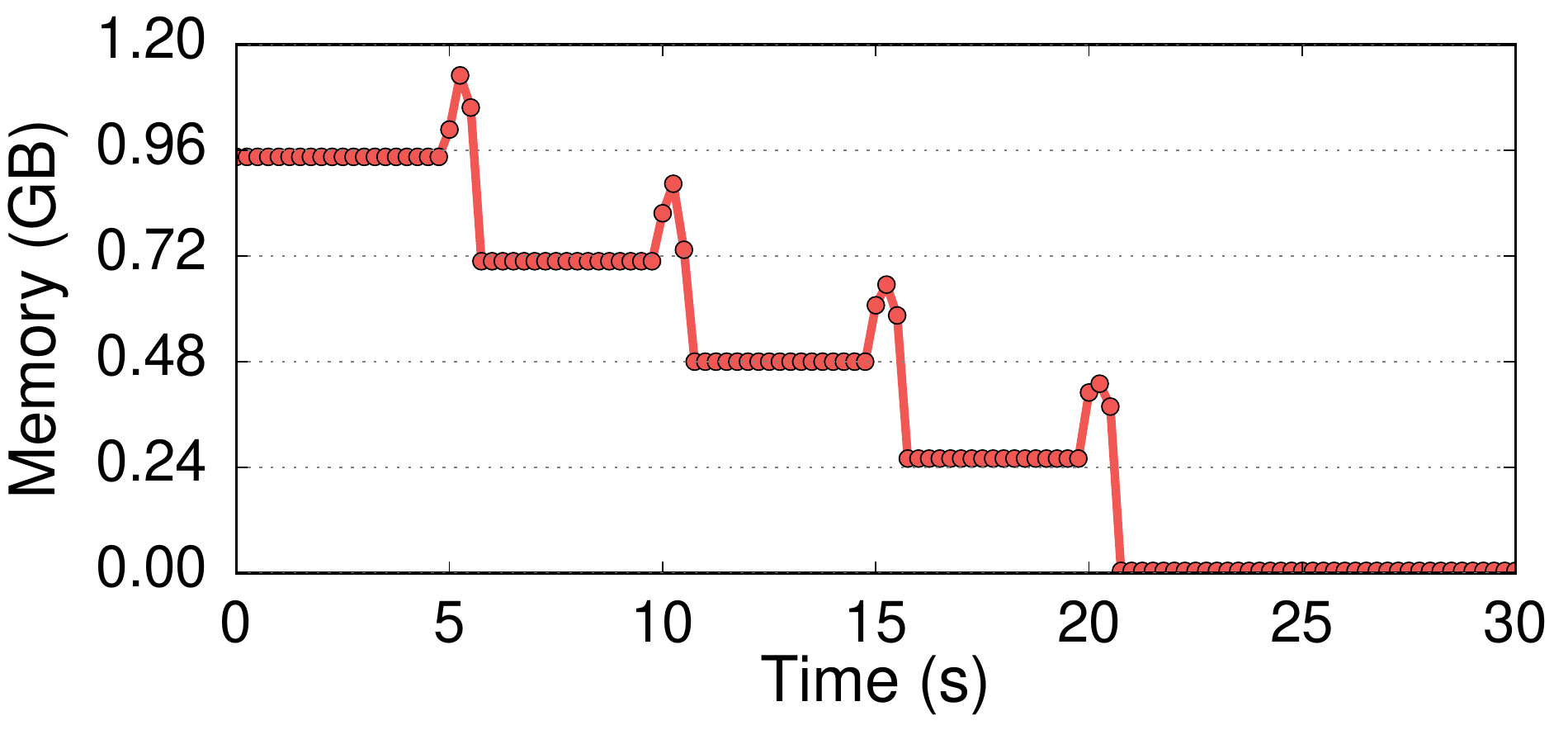}
        \label{figure:experiments:trace-memory}
    }
    \caption{
        Index reorg. performance (\textsc{Synthetic}-\textsc{Sigmoid}).
    }
    \label{figures:experiments:trace}
\end{figure}

\rev{
\mbox{\system}'s \mbox{\tree} can support online structure reorganization to re-optimize its efficiency. In this
experiment, we show how \mbox{\system}'s range lookup throughput and memory consumption changes during 
the process of index reorganization. We first create \mbox{\tree} on a table with 10 K tuples, 
and then insert another 19,990 K tuples to the table, yielding 20 million tuples in total. 
After that, we trigger structure reorganization every 5 seconds, each time reorganizing 1/4
of the structure (given our default $\mathit{node\_fanout}=8$, the reorganization procedure reorganizes 2 first-level subtrees each time). 
Note that this is an artificial scenario for testing purpose only. In real life scenarios, \mbox{\tree} can adjust its reorganization frequency based on the update rates, and the reorganization process would happen in parallel with updates.
During the test, each partial reorganization takes around 2 seconds to finish.
\mbox{\cref{figures:experiments:trace}} shows a 30-second trace of range lookup throughput 
(selectivity = 0.01\%) and memory consumption.
As we can see, the range lookup throughput remains stable during the reorganization. In general, reorganization reduces the sizes of outlier buffers, resulting in less number of direct pointer chasing during query processing. At the same time, it also produces more tree nodes, hence contributing to more precise characterization of the correlation. These two factors balance out during the process.
The memory consumption drops significantly thanks to the structure reorganization. However,
we also observed instant spike during the start of each reorganization. This is because 
the background thread needs to perform table scan and materialize corresponding 
data in order to compute linear function.
}

\subsection{Disk-Based RDBMSs}
\label{sec:experiments:diskdbms}
\begin{figure}[t!]
    \centering
    \subfloat[Lookup Throughput]{
        \includegraphics[width=0.48\columnwidth]
            {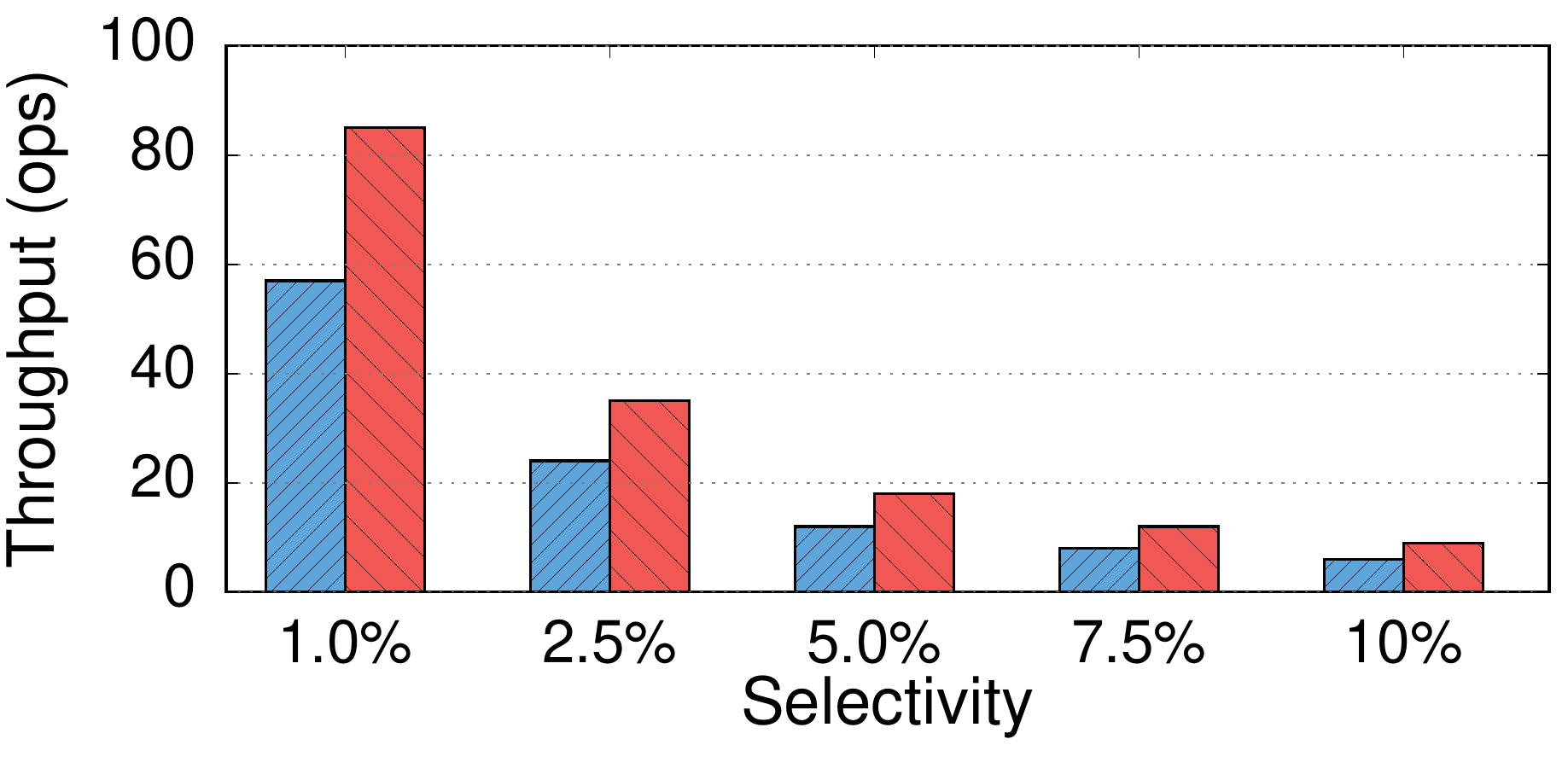}
        \label{figures:experiments:sensor-selectivity-postgres}
    }
    \subfloat[Performance Breakdown]{
        \includegraphics[width=0.48\columnwidth]
            {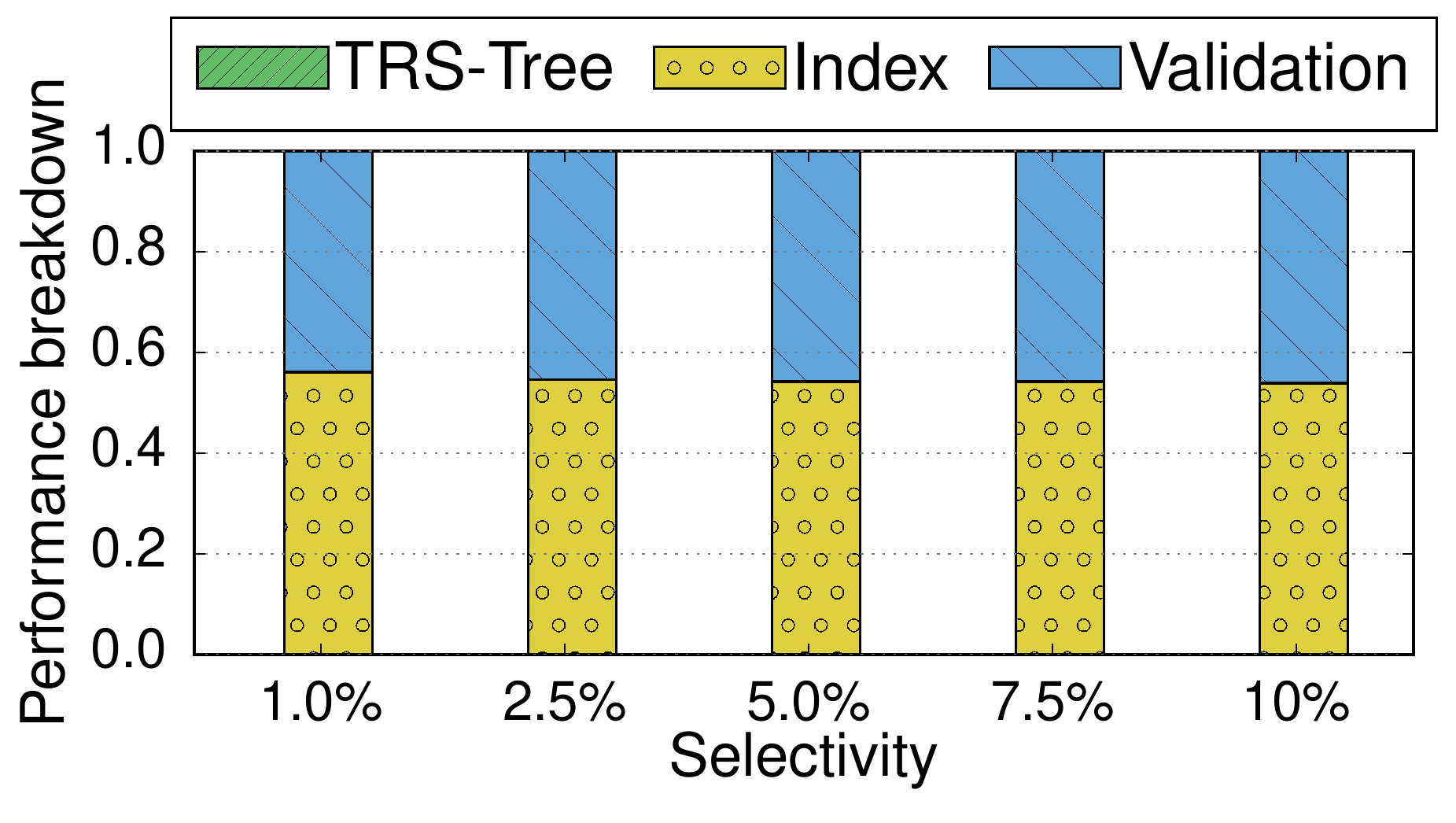}
        \label{figures:experiments:sensor-selectivity-breakdown-postgres}
    }
    \caption{
        Range lookup performance in PostgreSQL.
    }
    \label{figures:experiments:range-postgres}
\end{figure}

Now we integrate \system into a popular disk-based RDBMS, namely PostgreSQL. 
We use the \textsc{Sensor} application to compare 
\system's lookup performance with that of PostgreSQL's secondary indexing
mechanism (which is denoted as the baseline solution). 
Please note that PostgreSQL adopts
physical pointers for tuple identifiers. We implemented a PostgreSQL
client using libpqxx~\cite{libpqxx} to issue queries.
We still keep \system's \tree in memory.

\cref{figures:experiments:sensor-selectivity-postgres} shows \system's range lookup 
performance in PostgreSQL. Similar to what we observed before, the performance gap 
drops with the increase of the selectivity. When setting the selectivity to 1.0\%, 
\system is over 30\% slower than the baseline solution. The major reason is that 
fetching data from secondary storage is more expensive than fetching from main memory.
\cref{figures:experiments:sensor-selectivity-breakdown-postgres} further depicts the 
breakdown. Not surprisingly, \tree lookup is negligible compared to host index lookup 
in PostgreSQL. Validating false positives also take times, as our implementation 
materializes and then iterates the result set of the host index lookup.
One may optimize the performance by pushing the filter operator down to the index scan.

\section{Related Work}
\label{sec:related-work}

\textbf{Tree index structures.} 
B+-Tree~\cite{c79} is the textbook index for disk-oriented DBMSs and 
its structure is well-designed to reduce random disk accesses. 
With the decrease of main memory prices, 
researchers and practitioners have developed memory-friendly indexes that can 
efficiently leverage the larger main memory and fast random access speed.
Some pioneering works include T-tree~\cite{lc86} and cache-conscious indexes~\cite{rao2000making}. 
All these indexes use the hierarchical tree structure to return accurate query results in a timely manner. 
However, these solutions can lead to high memory consumptions, causing high
pressure to main-memory RDBMSs.

\textbf{Succinct index structures.}
Sparse indexes such as column imprints~\cite{lm13} and Hippo~\cite{ys16} only store pointers to disk pages (or column blocks) 
in parent tables and value ranges 
in each page (or column block) to reduce the space overhead. \rev{Column \mbox{Sketch~\cite{hki18}} indexes tables on a values-by-value basis but compresses the values into a lossy map.}
The tradeoff is that these structures can introduce false positives in query time.
BF-Tree~\cite{aa14}  is an approximate index designed for ordered or partitioned data. 
While generating unqualified results, it can largely reduce the space consumption by 
only recording approximate tuple locations.
The learned index~\cite{kbc+18} improves space efficiency by exploiting data distribution using machine 
learning techniques.
It yields good performance but requires a long training phase to generate the data structure. 
Zhang et al.~\cite{zhang2018surf} proposed a new range query filtering mechanism for log-structured 
merge trees.

Stonebraker~\cite{stonebraker1989case} introduced the partial index that 
stores only a subset of entries from the indexed columns to reduce the number of leaf nodes.
Idreos et al.~\cite{ikm09,idreos2011merging} developed a series of techniques called database cracking to adaptively generate indexes based on the query workload. Specifically, \rev{partial sideways \mbox{cracking~\cite{ikm09}} introduces an index called partial maps which consists of several self-organized chunks. These chunks can be automatically dropped or re-created according to the remaining storage space such that the maximum available space is exploited.}
Athanassoulis et al.~\cite{athanassoulis2016designing} later proposed the RUM conjecture to 
capture the relations among read, update, and memory overhead.

Compression techniques~\cite{grs98,zhn+06} drop redundant data information to 
save storage space.
However, these techniques require extra time for compressing data ahead of time and 
decompressing data at query time. 
This compromises the query performance and index maintenance speed. 
In addition, they still store the pointers for tuples such that the amount of saved memory is limited.

\textbf{Secondary index selection.}
Several works have also discussed how to select secondary indexes 
given a fixed amount of space budget. 
A group of researchers at Microsoft proposed a mechanism that 
analyzes a workload of SQL queries and suggests suitable 
indexes~\cite{chaudhuri1997efficient}. They further presented an end-to-end solution to address the 
problem of selecting materialized views and indexes~\cite{agrawal2000automated}.
Researchers at IBM modeled the index selection problem as a 
variant of the knapspack problem, and introduced an index 
recommendation mechanism into the DB2.
Most recently, Pavlo et al.~\cite{pavlo2017self} investigated 
this problem using a machine learning based approach.

\textbf{Column correlations.} 
BHUNT~\cite{brown2003bhunt} automatically discovers algebraic constraints between pairs of columns. 
By relaxing the dependency, 
CORDS~\cite{ilyas2004cords} uses sampling to discover correlations and soft functional dependencies 
between columns. In addition, CORDS recommends groups of columns on which to maintain certain 
simple joint statistics. 
Researchers on data cleansing also put lots of efforts on detecting functional dependencies including soft dependency and approximate dependency~\cite{cdp16,kn18}. 
CORADD~\cite{kimura2010coradd} proposes a correlation-aware database designer 
to recommend the best set of materialized views and indexes for given database size constraints. 
Correlation Maps~\cite{kimura2009correlation} (CM) is a data structure that 
expresses the mapping between correlated attributes for accelerating unindexed column access. While sharing a similar 
idea of leveraging column correlations to save space, \mbox{\system} does not require using clustered columns; more importantly, 
its ML-enhanced \mbox{\tree} structure can adaptively and dynamically model both complex correlations and outliers, hence yielding better performance 
in many cases. 


\textbf{Cardinality estimation.} Cardinality estimation plays a crucial role in RDBMS query optimizers. 
Column correlation is the most common reason that encumbers the estimation. 
Sample views~\cite{llz+07} and PSALM~\cite{zls09} use sampling methods to detect the column correlation. 
Recent projects~\cite{mbc07,lxy+15} start treating column semantics as a black box and 
use machine learning models to learn cardinalities from query feedbacks. \rev{\mbox{Kipf et al.~\cite{kkr+19}} opt to use deep learning techniques to learn cardinalities for join queries.}
\section{Conclusions}
\label{sec:conclusions}



We have introduced \mbox{\system}, a new secondary indexing mechanism 
that exploits column correlations to reduce index space consumption. 
\system utilizes \tree, a succinct, ML-enhanced tree structure to adaptively 
and dynamically capture complex correlations and outliers. 
Our extensive experimental study has confirmed \system's effectiveness in both 
main-memory and disk-based RDBMSs.

\clearpage

\newpage
\balance

\bibliographystyle{abbrv}
\small

\appendix
\balance

\section{Applications in the experiments}
\label{sec:appendix-applications}


\textbf{\textsc{Synthetic}}:
The synthetic data contains one single table with four 8-byte numeric columns, 
namely $col_A$, $col_B$, $col_C$, and $col_D$. 
Columns $col_B$ and $col_C$ are correlated, as $col_B$'s values are generated by 
a certain correlation function from $col_C$, i.e. $col_B = Fn(col_C)$.
We use two types of correlation functions: \textit{\textsc{Linear}} function 
and \textit{\textsc{Sigmoid}} function.
We also inject uniformly distributed 
noisy data to $col\_B$. By 
default, we inject 1\% noises (percent = $\frac{abnormal~tuples}{cardinality}$).
We have already built a primary index on $col_A$ and a secondary index on $col_B$.
The application frequently queries on $col_C$ to retrieve values on $col_D$.
Our experiments build indexes on $col_C$.

\textbf{\textsc{Stock}}:
This application records the market price of 100 stocks in the U.S. stock market 
over the last 60 years. 
We store over 15,000 rows containing datetime and daily highest and lowest prices of these 100 stocks in a wide table (201 columns in total).
We set the entries to \texttt{NULL} if certain readings are not available.
Each pair of the highest and lowest price columns forms a simple near-linear correlation.
We build a primary index on the datetime column, and a set of secondary indexes on each lowest price column.
The application continuously issues queries to those unindexed highest price columns.
The queries are like: 
``during which time periods do Stock \texttt{X}'s highest price fall between \texttt{Y} and \texttt{Z}?''.
Our experiments build indexes on all the unindexed columns and evaluate the performance.

\textbf{\textsc{Sensor}}:
This application monitors chemical gas concentration using 16 sensors.
We store 4,208,260 rows containing the timestamp, the 16 sensor readings, and the 
reading average in a single table 
(18 columns in total).
These 16 sensor reading columns and the average reading column form a non-linear correlation.
We have constructed one index on the average reading column.
The application continuously queries one of those 16 unindexed sensor reading columns.
The queries are like: 
``during which time period do the readings in Sensor \texttt{X} fall between \texttt{Y} and \texttt{Z}?''.
Our experiments build indexes on all the unindexed columns and evaluate the performance.






\section{Maintenance}
\label{sec:appendix-maintenance}

\mbox{\tree} can easily support concurrent insertions. 
As a single insert/delete/update operation only affects at most one leaf node, 
\mbox{\tree} can easily guarantee the structure consistency 
by using concurrent hash tables to implement the leaf nodes' outlier buffers.

\mbox{\tree} also enables \mbox{\textit{online}} structure reorganization at runtime 
without incurring much overhead to any concurrent operations. 
Unlike conventional concurrent tree-based structures, 
\mbox{\tree} does not implement latch coupling, 
which can be overly complicated and expensive for \mbox{\tree}.
Instead, it adopts a coarse-grained latching protocol to maximize concurrency. 
The intuition behind this decision is that insert/delete/update 
operations in \mbox{\tree} never trigger cascading node modifications, and the reorganization  
happens infrequently and can be processed with low latency.

\tree uses a flag to identify the reorganization phase. 
A dedicated background thread starts the reorganization by setting the flag to true.
When observing this flag, 
any concurrent insert / delete / update operations append their modifications to a temporal 
buffer to avoid phantom problems. 
The background thread then scans and retrieves all corresponding entry pairs 
and subsequently creates the new tree nodes. Before installing these nodes to \tree, 
the background thread further holds a coarse-grained latch on the entire tree and applies all the 
changes in a temporal buffer. The latch is released once the new nodes are installed.

\section{Comparison}
\label{sec:appendix-comparison}

\textbf{Compare with Correlation Maps (CM).}
\tree in \system is a {ML-enhanced} tree index.
CM~\cite{kimura2009correlation} adopts a map-like structure which stores the bucket mappings between correlated columns. Both CM and HERMIT leverage column correlations to save space. But we find that these two proposals are drastically different.

	{$\bullet$} \tree captures correlations using tiered curve fitting and {handles outliers}. This makes it robust to noisy data which is prevalent in real-world applications. In contrast, CM does not include 
	any scheme to handle outliers, and hence its performance can drop when confronting sparsely distributed noisy data.
	
	{$\bullet$} \tree adaptively constructs its internal structures and automatically decides the partition granularity. It {dynamically} maintains its internal structures, and performs reorganization in the presence of large amounts of insert/delete/update operations. In comparison, CM relies on its tuning advisor to decide the granularity for its single-layer buckets by building multiple histograms beforehand. It is unknown how CM adapts to dynamic workload where the underlying data drastically change over time.
	
	
	
	{$\bullet$} \system is a {general} secondary indexing mechanism and can exploit multiple correlations on the same table. In contrast, 
	CM can only exploit correlations when there is a clustered index on the host column. At most one clustered index can exist in a table.
	
	We also empirically compare \system with CM in Appendix~\ref{sec:appendix-experiments}.
	

\textbf{Compare with BF-Tree.}
BF-Tree~\cite{aa14} is an approximate index that exploits implicit ordering and clustering in the underlying data to reduce storage overhead. It adopts the same tree structure with B+ Tree but stores a set of Bloom Filters in its leaf nodes. Those Bloom Filters record the approximate physical locations of values.
Although both BF-Tree and \system tend to reduce index size by introducing false positives, they act very differently.

{$\bullet$} BF-Tree exploits implicit ordering and clustering in the underlying data to reduce the index size. \system leverages the correlation between the host column and the indexed column to shrink its size on the indexed column.

{$\bullet$} BF-Tree requires that the underlying data should be ordered or at least have some clusterings. \system does not have any specific requirement for the data distribution. It works for any data order.

{$\bullet$} BF-Tree stores Bloom Filters and disk page ranges in its leaf nodes. For every key lookup on the indexed column, these Bloom Filters may return "true" for some non-existing keys and thus result in page scans on the false positive disk pages. The TRS-Tree in \system stores ML models (linear regression in our paper) in leaf nodes. This range may include some false positive values which need to be pruned later.
\section{More Discussions}
\label{sec:appendix-discussion}

\subsection{Correlation Discovery}
\label{sec:appendix-discussion:correlation}

\system fully relies on the underlying RDBMS or users to perform correlation discovery. There has been a flurry of systems that addressed the problem of correlation (or functional dependency) discovery in different ways. In the past two decades, researchers extensively studied how to automatically find all functional dependencies (including those among composite columns) in a database. To accelerate the discovery, they~\cite{km95,ilyas2004cords,kn18,pem+15,cdp16} opt to leverage specific rules to prune columns or compute approximate coefficients based on samples.

In practice, most commercial RDBMSs still largely rely on ``human knowledge''
to discover \textit{possibly} correlated columns because of the huge search space of column combinations. A database administrator (DBA) can identify candidate columns that exhibit semantic relationships, then evaluate the correlation using different correlation coefficients, including \textit{Pearson coefficient}
and \textit{Spearman coefficient}~\cite{well2003research}. Once the coefficient reaches a certain 
predefined threshold, the DBA can recommend this correlation information to the database optimization module.

\begin{figure}[t!]
    \centering
    \subfloat[$y=x$]{
        \includegraphics[width=0.3\columnwidth]
            {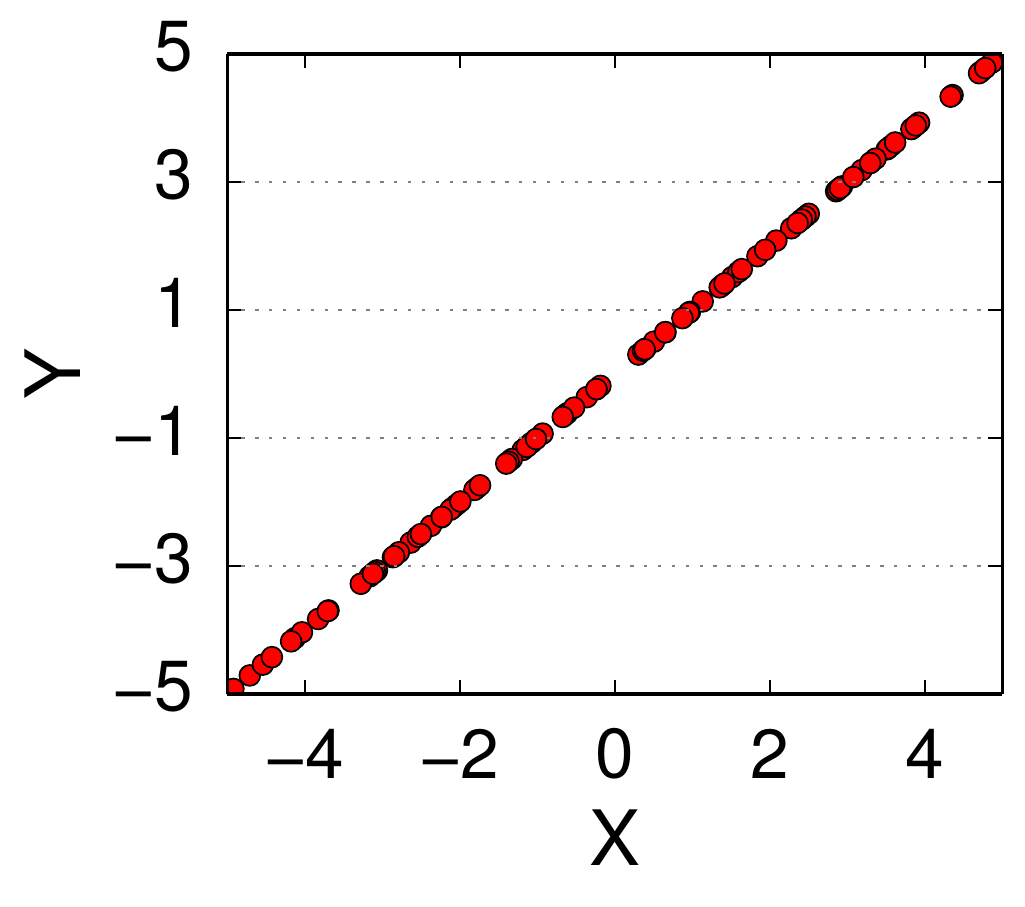}
        \label{figures:experiments:linear}
    }
    \subfloat[$y=sigmoid(x)$]{
        \includegraphics[width=0.3\columnwidth]
            {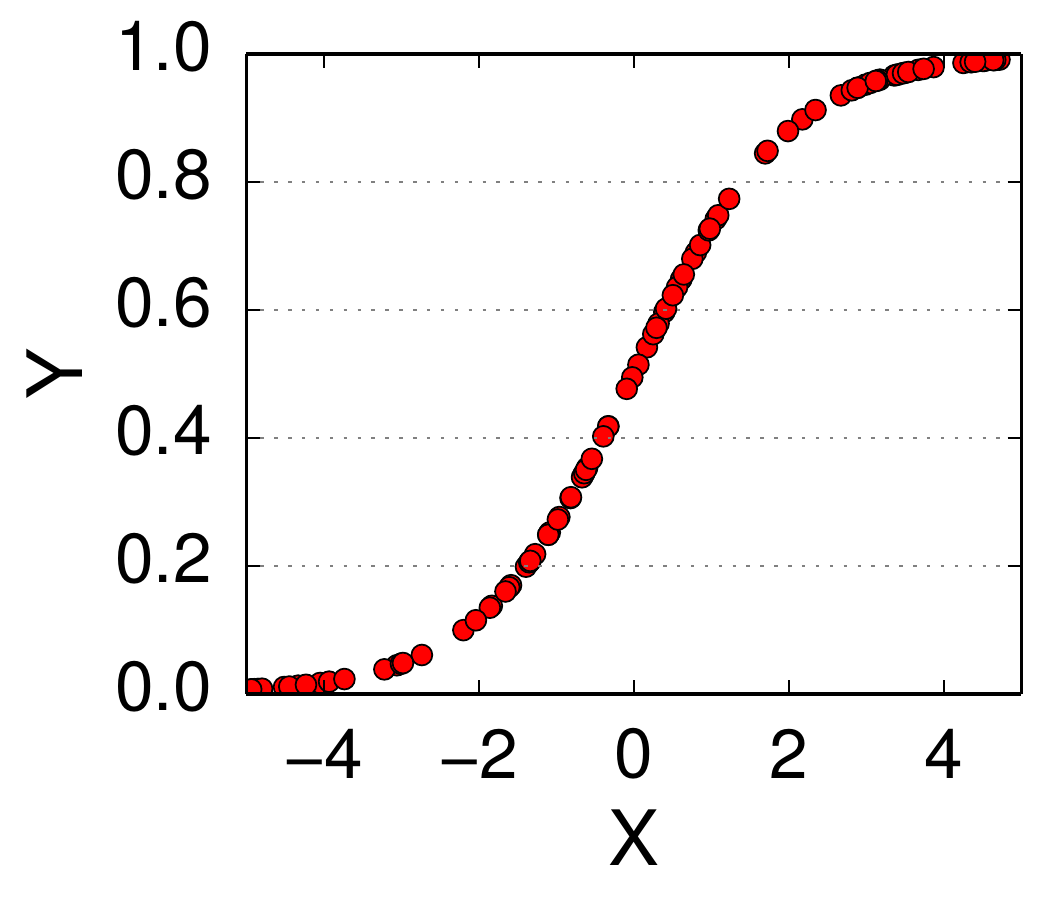}
        \label{figures:experiments:sigmoid}
    }
    \subfloat[$y=sin(x)$]{
        \includegraphics[width=0.32\columnwidth]
            {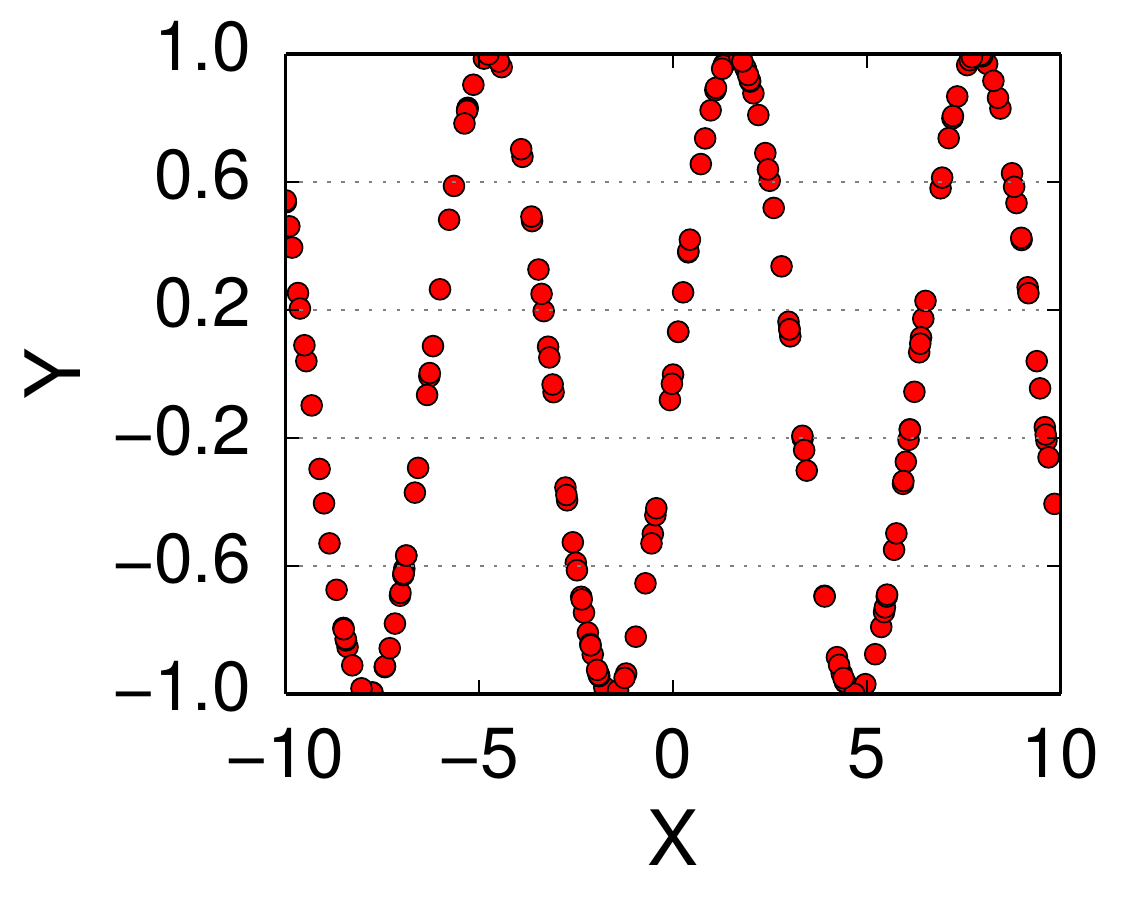}
        \label{figures:experiments:sin}
    }
    \caption{
        Three different correlation functions. 
    }
    \label{figures:experiments:functions}
\end{figure}

Now let us discuss correlation types \system may capture.
\cref{figures:experiments:functions} shows three different functions: 
(1) linear (e.g., $y = x$) (2) monotonic (e.g., $y=sigmoid(x)$) (3) non-monotonic (e.g., $y = sin(x)$).
We do not consider noisy data at this moment.
\tree can perfectly index the correlations in both (1) and (2) since it directly navigates a key lookup to a single value on the indexed host column. 
A DBA can easily capture these two correlations using Pearson coefficient and Spearman 
coefficient, respectively (coefficient = 1). 
However, \system cannot yield good performance for non-monotonic correlation 
like $sin$ function because a single value on the host column can be mapped to 
many values on the target column. Consequently, \tree can generate many false positives, 
resulting in low performance.
The DBA can detect a non-monotonic correlation using Spearman coefficient (coefficient = 0).




\begin{figure}[t!]
    \centering
    \subfloat[Stocks over year]{
        \includegraphics[width=0.48\columnwidth]
            {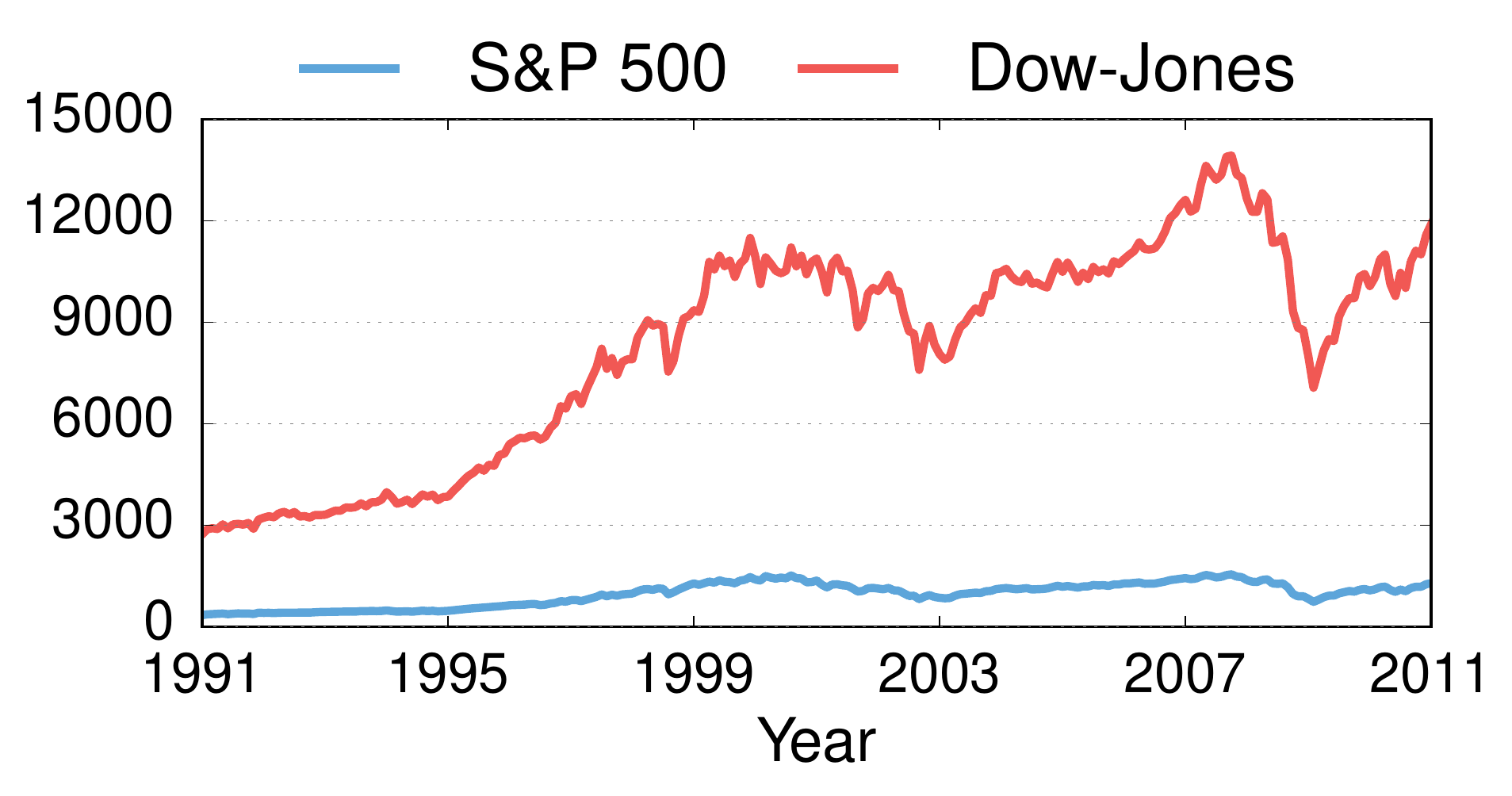}
        \label{figure:dji-sp-0}
    }
    \subfloat[Correlation function]{
        \includegraphics[width=0.48\columnwidth]
            {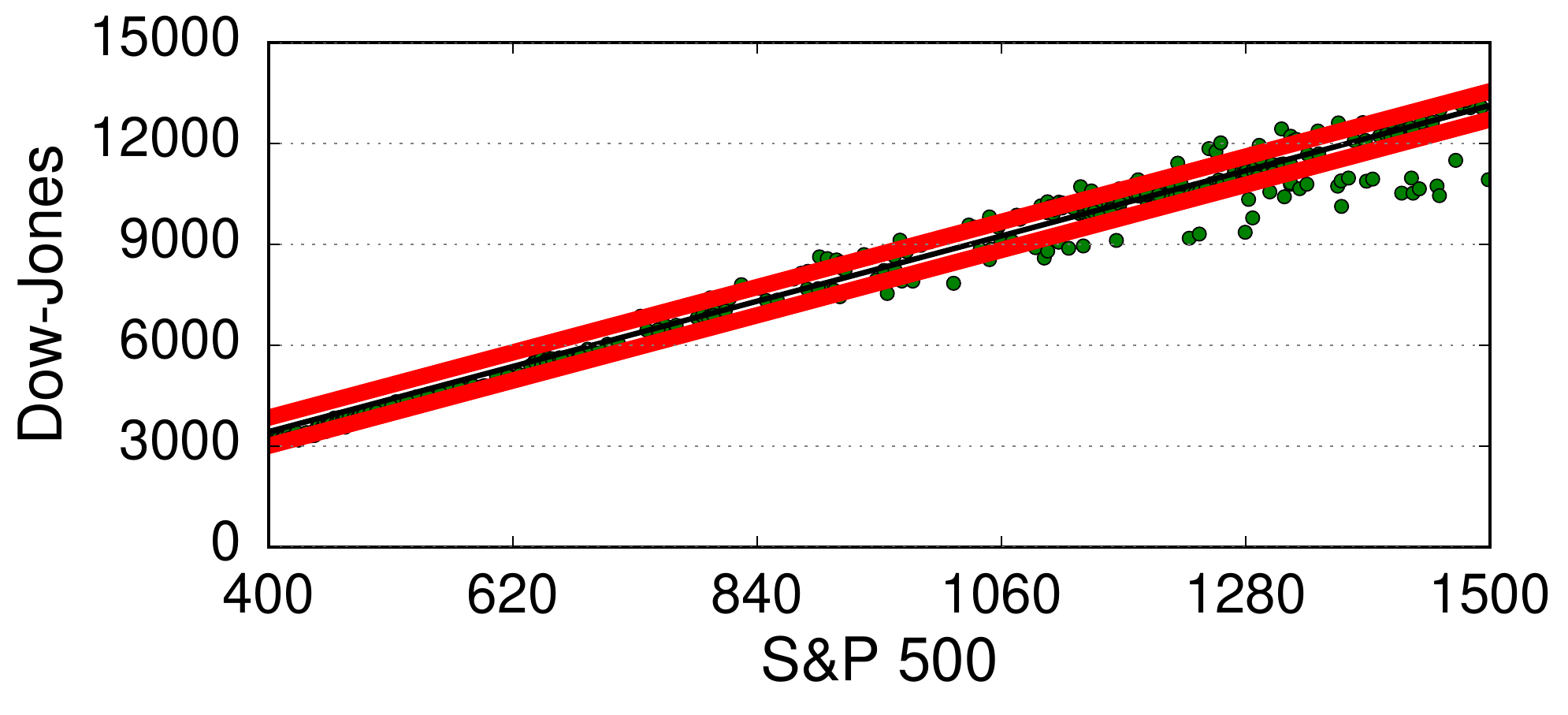}
        \label{figure:dji-sp-1}
    }
    \caption{
      Dow-Jones and S\&P 500 values. In (b),
      Green dots falling beyond red lines are identified as outliers. 
    }
    \label{figure:dji-sp}
\end{figure}

\system captures correlations and handles outliers. This makes 
\system applicable to cases where nice difference bounding does not exist. 
\cref{figure:dji-sp-0} shows the value trace of 
Dow-Jones and S\&P 500 during the years 1991 to 2011.
While the two indices are correlated in most years, we observe 
some major shifts in certain months or years.
\system handles them by directly identifying and maintaining them 
in outlier buffers, and hence eliminate some false positives.


%

\subsection{Optimization}
\label{sec:appendix-discussion:optimization}

\textbf{Sampling-based outlier estimation.} 
During the index construction, a \tree needs to compute the linear function
parameters, namely slope $\beta$ and intercept $\alpha$, of the leaf nodes
using the standard linear regression formula (see \cref{subsec:internal}).
Sometimes the default construction algorithm shown in \cref{alg:indexconstruction}
unnecessarily computes these parameters even if the corresponding node 
later splits due to too many outliers.
We adopt a sampling-based strategy to avoid this problem. 
Before performing the parameter computation, our optimization algorithm randomly 
samples the data (by default 5\%) covered by the range 
and runs the simple linear regression on them. 
Within the sample set, if the number of outliers has already exceeded the 
pre-defined threshold, then the construction directly partitions the range. 
This helps us make the decision quickly.

\textbf{Multi-threaded index construction.} 
\cref{alg:indexconstruction} shows how we construct \tree using a single thread.
Observing the popularity of massively parallel processors, we can now leverage
multithreading to speed up the construction algorithm.
Different from B-tree, \system constructs its internal and leaf nodes 
following a top-down scheme. This means that we can parallelize 
the tree construction without confronting any synchronization points.
Assuming that the index fanout is set to $k$, we can easily parallelize 
the splitting and computation of \tree's every single node using $k$ threads. These threads
proceed independently without incurring any inter-thread communication.


\subsection{Complex Machine Learning Models}
\begin{table}[th!]
    \caption{Training time for different ML models}
    \label{tbl:mlmodels}
    \begin{scriptsize}
        \begin{center}
            
            \begin{tabular}{ c | c | c | c }
                \hline
                Number of tuples & 1 K & 10 K & 100 K \\ \hline
                Linear regression & 0.42 ms & 0.81 ms & 3.2 ms \\
                SVR (RBF)    & 0.09 s & 4.5 s & > 60 s \\
                SVR (linear) & 0.28 s & 29 s & > 60 s \\
                SVR (polynomial)   & 0.29 s & 24 s & > 60 s \\
                \hline
            \end{tabular}
        \end{center}
    \end{scriptsize}
\end{table}

\system performs a linear regression in each \tree node which costs a scan of corresponding tuples. Actually, \tree's structure is also flexible enough to adopt more complex models such as Support Vector Regression (SVR) and neural networks. However, although these models may yield less false positives, training these models takes tremendous time (orders of magnitude slower than linear regression) if the table size increases significantly.

To prove that, we run a set of machine learning models on different scales of data and report the training time in Table~\ref{tbl:mlmodels}. As depicted in the table, training linear regression models only takes several milliseconds while training SVR models with different kernels including RBF, linear and polynomial is at least 200 times slower. Having said that, we believe that researchers and practitioners still can easily extend \system to incorporate other models and fit in their specific scenarios.

In addition, a recent paper featuring learned indexes~\cite{kbc+18} discusses the cases of using complex machine learning models such as neural networks and multivariate regression models to predict locations of keys. As opposed to learned indexes, \system models the correlation between two columns and leverages the curve-fitting technique to adaptively create simple yet customized ML models for different regions (\tree tree nodes).

\subsection{\system on Secondary Storage}

Although the storage overhead of an index may not seem too expensive on traditional Hard Disk Drives (HDDs), the dollar cost increases dramatically when the index is deployed on modern storage devices (e.g., Solid State Drives and Non-Volatile Memory) because they are still more than an order of
magnitude expensive than HDDs. 
This is also an important issue when deploying RDBMSs on the cloud, where the customers 
are charged on a pay-as-you-go basis.
\system is a generic indexing mechanism designed for both main-memory and disk-based RDBMSs. Deploying it on precious secondary storage such as SSD can save considerable storage budget and still exhibit comparable lookup performance as opposed to B+Tree. Our experiment in \cref{sec:experiments:diskdbms} also confirmed this claim.

\subsection{Complex SQL Queries}

As a replacement 
of classic B+-tree-based secondary indexes, \system is general enough and can be used 
in any complex queries whenever a classic secondary index is used.
Even for complex queries like join, RDBMSs can still use \system to execute local 
predicate, consequently accelerating the processing of the entire query plan.
\section{Experiments}
\label{sec:appendix-experiments}

\begin{figure*}[t!]
    \centering
     \fbox{
     \includegraphics[width=0.9\columnwidth]
         {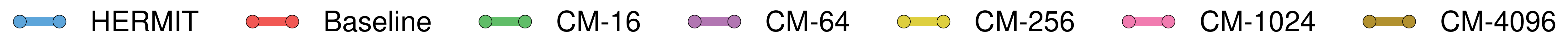}
     }
    \subfloat[Host bucket size = $2^4$]{
        \includegraphics[width=0.4\columnwidth]
            {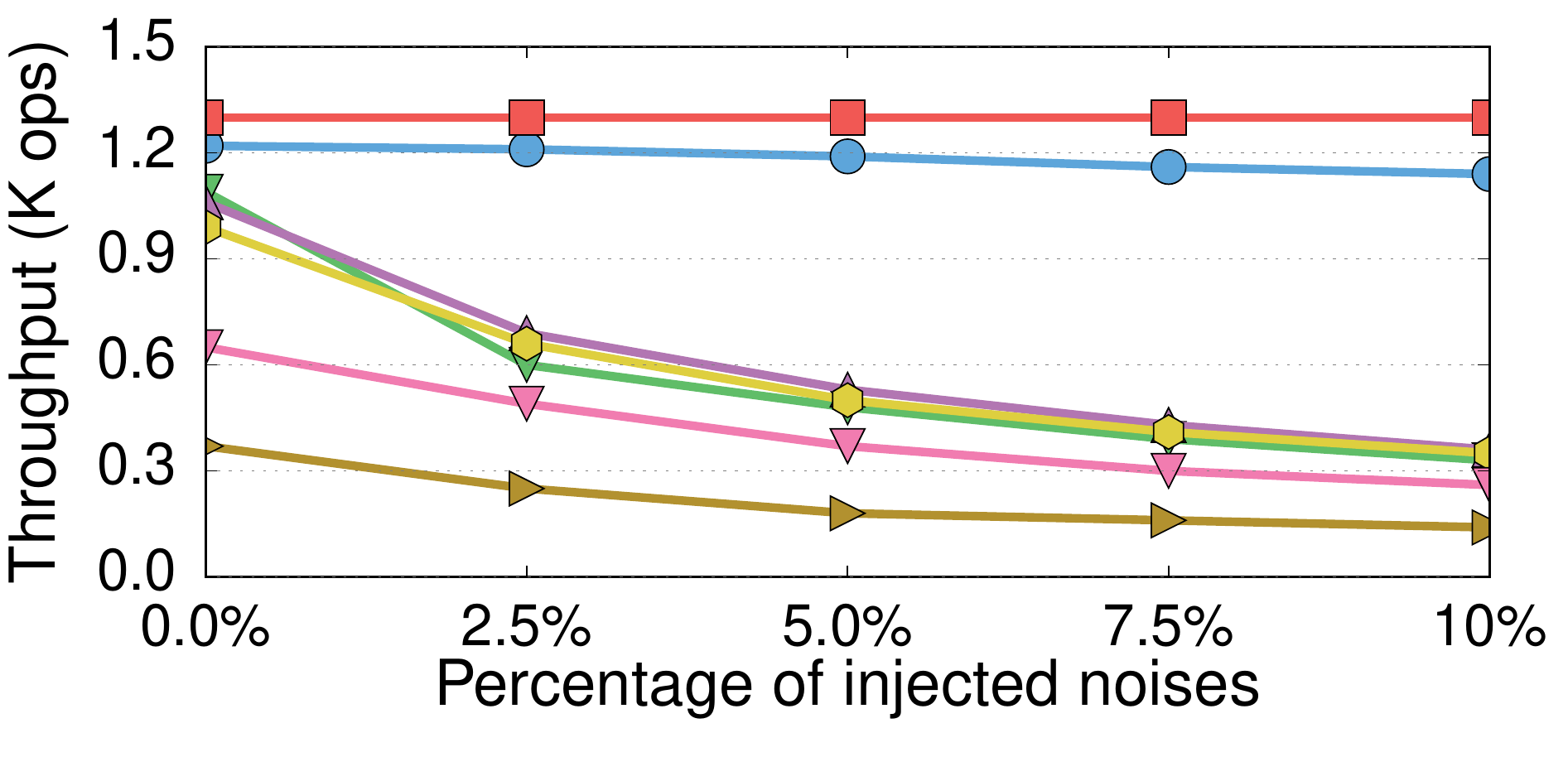}
        \label{figure:experiments:cm-range-16-linear-throughput}
    }
    \subfloat[Host bucket size = $2^6$]{
        \includegraphics[width=0.4\columnwidth]
            {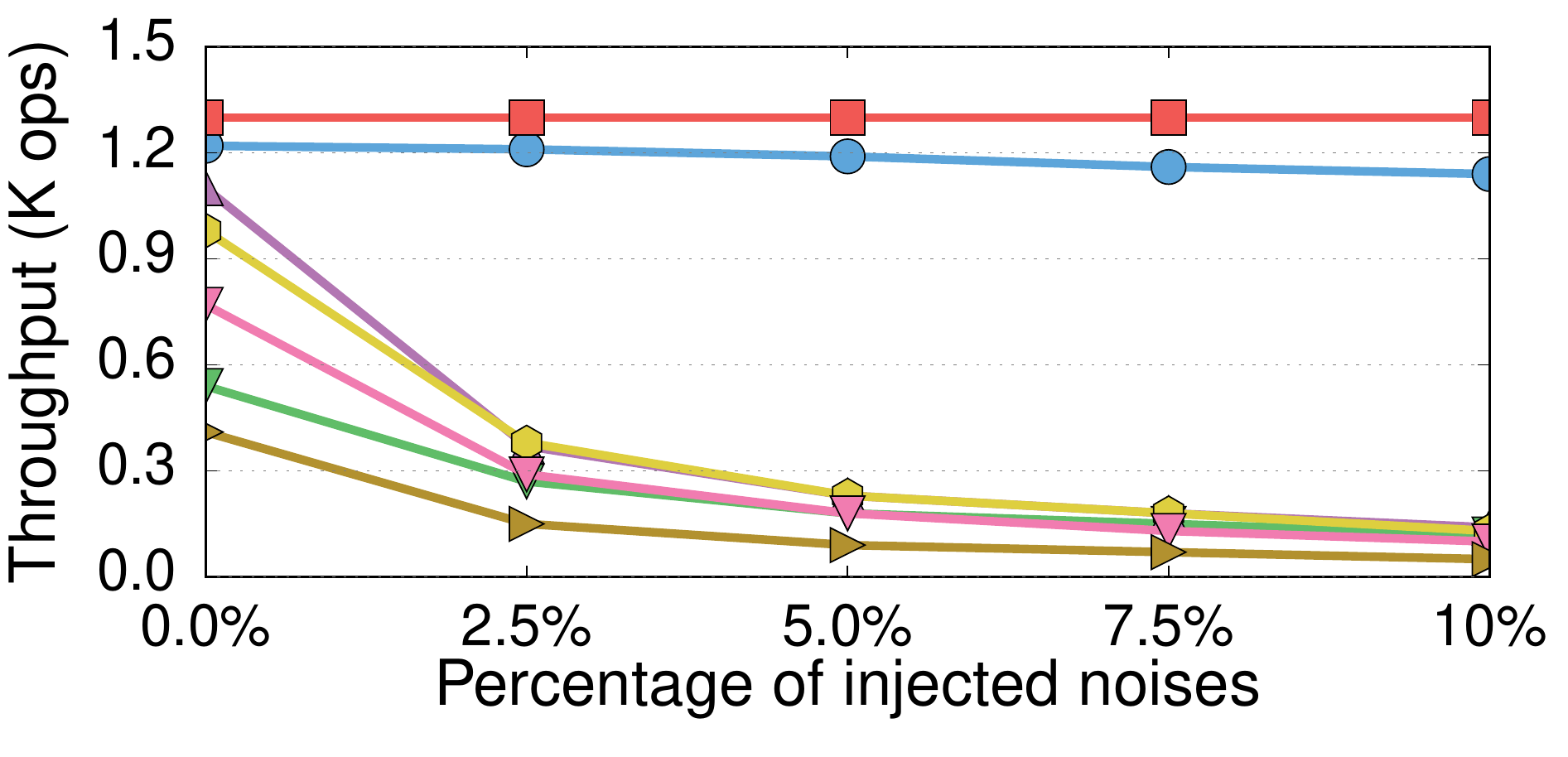}
        \label{figures:experiments:cm-range-64-linear-throughput}
    }
    \subfloat[Host bucket size = $2^8$]{
        \includegraphics[width=0.4\columnwidth]
            {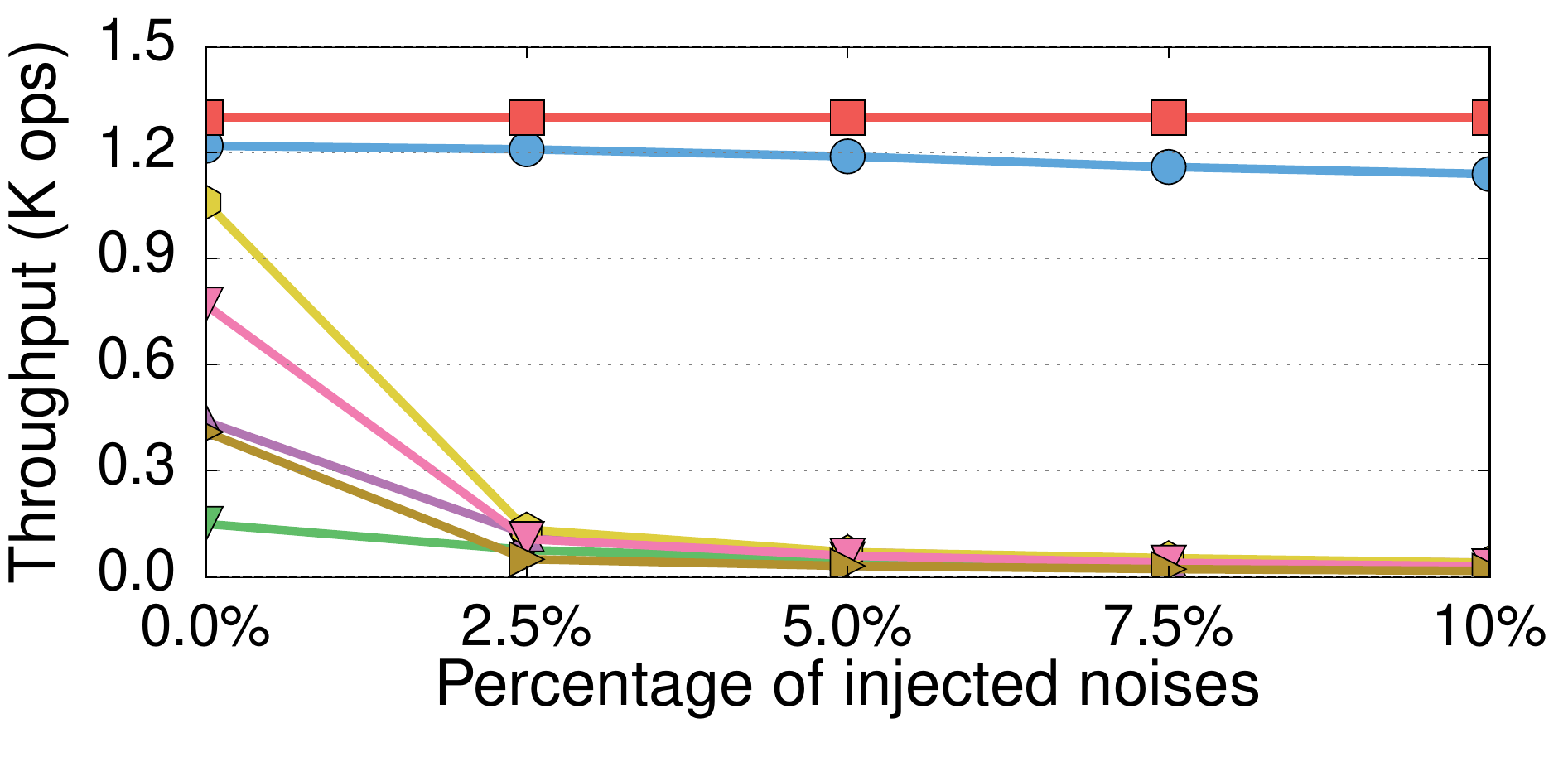}
        \label{figures:experiments:cm-range-256-linear-throughput}
    }
    \subfloat[Host bucket size = $2^{10}$]{
        \includegraphics[width=0.4\columnwidth]
            {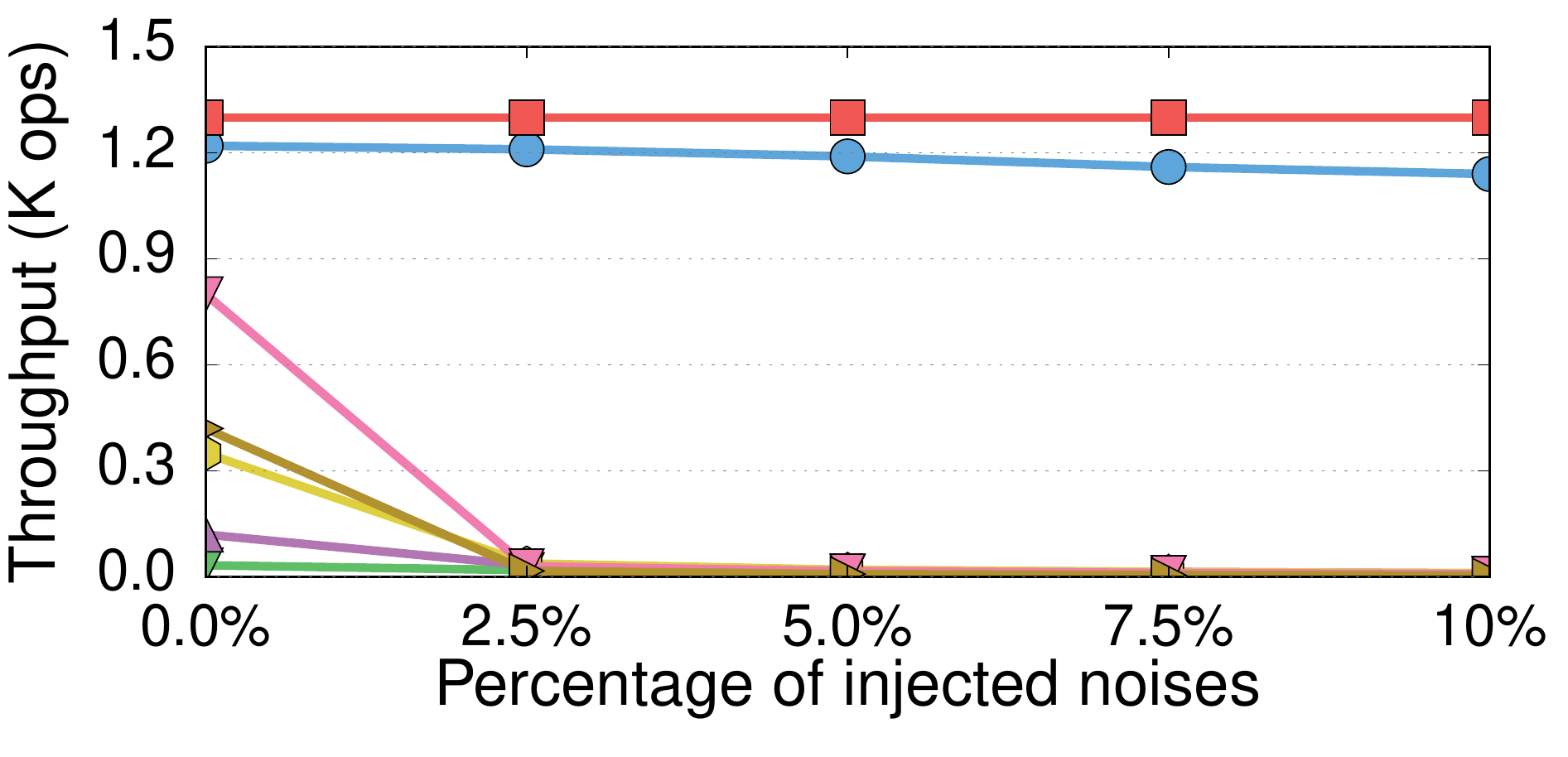}
        \label{figures:experiments:cm-range-1024-linear-throughput}
    }
    \subfloat[Host bucket size = $2^{12}$]{
        \includegraphics[width=0.4\columnwidth]
            {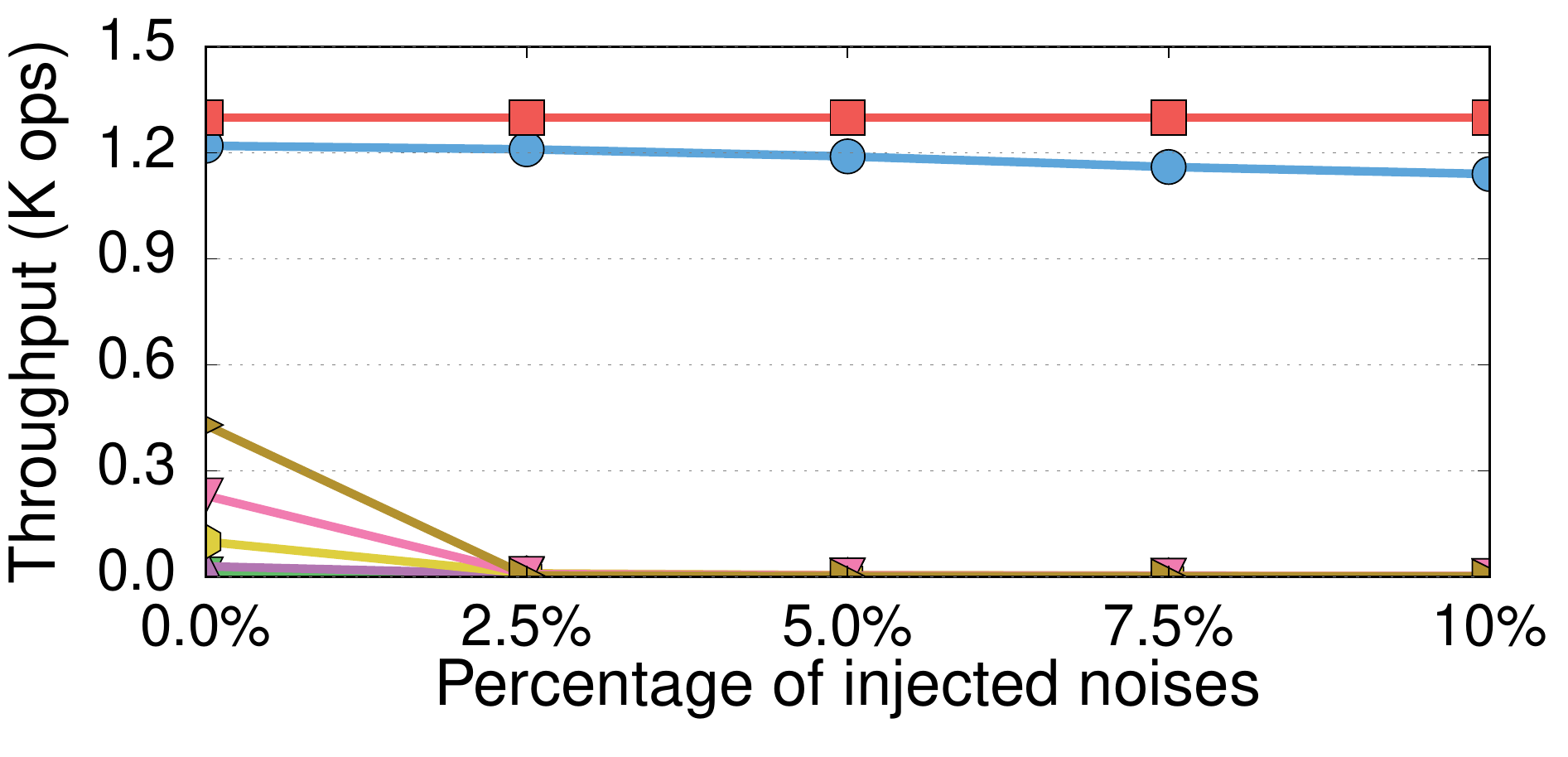}
        \label{figures:experiments:cm-range-4096-linear-throughput}
    }
    \caption{
        Range lookup throughput with different percentage of injected noises (\textsc{Synthetic}-\textsc{Linear}).
    }
    \label{figures:experiments:cm-range-linear-throughput}
\end{figure*}

\begin{figure*}[t!]
    \centering
    \subfloat[Host bucket size = $2^4$]{
        \includegraphics[width=0.4\columnwidth]
            {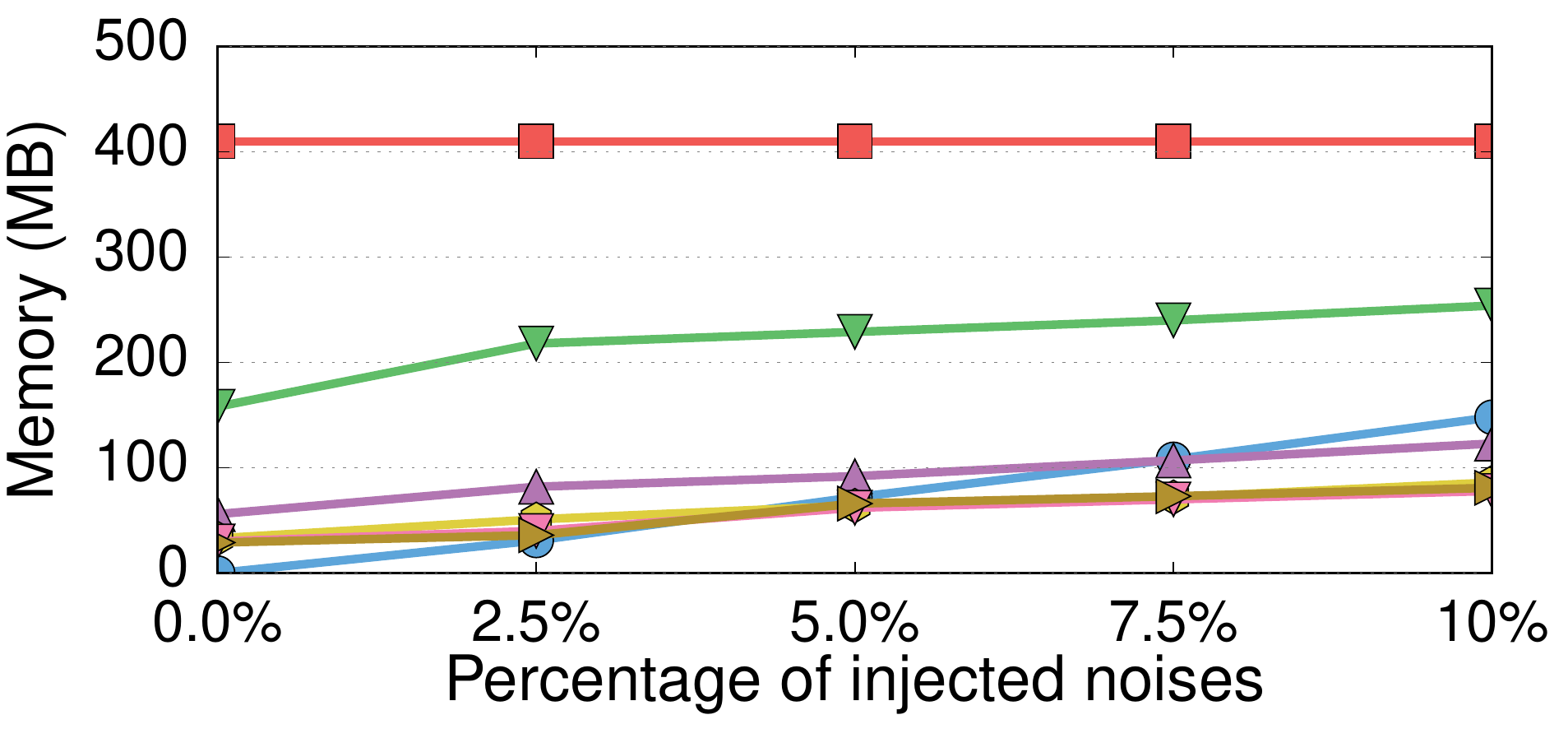}
        \label{figure:experiments:cm-range-16-linear-memory}
    }
    \subfloat[Host bucket size = $2^6$]{
        \includegraphics[width=0.4\columnwidth]
            {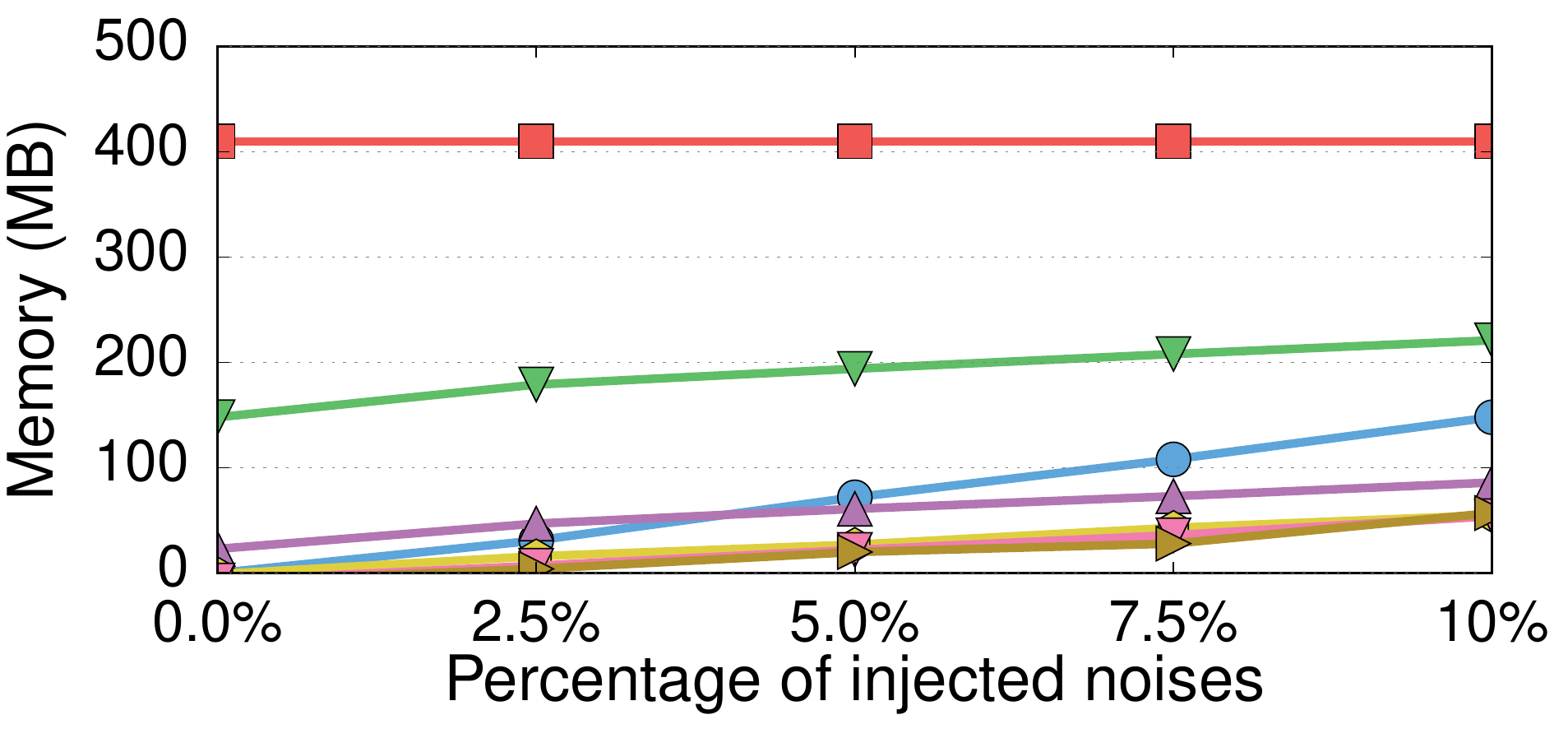}
        \label{figures:experiments:cm-range-64-linear-memory}
    }
    \subfloat[Host bucket size = $2^8$]{
        \includegraphics[width=0.4\columnwidth]
            {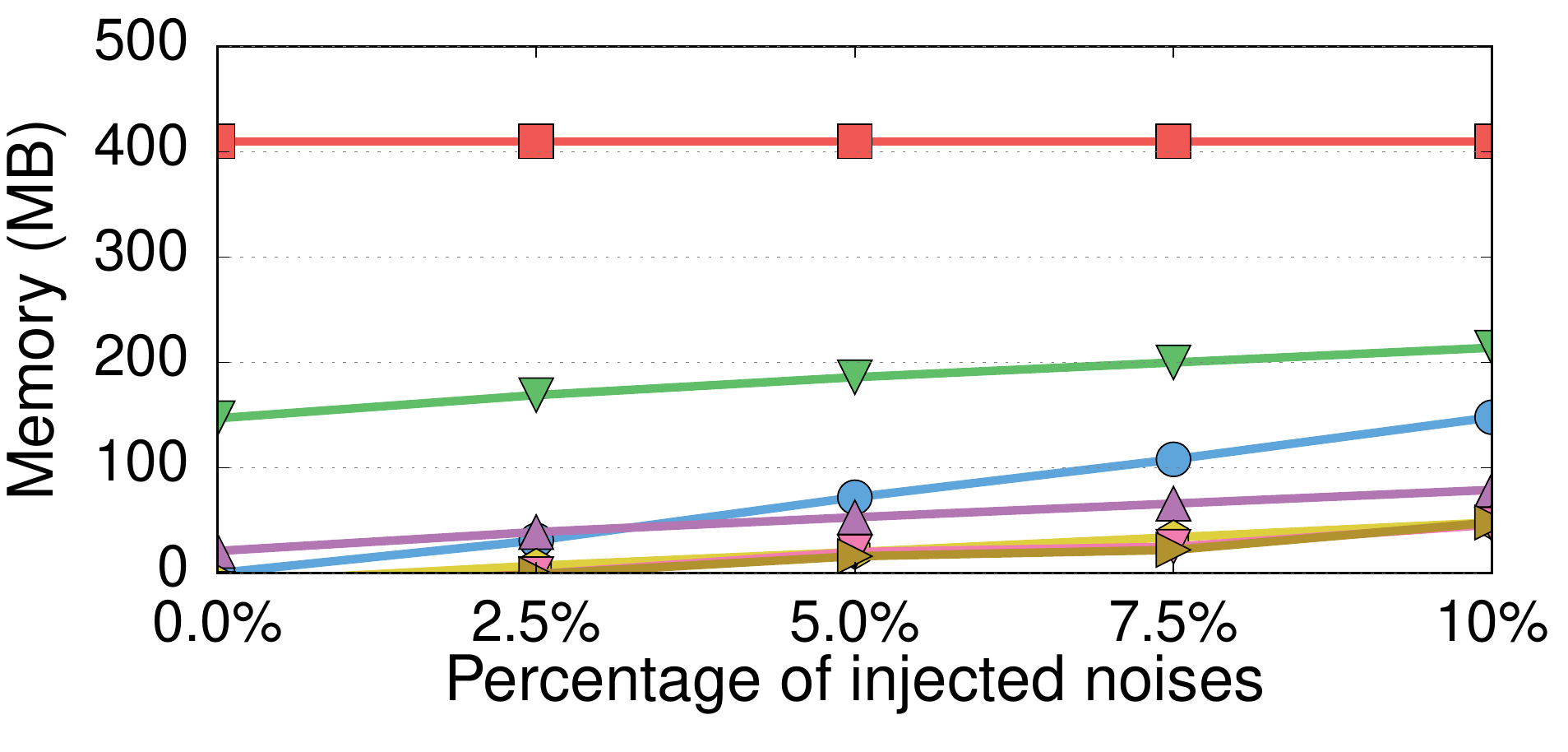}
        \label{figures:experiments:cm-range-256-linear-memory}
    }
    \subfloat[Host bucket size = $2^{10}$]{
        \includegraphics[width=0.4\columnwidth]
            {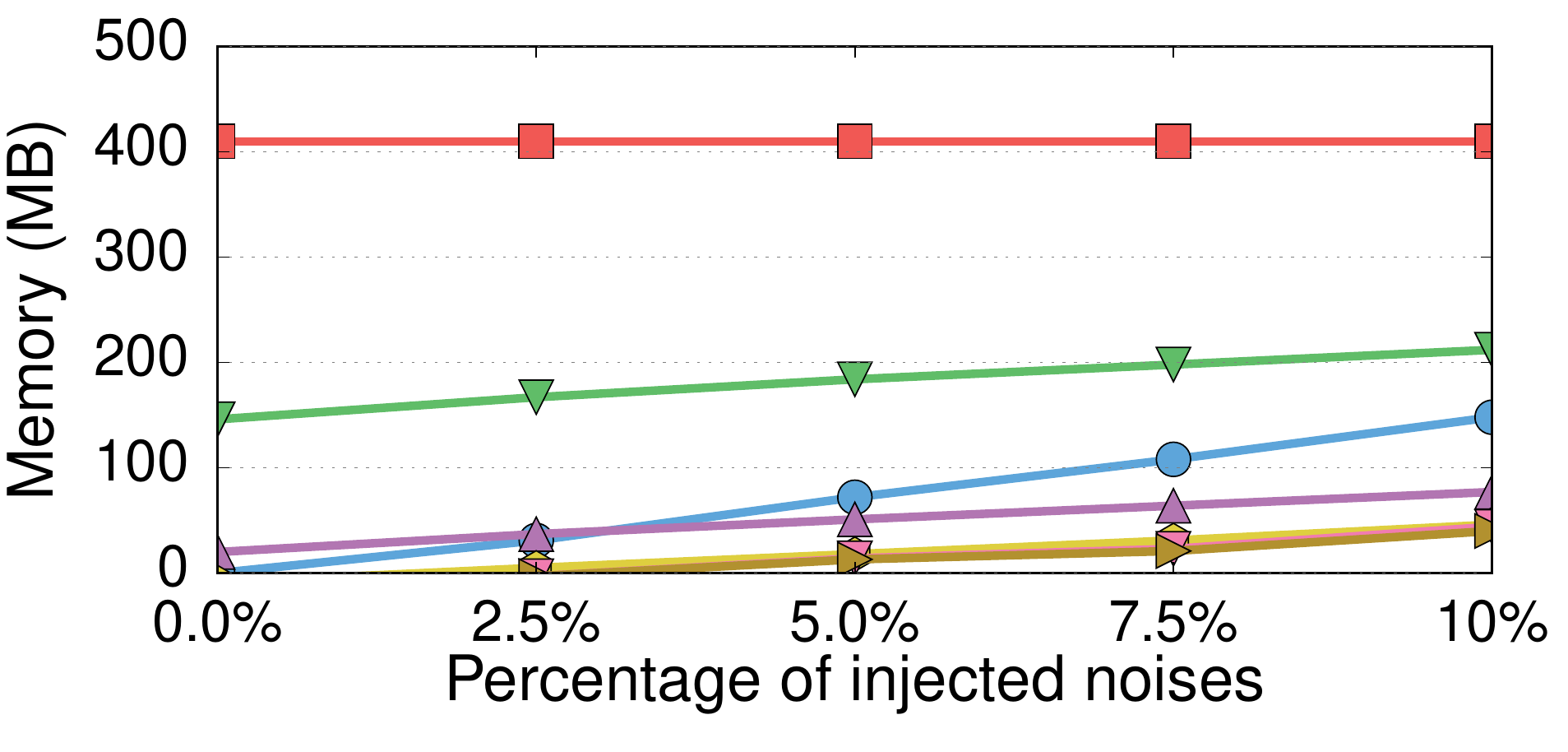}
        \label{figures:experiments:cm-range-1024-linear-memory}
    }
    \subfloat[Host bucket size = $2^{12}$]{
        \includegraphics[width=0.4\columnwidth]
            {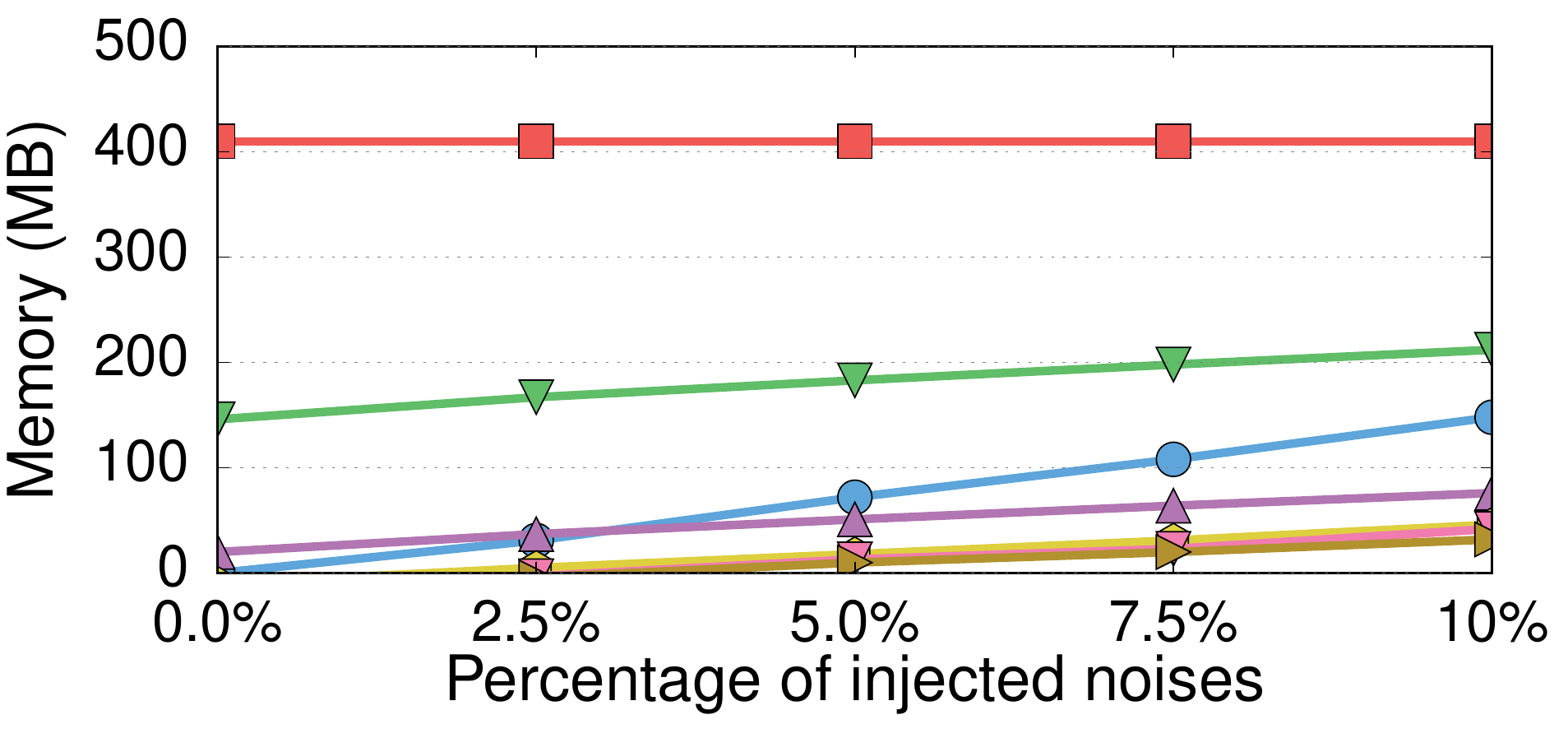}
        \label{figures:experiments:cm-range-4096-linear-memory}
    }
    \caption{
        Memory consumption with different percentage of injected noises (\textsc{Synthetic}-\textsc{Linear}).
    }
    \label{figures:experiments:cm-range-linear-memory}
\end{figure*}

\begin{figure*}[t!]
    \centering
    \subfloat[Host bucket size = 16]{
        \includegraphics[width=0.4\columnwidth]
            {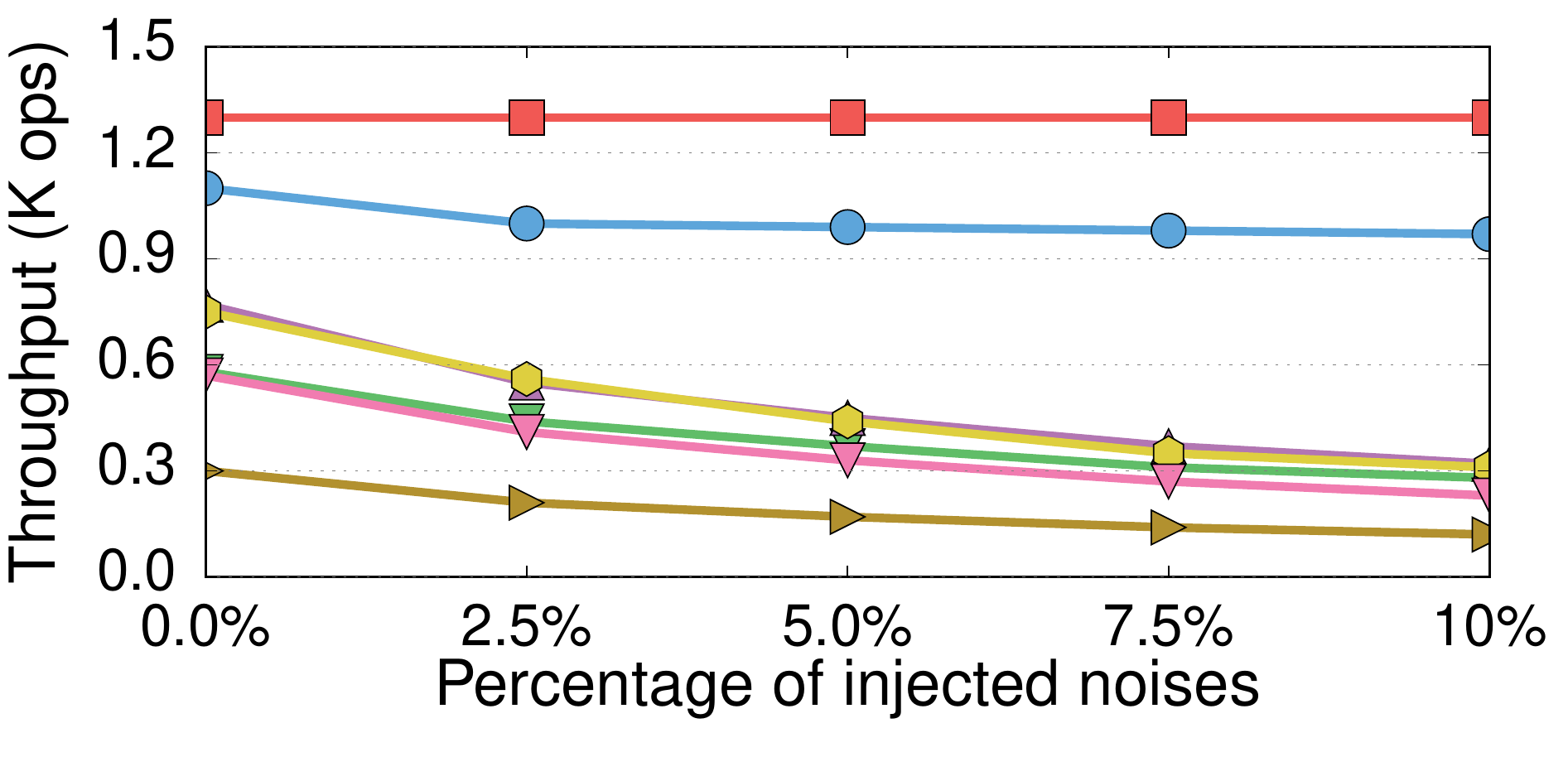}
        \label{figure:experiments:cm-range-16-sigmoid-throughput}
    }
    \subfloat[Host bucket size = $2^6$]{
        \includegraphics[width=0.4\columnwidth]
            {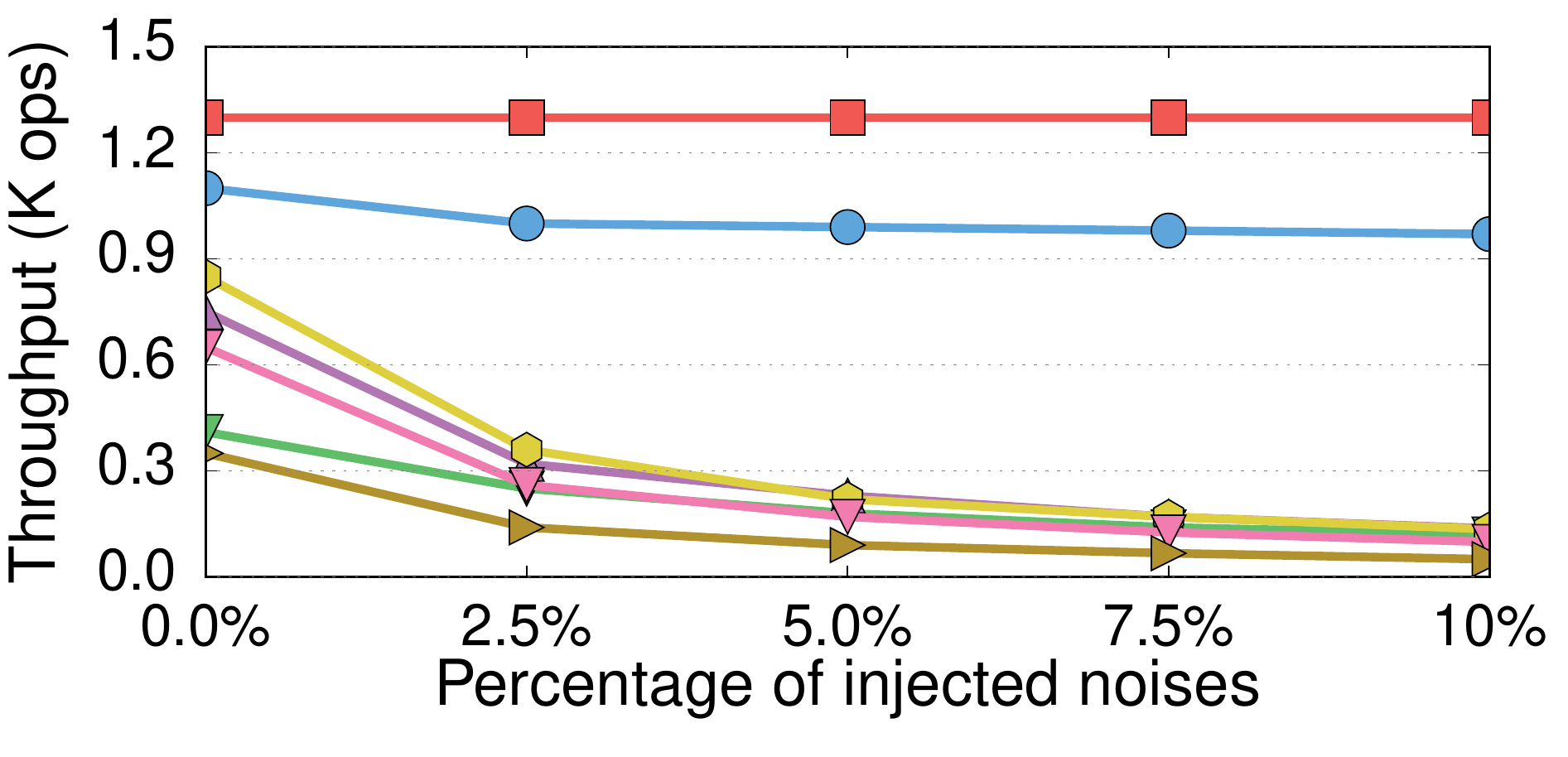}
        \label{figures:experiments:cm-range-64-sigmoid-throughput}
    }
    \subfloat[Host bucket size = $2^8$]{
        \includegraphics[width=0.4\columnwidth]
            {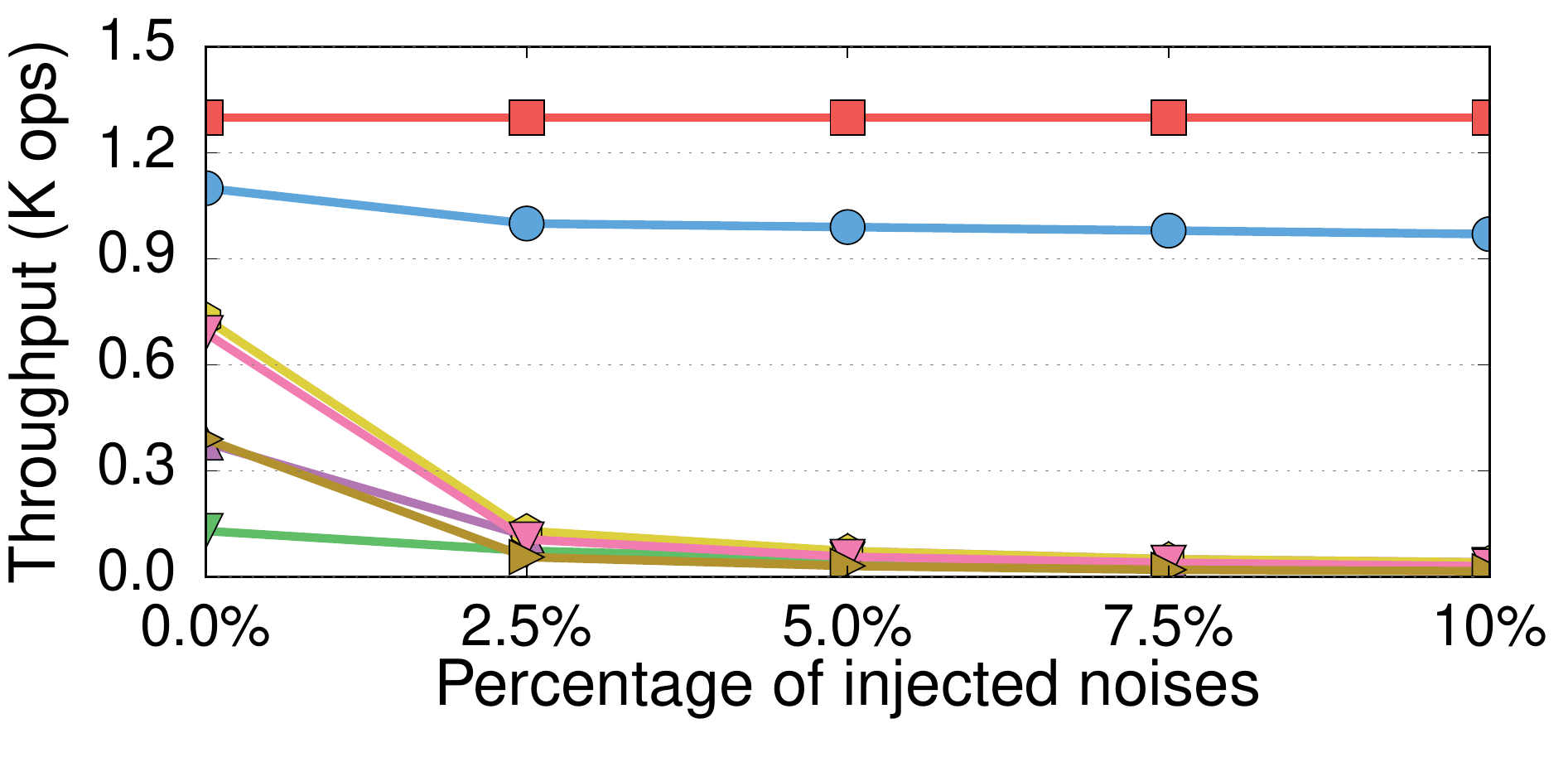}
        \label{figures:experiments:cm-range-256-sigmoid-throughput}
    }
    \subfloat[Host bucket size = $2^{10}$]{
        \includegraphics[width=0.4\columnwidth]
            {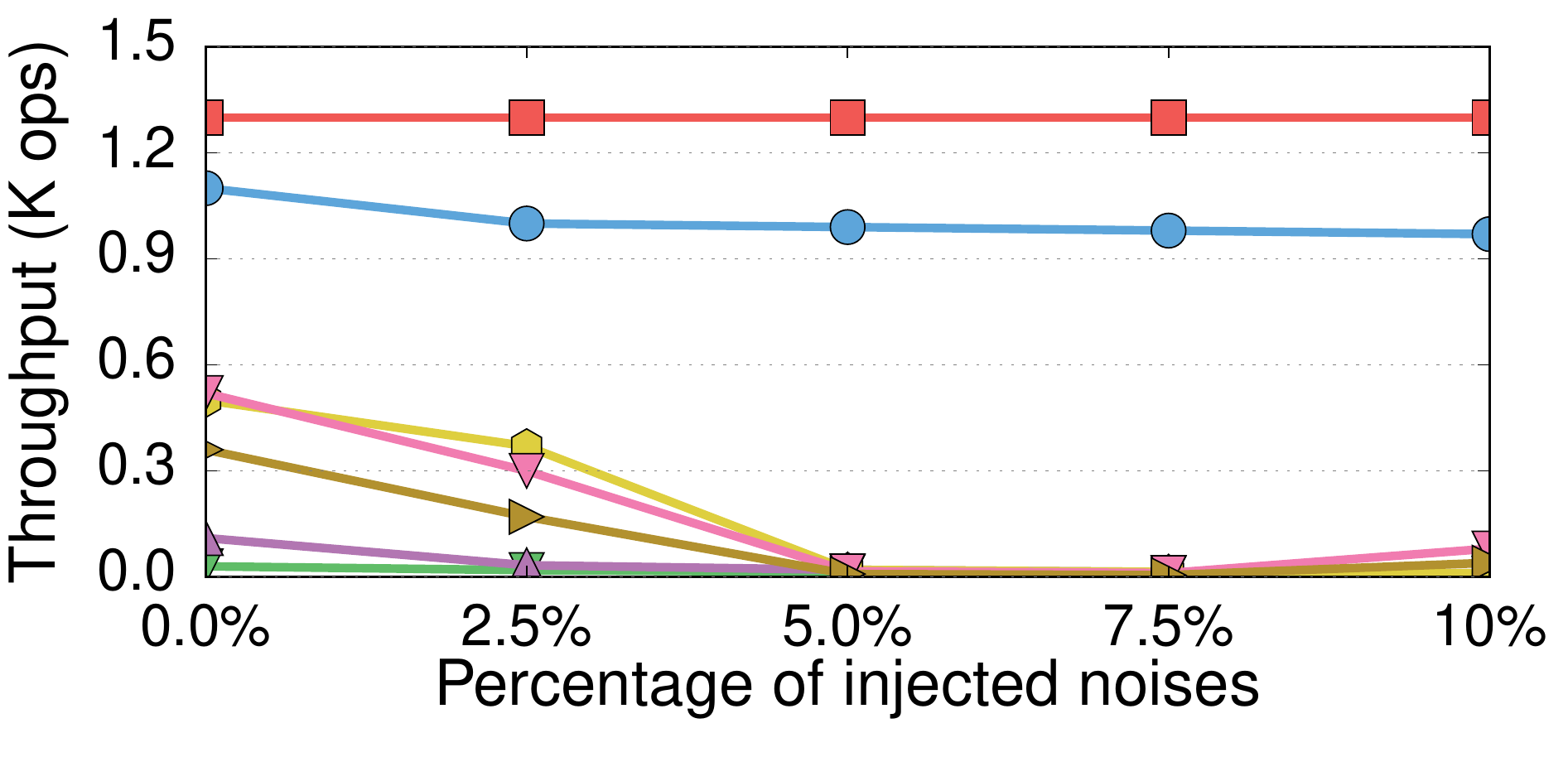}
        \label{figures:experiments:cm-range-1024-sigmoid-throughput}
    }
    \subfloat[Host bucket size = $2^{12}$]{
        \includegraphics[width=0.4\columnwidth]
            {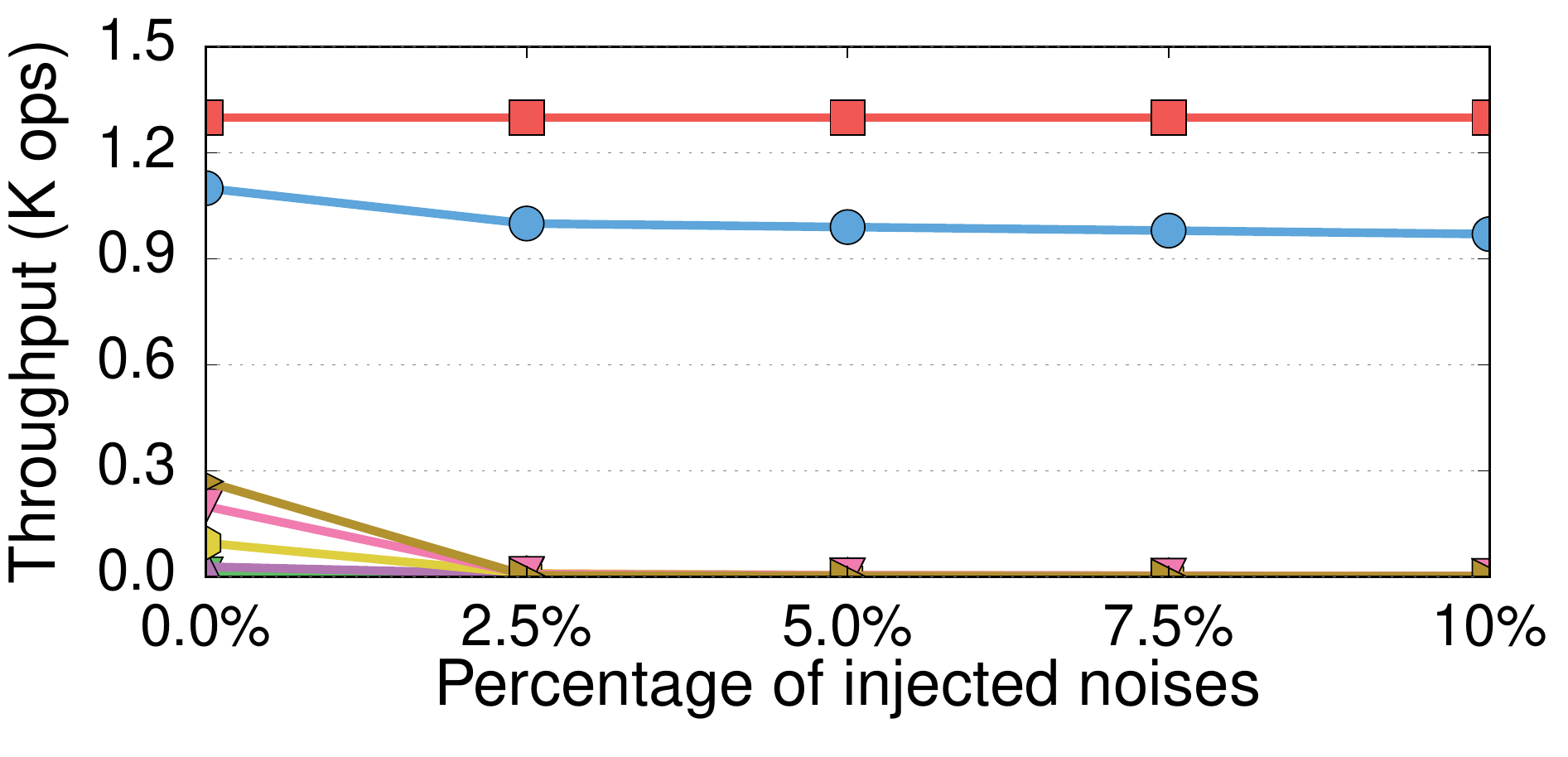}
        \label{figures:experiments:cm-range-4096-sigmoid-throughput}
    }
    \caption{
        Range lookup throughput with different percentage of injected noises (\textsc{Synthetic}-\textsc{Sigmoid}).
    }
    \label{figures:experiments:cm-range-sigmoid-throughput}
\end{figure*}

\begin{figure*}[t!]
    \centering
    \subfloat[Host bucket size = 16]{
        \includegraphics[width=0.4\columnwidth]
            {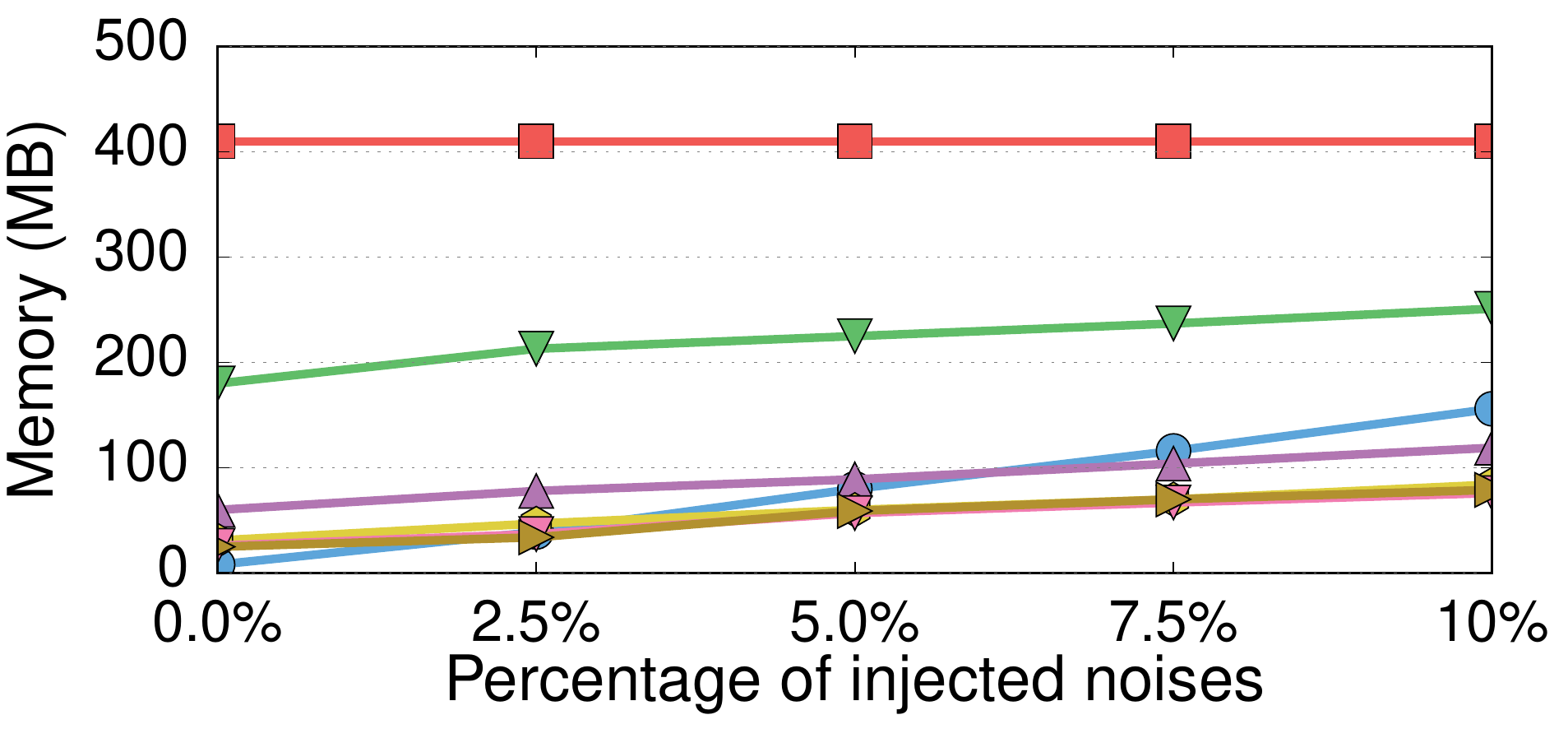}
        \label{figure:experiments:cm-range-16-sigmoid-memory}
    }
    \subfloat[Host bucket size = $2^6$]{
        \includegraphics[width=0.4\columnwidth]
            {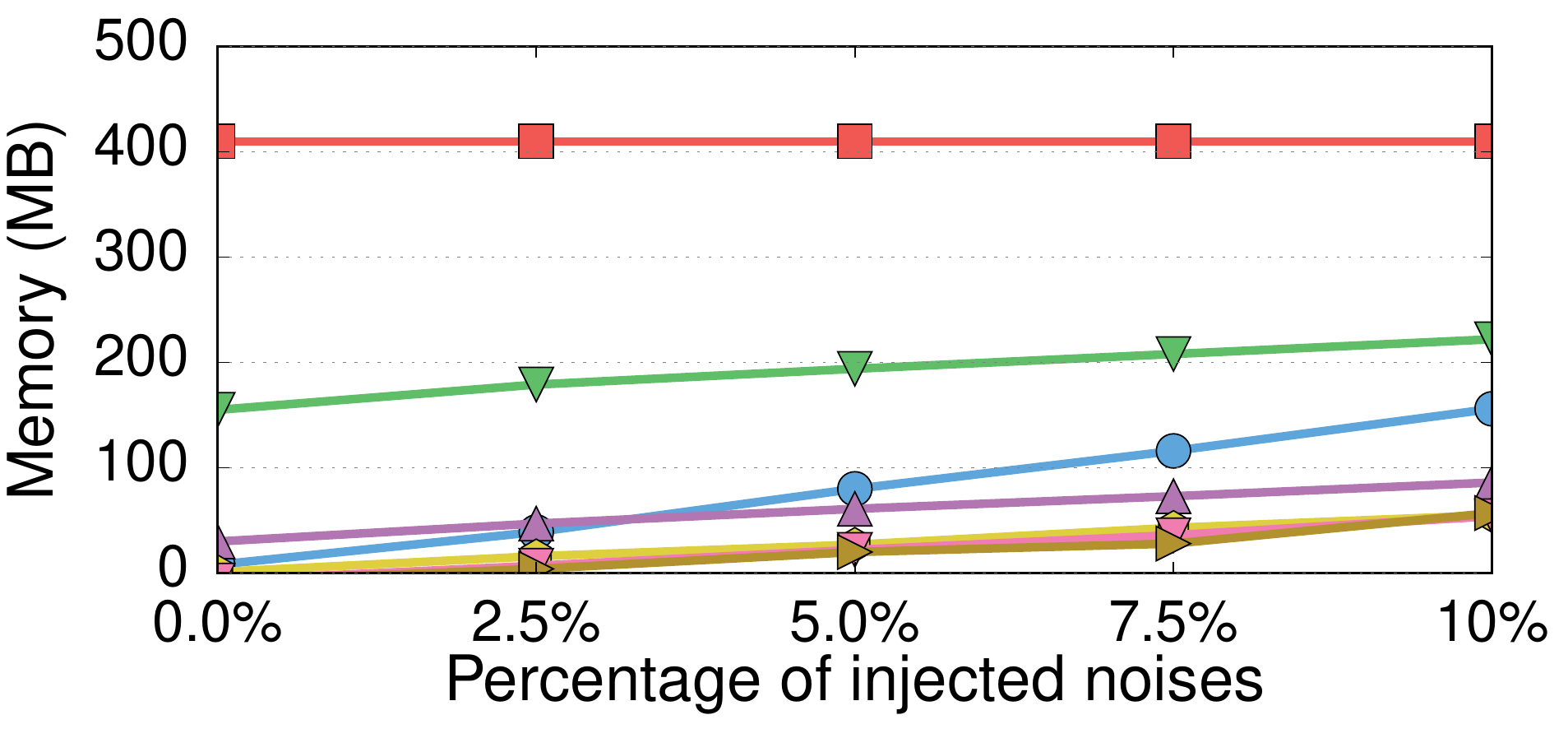}
        \label{figures:experiments:cm-range-64-sigmoid-memory}
    }
    \subfloat[Host bucket size = $2^8$]{
        \includegraphics[width=0.4\columnwidth]
            {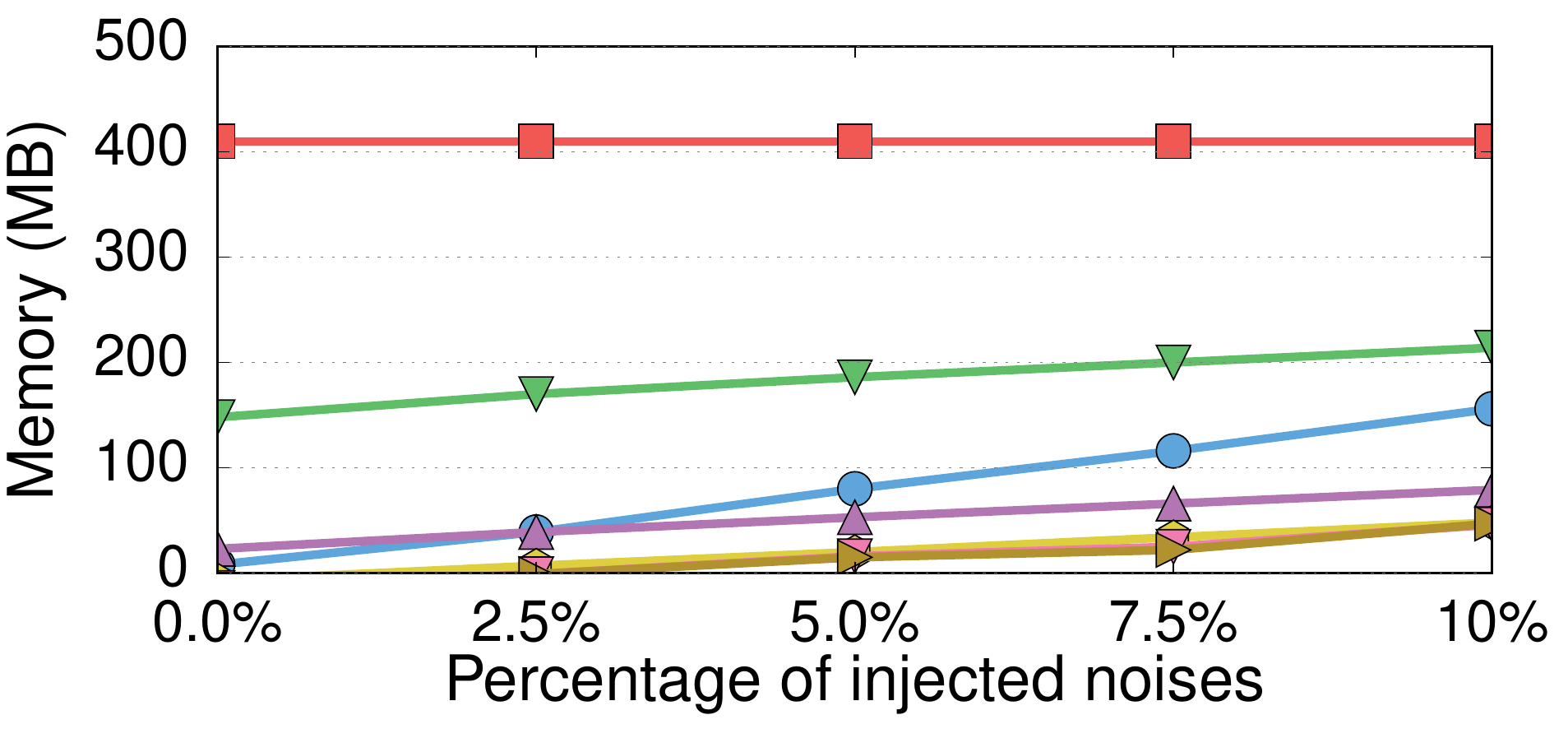}
        \label{figures:experiments:cm-range-256-sigmoid-memory}
    }
    \subfloat[Host bucket size = $2^{10}$]{
        \includegraphics[width=0.4\columnwidth]
            {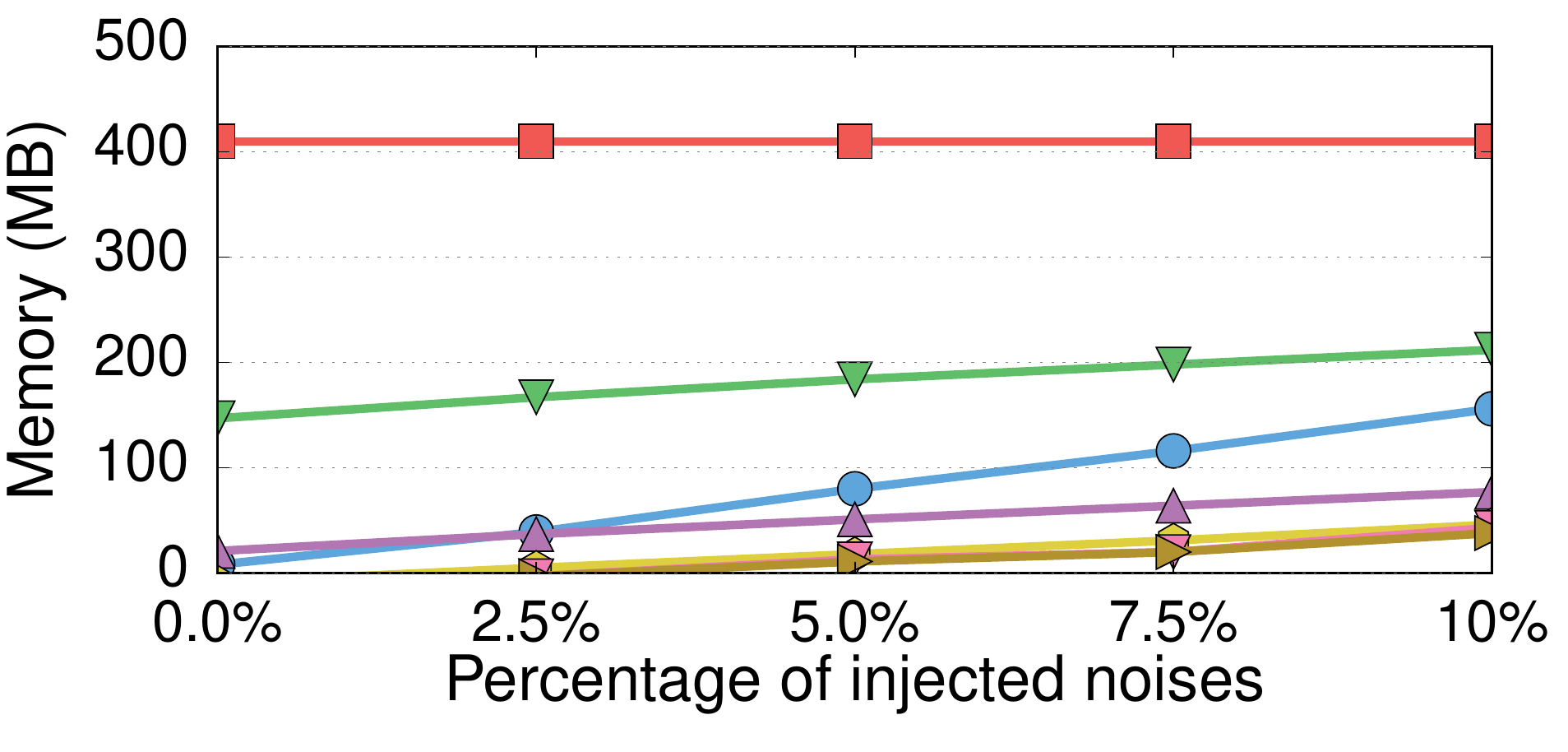}
        \label{figures:experiments:cm-range-1024-sigmoid-memory}
    }
    \subfloat[Host bucket size = $2^{12}$]{
        \includegraphics[width=0.4\columnwidth]
            {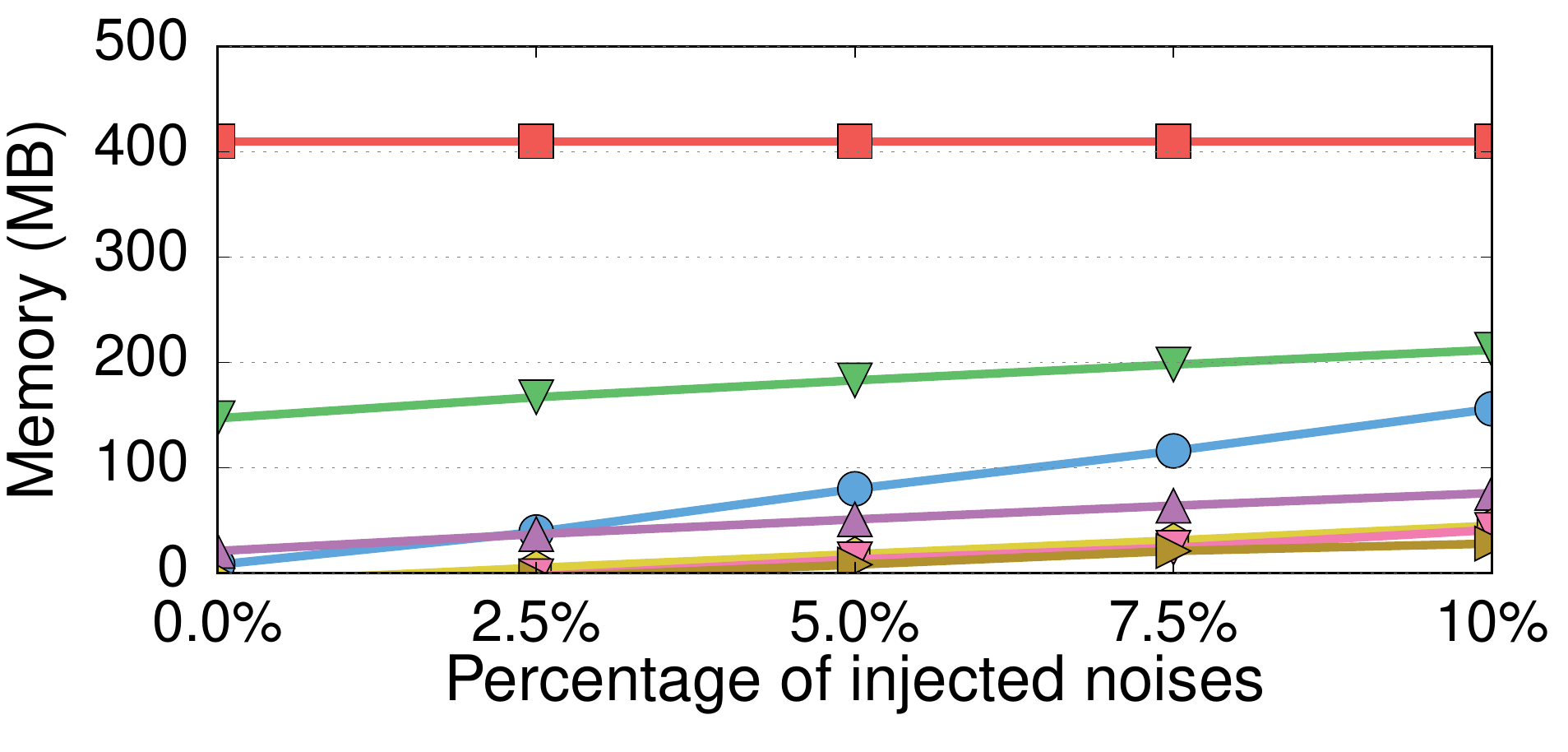}
        \label{figures:experiments:cm-range-4096-sigmoid-memory}
    }
    \caption{
        Memory consumption with different percentage of injected noises (\textsc{Synthetic}-\textsc{Sigmoid}).
    }
    \label{figures:experiments:cm-range-sigmoid-memory}
\end{figure*}

We compare \system with an existing solution, 
namely Correlation Maps (CM). 
We implemented CM faithfully based on its original paper.
Instead of implementing CM's tuning advisor, we performed parameter sweeping 
and tuned the bucket size in both target and host columns to evaluate the performance. Throughout this section, we use \textit{host bucket size} to refer to the bucket size in host column. We use \textit{CM-X} to denote CM with the bucket size in target column set to X (e.g., CM-16 means the bucket size in target column is 16).

\cref{figures:experiments:cm-range-linear-throughput} and 
\cref{figures:experiments:cm-range-linear-memory} show the 
range lookup throughput (selectivity=0.01\%) and 
memory consumption of \system, CM, and the B+-tree-based baseline solution 
using \textsc{Synthetic}-\textsc{Linear}. We change the percentage of 
injected noises from 0\% to 10\%. Since CM was designed for disk-based RDBMSs, \cite{kimura2009correlation} showed that CM usually performs better with a smaller bucket size. Now CM is adapted to in-memory databases, this does not always hold true any more. The host index look up and base table access are much faster now in memory, thus the overhead of accessing the CM structure itself plays a bigger role in the overall performance. A smaller bucket size means more buckets in CM and accordingly more overhead for accessing the CM structure. 
We also observed a compute-storage tradeoff in CM. While CM's lookup throughput 
drops with the increase of bucket size, its memory consumption is actually reduced. 
Another key observation in these figures is that CM's performance can drop 
significantly with the increase of percentage of injected noises. 
This is because CM has to maintain mappings among buckets for every single entry 
pairs in the target and host columns. Even with small amount of sparsely distributed
outliers, CM in the extreme case may have to scan the entire table to remove outliers (just consider the case where the outliers are scattered to every single bucket in the target column).
Compared to CM, \system can sustain a high throughput even when the noise 
percentage is increased to 10\%. This is because CM can identify and maintain 
outliers in leaf nodes' outlier buffers. The tradeoff is that its memory consumption
can increase. Overall, \system and CM both can reduce memory consumption 
compared to B+-tree. However, \system can achieve much better performance 
in the presence of outliers, and it saves much more memory as the correlation 
information is well captured by the tiny linear regression models.

\cref{figures:experiments:cm-range-sigmoid-throughput} and 
\cref{figures:experiments:cm-range-sigmoid-memory} further show 
the experiment results with \textsc{Synthetic}-\textsc{Sigmoid}.
We also observed similar results in this set of experiments. The only difference 
is that \system needs to spend more memory to capture correlations. 

\end{document}